\documentclass[useAMS, usenatbib]{mn2e}
\usepackage{amsmath}
\usepackage{amssymb}
\usepackage{graphicx}
\usepackage{subfigure}
\usepackage[letterpaper,margin=0.65in]{geometry}
\usepackage{natbib}
\usepackage{hyperref}
\usepackage{enumitem}
\makeatletter

\renewcommand{\@thesubfigure}{\alph{subfigure}\hskip\subfiglabelskip}
\renewcommand{\@@thesubfigure}{(\alph{subfigure})}
\makeatother

\title{Pre-Processing and Post-Processing in Group-Cluster Mergers}
\author[R. Vijayaraghavan \& P. M. Ricker]
       {R. Vijayaraghavan$^1$\thanks{E-mail: vijayar2@illinois.edu} \& P. M. Ricker$^1$ \\
        $^1$Department of Astronomy, University of Illinois, Urbana, IL 61801}

\begin{document}

\date{\today}
\maketitle

\begin{abstract}
  Galaxies in clusters are more likely to be of early type and to have lower star formation rates than galaxies in the field. Recent observations and simulations suggest that cluster galaxies may be `pre-processed' by group or filament environments and that galaxies that fall into a cluster as part of a larger group can stay coherent within the cluster for up to one orbital period (`post-processing'). We investigate these ideas by means of a cosmological $N$-body simulation and idealized $N$-body plus hydrodynamics simulations of a group-cluster merger. We find that group environments can contribute significantly to galaxy pre-processing by means of enhanced galaxy-galaxy merger rates, removal of galaxies' hot halo gas by ram pressure stripping, and tidal truncation of their galaxies. Tidal distortion of the group during infall does not contribute to pre-processing. Post-processing is also shown to be effective: galaxy-galaxy collisions are enhanced during a group's pericentric passage within a cluster, the merger shock enhances the ram pressure on group and cluster galaxies, and an increase in local density during the merger leads to greater galactic tidal truncation. 
\end{abstract}

\section{Introduction}
\label{sec:intro}

Galaxies in cluster environments are more likely to be elliptical or spheroidal compared to field galaxies (\citealt{Dressler80}, \citealt{Postman84}) and to have systematically older (or redder) stellar populations and lower star formation rates. These phenomena, respectively known as morphological segregation and star formation quenching, are observed in both central and satellite galaxies of massive halos. They are widely interpreted as being due to the interaction of galaxies with their environments.

Several environmental processes that transform galaxies in high-density environments have been identified, either observationally or in simulations. These include galaxy-galaxy mergers (\citealt{Richstone76}, \citealt{Barnes92}, \citealt{Gnedin03a}); galaxy harassment, or repeated high-speed encounters between galaxies (\citealt{Moore96}, \citealt{Moore99}, \citealt{Gnedin03a}, \citealt{Mastropietro05}); tidal stripping by the host group or cluster's gravitational potential (\citealt{Gnedin03b}); ram pressure stripping of cold gas by the hot gas of the diffuse intra-cluster or intra-group medium (ICM or IGM) (\citealt{Gunn72}); strangulation, or the ram pressure-driven removal of diffuse galactic halo gas, reducing the fuel available for later star formation (\citealt{Larson80}, \citealt{McCarthy08}); and mechanical and thermal feedback due to active galactic nuclei (AGN) whose feeding rate can be influenced by environmental effects (\citealt{Sijacki07}, \citealt{Dubois13}). The rates of these processes depend on the velocities of galaxies relative to their environments, the ambient gas and dark matter density, and/or the external potential gradient. Thus the degree to which they have had time to affect any given group or cluster galaxy depends on where that galaxy originated.

While observationally the link between galaxy properties and cluster membership is well-established, it is not as clear how much of this correlation is due to processes operating on galaxies within their observed hosts versus previous environments they may have experienced. Quantitative investigation of this question is necessary in the light of several recent studies that have shown that galaxies outside the virial radii of clusters, where effects due to the cluster environment are presumably still weak, nevertheless show modification compared to field galaxies. \citet{Lewis02} studied a sample of galaxies in the fields of 17 known clusters at redshifts $0.05 < z < 0.1$ in the 2dFGRS and found that star formation is suppressed relative to the field at distances up to 3 cluster virial radii. \citet{Gomez03}, using SDSS data,  found that the star formation rate of cluster galaxies starts to differ significantly from that of field galaxies at cluster-centric distances of 2--3 virial radii. \citet{Lu12}, using optical and UV data from the CFHTLS and GALEX, found that the fraction of galaxies with detectable star formation is lower than the field value at distances up to 7 Mpc from cluster centers. \citet{Rasmussen12}, in a study of 23 optically selected and spectroscopically confirmed galaxy groups, found that star formation is suppressed even in galaxies out to $2R_{200}$\footnote{We define $R_{200}$ as the radius within which the mean density of the group or cluster halo is equal to 200 times $\rho_{\rm crit}$, the critical density of the Universe.} from the centers of groups, similar to the trends observed in massive clusters. The above observations suggest that physical processes that suppress star formation in galaxies in massive halos begin to act before these galaxies fall into a cluster, and that these galaxies may therefore undergo some degree of pre-processing outside the cluster environment. 

In the current cold dark matter (CDM) paradigm of hierarchical structure formation, smaller galaxies and their dark matter halos tend to form earlier in the history of the Universe, then merge to form larger groups ($\sim 10^{13} - 10^{14} \mbox{M}_{\odot}$) and clusters ($\gtrsim 10^{14} \mbox{M}_{\odot}$) of galaxies. Thus even before they become cluster members many galaxies experience high-density environments, either as members of smaller groups or by forming within large-scale filaments. We use the term `pre-processing' to refer to any of the transformation processes described above when operating in the context of high-density environments experienced before a galaxy's infall into a cluster. However, the concept originally referred specifically to the fact that outright mergers of galaxies (as opposed to high-speed, non-merger collisions) should be far more common in host systems with low velocity dispersions ($\lesssim 400$~km~s$^{-1}$), namely groups, than in clusters. This idea is supported observationally by the fact that galaxy populations in groups have morphological type and star formation rate distributions intermediate between those of clusters and the field (\citealt{Zabludoff98}, \citealt{Balogh00}, \citealt{Hoyle12}). 

Pre-processing only makes sense, of course, if a significant fraction of cluster galaxies actually experiences high-density environments prior to cluster infall. Cosmological $N$-body simulations can directly provide the fraction of cluster subhalos that were previously subhalos of groups. For example, \citet{Berrier09} and \citet{White10} found using $\Lambda$CDM $N$-body simulations that $\sim 30$\% of all infalling cluster subhalos were members of larger halos on infall. However, determining the fraction of cluster {\it galaxies} that were previously group members requires association of galaxies with halos, a step that is still model-dependent. \citet{McGee09} studied halo merger trees and a semi-analytic galaxy catalog constructed from the Millennium Simulation (\citealt{Springel05}) and found that $\sim 40 \%$ of galaxies in a $10^{14.5} h^{-1}  ~\mbox{M}_{\odot}$ cluster at $z = 0$ accreted from group-scale halos of mass greater than $10^{13}  ~\mbox{M}_{\odot}$.  \citet{DeLucia12} used $N$-body merger trees (also from the Millennium Simulation) together with a different semi-analytic model and found that $\sim 44\%$ of galaxies in clusters with stellar mass greater than $10^{11}  ~\mbox{M}_{\odot}$ are accreted as central or satellite galaxies of halos more massive than $10^{13}  ~\mbox{M}_{\odot}$, and about half of all galaxies of stellar mass lower than $10^{11} ~\mbox{M}_{\odot}$ are accreted as satellite galaxies of more massive halos. Overall, these simulations suggest that up to half of all cluster galaxies could have been subject to transformation processes in group environments. 

Galaxies that do fall into clusters as members of groups are not immediately dissociated from each other and virialized (\citealt{White10}, \citealt{Cohn12}). They remain correlated in velocity and position for some time (as much as several Gyr) after infall. Observational evidence that group-scale subhalos persist inside clusters is provided by optical detections of galaxy substructure in position (\citealt{Fitchett87}) and velocity space (e.g., \citealt{Dressler88}, \citealt{Aguerri10}, \citealt{Einasto10}) as well as gravitational lensing (e.g., \citealt{Okabe10}, \citealt{Richard10}, \citealt{Coe10}). These substructures also contribute to detectable features in the hot gas distribution (e.g., \citealt{Markevitch00}, \citealt{Kraft06}, \citealt{OHara06}, \citealt{Andrade-Santos13}). Thus interaction rates computed assuming a virialized galaxy population should not immediately apply to these galaxies. Moreover, dark matter and gas associated with an infalling group interact with those of the cluster and thus affect the local environment experienced by group member galaxies. We collectively refer to these effects as `post-processing.'

In this paper, we focus on the dynamics of groups that merge with clusters to qualitatively and quantitatively understand the importance of pre-processing and post-processing. We study some of the physical processes that affect galaxies within a group environment before merging with a cluster as well as some processes that are a result of the group-cluster merger itself. We concentrate on the effect that the large-scale group and cluster environments have on the dynamics of model galaxies (or galaxy particles, which are randomly chosen dark matter particles within the cluster whose orbits are assigned to model galaxies). Initially we study the merger of a group and cluster in an $N$-body cosmological simulation; we then perform an idealized resimulation of the merger including adiabatic gasdynamics. We also perform idealized resimulations of the group and cluster in isolation and compare the results of these simulations to the merger in order to isolate the effects of the merger. We study the velocity coherence of the merging group's bound and stripped components, the impact of this coherence on group and cluster interaction rates, the evolution of ram pressure due to the merger and the importance of ram pressure on stripping of the hot gaseous halos of model galaxies, and finally, the tidal truncation of galaxy subhalos due to the gravitational fields of the group and cluster. 

Our models contribute toward understanding the distinct effects that a group-cluster merger has on the evolution of the galactic constituents of the group and cluster. Idealized cluster mergers have been studied in the literature before but have mainly focused on cluster-scale effects (e.g., \citealt{Roettiger97}, \citealt{Ricker01}, \citealt{Poole06}, \citealt{Poole08}, \citealt{ZuHone11}). \citet{Bekki99} studied the dynamical evolution of a gas-rich spiral galaxy in a group-cluster merger and found that the merger can trigger a starburst for a brief period within the galaxy. However, this paper did not include any other galaxy transformation mechanisms, particularly gas removal and interactions with other galaxies. The effects of a cluster environment on the evolution of individual galaxies have also been studied using numerical simulations; however, these simulations consider the effect of a relaxed cluster in equilibrium on its galaxies (\citealt{Moore98}, \citealt{Gnedin03a}, \citealt{Roediger06}). \citet{White10} and \citet{Cohn12} studied the coherence of infalling galaxy subgroups within clusters in the context of a cosmological simulation to investigate the detectability of this substructure, but they did not consider the impact of such a process on the evolution of galaxies in the merging group and cluster.

The paper is organized as follows. In \S\ref{sec:method}, we describe the properties of the cosmological simulation and the initial conditions of the idealized resimulations. In \S\ref{sec:results-cosmo}, we describe some of the interesting results from studying a group-cluster merger in a cosmological context that motivated us to perform an idealized resimulation. In \S\ref{sec:results-ideal}, we describe the results of our idealized merger resimulation and compare these to simulations of the isolated group and cluster. In \S\ref{sec:discussion}, we discuss the implications of these results, including understanding the contribution of pre-processing towards cluster galaxy evolution and the impact of the merger itself on the evolution of group and cluster galaxies. Finally, we summarize our important results in \S\ref{sec:conclusions}. 

\section{Method}
\label{sec:method}
The simulations described in this paper were run using FLASH 3 (\citealt{Fryxell00}, \citealt{Dubey08}), a parallel $N$-body plus adaptive mesh refinement (AMR) hydrodynamics code. In FLASH 3 a direct multigrid solver (\citealt{Ricker08}) is used to calculate the gravitational potential on the mesh, and cloud-in-cell (CIC) mapping is used to interpolate between particles and the mesh. For Euler's equations we use the directionally split piecewise parabolic method (PPM, \citealt{Colella84}). AMR is implemented using PARAMESH 4 (\citealt{MacNeice00}). The same code is used for both cosmological and idealized simulations except for the initial and boundary conditions setups and the use of comoving or proper coordinates as appropriate.

\subsection{Cosmological simulation}
Details of our cosmological simulation appear in \citet{Sutter10}; here we summarize the main features. This was a uniform-mesh dark matter-only simulation in a cubic $50 h^{-1}$ Mpc box with $512^3$ particles and $1024$ zones per side. It used cosmological parameters $\Omega_{M,0} = 0.238$, $\Omega_{\Lambda,0} = 0.762$, $H_0 = 100h = 73.0$~km~s$^{-1}$~Mpc$^{-1}$, and $\sigma_8 = 0.74$. Each particle had a mass of $6.12 \times 10^7 h^{-1}  ~\mbox{M}_{\odot}$, and the spatial resolution in the box was $48.8 h^{-1}$ kpc. 

We identified halos, subhalos, and subsubhalos, and generated merger trees within the cosmological simulation using the AMIGA halo finder (AHF, \citealt{Gill04} and \citealt{Knollmann09}). AHF uses a recursively refined grid to identify density peaks in the simulation box and generates a tree connecting parent and child halos. In addition, it iteratively removes particles that are not gravitationally bound to a density peak and calculates halo properties based on the remaining particles. The virial radius of each halo is taken to be the halo's $R_{200}$. Each halo is required to have at least 40 particles, corresponding to a minimum halo mass of $M_{\rm h,min} = 2.4 \times 10^9 h^{-1}  ~\mbox{M}_{\odot}$. Given the spatial resolution used in this run, however, we expect our halo statistics to be complete only for halos containing more than $\sim$300 particles, or $1.8 \times 10^{10} h^{-1} ~\mbox{M}_{\odot}$ (\citealt{Lukic07}).

\subsection{Idealized resimulation}

\begin{table*}
\begin{center}
  \begin{tabular}{c c c c c c c c c c}
    \hline 
    Halo & $M_{200} (\mbox{M}_{\odot}$) & $R_{200}$ (kpc) & $N_{\rm part}$ & $r_{\rm s}$ (kpc) & $\rho_{\rm s}$ ($\mbox{M}_{\odot}$ kpc$^{-3}$) & $f_{\rm g}$ & $S_0$ (keV cm$^2$) & $S_1$ (keV cm$^2$)\\
    \hline
    Cluster & $1.17 \times 10^{14}$ & $880$ & 1,000,000 & 186 & $1.6 \times 10^6$  & 0.091 & 4.8 & 90.0\\
    Group & $3.23 \times 10^{13}$ & $551$ & 199,321 & 108 & $2 \times 10^6$ & 0.07 & 2.0 & 40.0\\
    \hline 
  \end{tabular}
  \caption{Group and cluster parameters in the idealized merger resimulation. \label{table1}} 
\end{center}
\end{table*}

We performed an idealized resimulation of a group-cluster merger observed in our cosmological simulation beginning around redshift $z=0.2$ (the cosmological merger is described in detail in \S\ref{sec:results-cosmo}). To constrain the resimulation, we used the mass $M_{200}$ within radius $R_{200}$ for each halo at $z=0.2$, where
\begin{equation}
  M_{200} = \frac{4}{3}\pi(200\rho_{\rm crit})R_{200}^3 .
\end{equation}
Parameters used in the resimulation are summarized in Table~\ref{table1}. Because the cosmological merger had a small impact parameter, we treated the idealized case as head-on.

We used the cluster initialization technique developed by \citet{ZuHone11} to construct the initial conditions for the resimulation. The group and cluster were initialized as spherically symmetric dark matter halos in equilibrium with a diffuse gas component. The total density profile of each halo is specified using a Navarro-Frenk-White profile (NFW, \citealt{Navarro97}) for $r \leq R_{200}$ with an exponential fall-off at $r > R_{200}$:
\begin{equation}
  \rho_{\rm tot}(r) = \begin{cases}
    \frac{\rho_{\rm s}}{r/r_{\rm s} (1 + r/r_{\rm s})^2} & r \leq R_{200},\\
    \frac{\rho_{\rm s}}{c_{200}(1 + c_{200})^2}\left(\frac{r}{R_{200}}\right)^{\kappa} \exp\left(-\frac{r-R_{200}}{r_{\rm decay}}\right)   & r > R_{200}.
  \end{cases}
\end{equation}
Here $r_{\rm decay} = 0.1R_{200}$, and $\kappa$ is chosen such that the density and the slope of the density profile are continuous at $R_{200}$:
\begin{equation}
  \kappa = \frac{R_{200}}{r_{\rm decay}} - \frac{3c_{200}+1}{1+c_{200}} .
\end{equation}
$c_{200}$\footnote{$c_{200}$ is determined from the concentration-mass relationship in \citet{Prada12}. This relationship exhibits a $\sim 20\%$ discrepancy with other relations (see \citealt{Kwan13} for further details), but given the large scatter in observed c-M relations, the discrepancy does not significantly affect our conclusions on the importance of group and cluster environments during a merger in galaxy evolution} is the concentration parameter, $r_{\rm s}$ is the NFW scale radius, and $\rho_{\rm s}$ is the NFW scale density. These parameters are related via
\begin{align}
  r_{\rm s} &= \frac{R_{200}}{c_{200}} \\
  \rho_{\rm s} &= \frac{200}{3}\rho_{\rm crit} \frac{c_{200}^3} {\log(1 + c_{200}) - c_{200}/(1 + c_{200})}.
\end{align}

The gas fraction of each halo within its $R_{200}$, $f_{\rm g}$, is determined using the observed relation (\citealt{Vikhlinin09}):  
\begin{equation}
  f_{\rm g}(h/0.72)^{1.5} = 0.125 + 0.037\log_{10} (M/10^{15} ~\mbox{M}_{\odot}).
\end{equation}
The gas is constrained to be in hydrostatic equilibrium with the halo's total gravitational potential $\Phi$ using
\begin{equation}
  \frac{dP}{dr} = -\rho_{\rm g}\frac{d\Phi}{dr},
\end{equation}
where the gas pressure, $P$, the gas density, $\rho_{\rm g}$, and the temperature, $T$, are related in the usual ideal gas form,
\begin{equation}
  P = \frac{k_B}{\mu m_p}\rho_{\rm g} T ,
\end{equation}
with $\mu\approx 0.59$ for a fully ionized hydrogen plus helium plasma with cosmic abundances.
The corresponding adiabatic index is $\gamma = 5/3$.
The equation of hydrostatic equilibrium is solved to initialize the gas density profile, assuming that the cluster and group are relaxed, cool-core systems\footnote{The cool core assumption is justified in \citealt{ZuHone11} and references therein.}, with small core entropies  and a given radial entropy profile $S(r) \equiv k_B T(r) n_e(r)^{-2/3}$, where $n_e$ is the electron number density. The entropy profile of each halo is based on observations by \citet{Cavagnolo09} and is of the form
\begin{equation}
  S(r) = S_0 + S_1\left(\frac{r}{R_{200}}\right)^{1.1}.
\end{equation}
We also impose the condition that the `virial temperature,' $T(R_{200})$, is
\begin{equation}
  T(R_{200}) = \frac{1}{2}T_{200} ,
\end{equation}
where $T_{200}$ is given by
\begin{equation}
  k_B T_{200} \equiv \frac{G M_{200} \mu m_p}{2 R_{200}} .
\end{equation}

The dark matter density profile, $\rho_{\rm DM} = \rho_{\rm tot} - \rho_{\rm g}$, determines the distribution of dark matter particles. We use the procedure outlined in \citet{Kazantzidis04} to initialize the positions and velocities of dark matter particles. For each particle we draw a uniform random deviate $u$ in $[0,1)$ and choose the particle's halo-centric radius, $r$, by inverting the function 
\begin{equation}
u = \frac{\int_0^r \rho_{\rm DM}(r) r^2 dr}{\int_0^\infty \rho_{\rm DM}(r) r^2 dr} .
\end{equation}
To calculate particle velocities, we use the Eddington formula for the distribution function (\citealt{Eddington16}, \citealt{Binney08}):
\begin{equation}
  f(\mathcal{E}) = \frac{1}{\sqrt{8}\pi^2}\left[\int_0^{\mathcal{E}} \frac{d^2\rho}{d\Psi^2}\frac{d\Psi}{\sqrt{\mathcal{E} - \Psi}} + \frac{1}{\sqrt{\mathcal{E}}}\left(\frac{d\rho}{d\Psi}\right)_{\Psi=0} \right].
\end{equation}
Here, $\Psi = -\Phi$ is the relative potential of the particle and $\mathcal{E} = \Psi - \frac{1}{2}v^2$ is the relative energy. Using an acceptance-rejection technique, we choose random particle speeds $v$ given $f(\mathcal{E})$. 

We ran our idealized merger resimulation in a cubic box of side 6.48 Mpc with a minimum of 4 levels of refinement (corresponding to a minimum resolution of 101.3 kpc) and a maximum of 8 levels of refinement (corresponding to a maximum resolution of 6.33 kpc). The simulation box had outflow (zero-gradient) boundary conditions, so matter was allowed to leave the system. Over the course of the simulation, $0.87 \%$ of the total mass was lost through these boundaries.

In addition to a merger resimulation, we also performed simulations of the group and cluster at rest in isolation within the same simulation box and with the same refinement criteria. These isolated simulations served two purposes. First, they enabled us to check the stability of our initial conditions: these halos should evolve quiescently and retain their dark matter and fluid profiles for many dynamic timescales in the absence of processes like mergers and cooling. Second, they enabled us to determine the effects of the merger itself on the rates of different galaxy transformation processes. The isolated runs and merger resimulation were each run for a total of 6.34 Gyr, corresponding to 2.7 dynamical times.

\section{Results: A Group-Cluster Merger in a Cosmological Simulation}
\label{sec:results-cosmo}
\subsection{The merger}
\begin{figure*}  
  \begin{center}
    \includegraphics[width=6in]{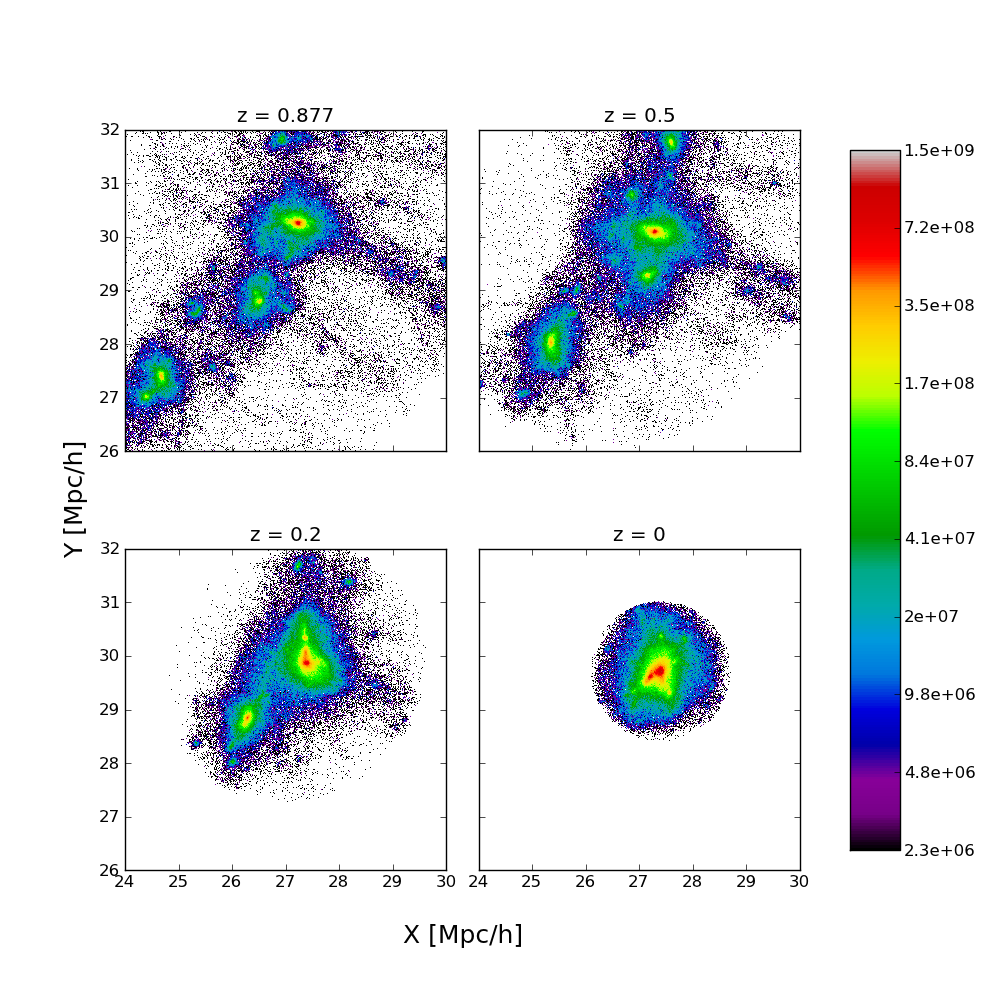}
    \caption{Projected mass densities of the group and cluster during the cosmological merger, in units of $ ~\mbox{M}_{\odot}$ kpc$^{-2}$.
    \label{fig:fig1}}
  \end{center}  
\end{figure*}

We identified two cluster-sized halos ($M > 10^{14}  ~\mbox{M}_{\odot}$) in the cosmological simulation box at $z = 0$. One of these clusters (of mass $M_{\rm c} = 1.2 \times 10^{14}  ~\mbox{M}_{\odot}$) merged with a group-sized halo (of mass $M_{\rm g} = 3.2 \times 10^{13}  ~\mbox{M}_{\odot}$) beginning at $z = 0.2$. The projection of the merger axis onto the simulation volume's $xy$ plane forms a nearly 45 degree angle with the $x$ and $y$ axes. The progress of this merger, projected into the $xy$ plane, is seen in Figure~\ref{fig:fig1}. This figure shows the two-dimensional surface density of all the redshift 0 cluster particles at three earlier redshifts: $z = 0.2, 0.5,$ and 0.877. When the group and cluster halo centers are separated by $\sim$ 3.5~$h^{-1}$~Mpc at $z = 0.5$, the group begins to appear tidally distorted. The group's `stretching' increases as it falls into the cluster along an overdense filament and is affected by the cluster's tidal field. At $z = 0.2$, shortly after the two halos' virialized regions have begun to overlap, the group has developed a relative velocity of $677$ km s$^{-1}$ with respect to the cluster; its approach is nearly head-on. The group's first pericentric passage is at $z \sim 0$, when we see two distinct density peaks near the cluster center. We also see smaller density peaks correspoding to smaller subhalos that fall into the cluster. 

\subsection{Group subhalos}
\begin{figure*} 
  \begin{center}
    \subfigure[$z=0.5$]{\includegraphics[width=2.3in]{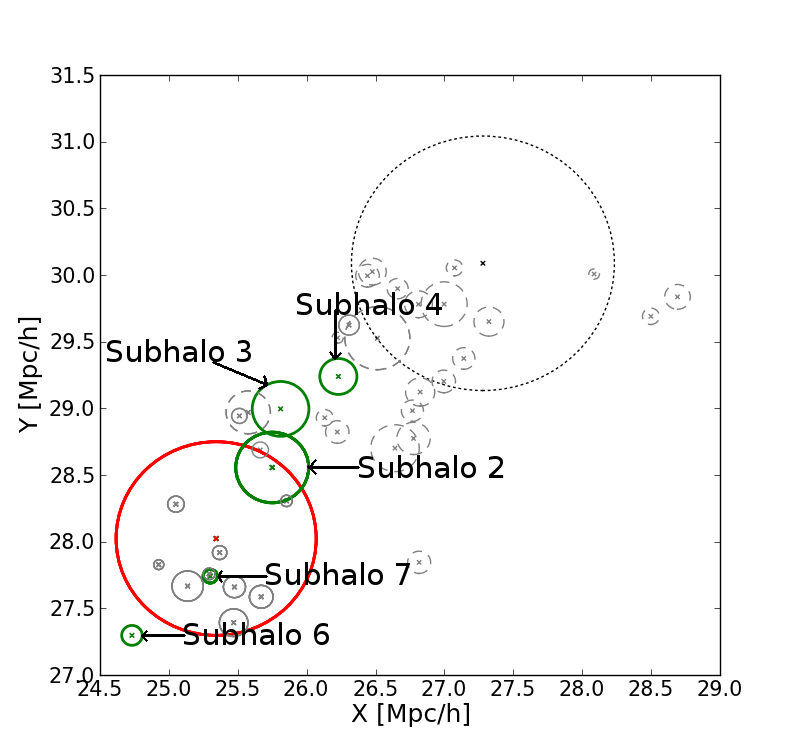}\label{fig:fig2a}}
    \subfigure[$z=0.2$]{\includegraphics[width=2.3in]{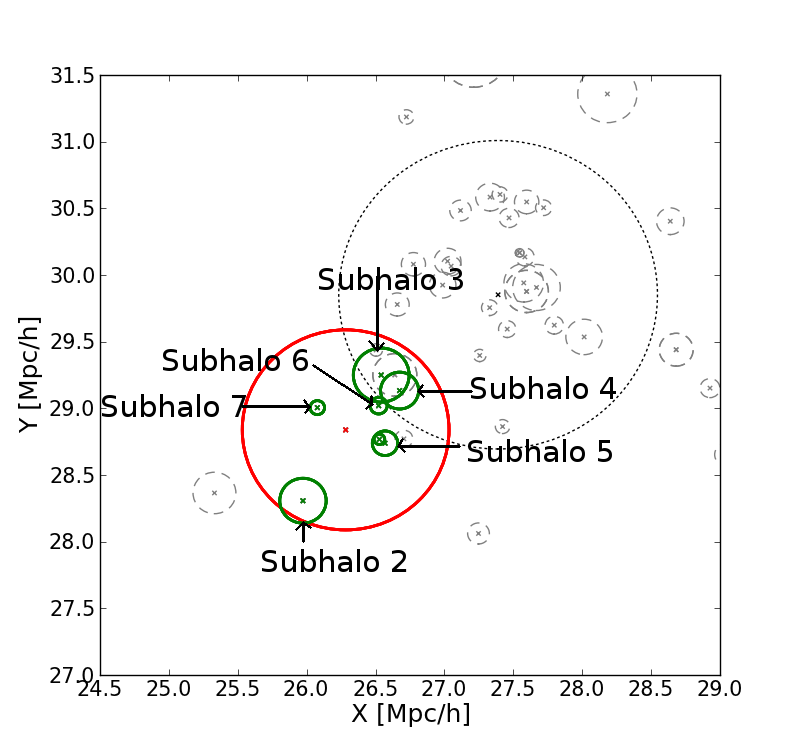}\label{fig:fig2b}}
    \subfigure[$z=0$]{\includegraphics[width=2.3in]{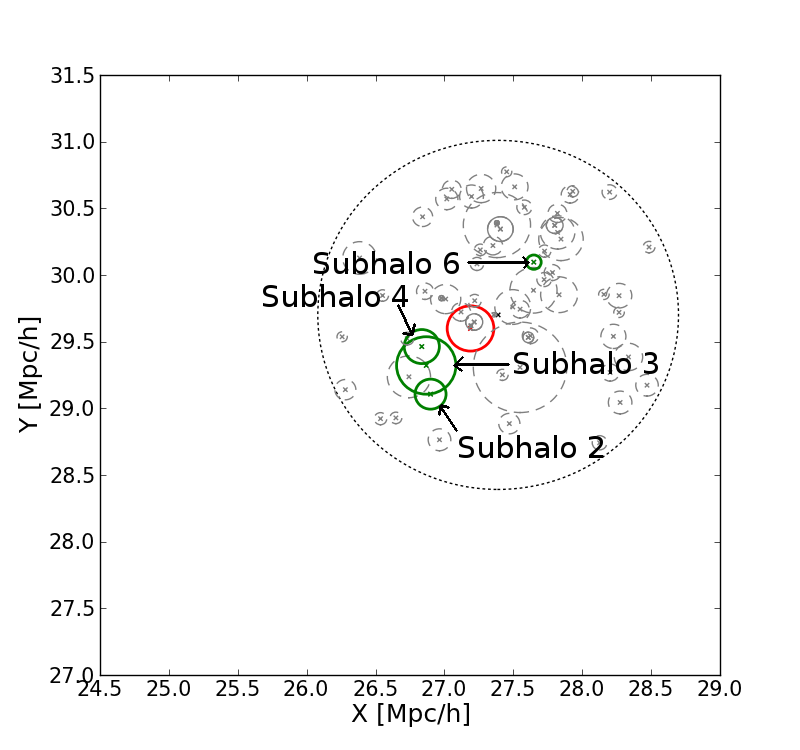}\label{fig:fig2c}}
    \caption{Subhalo `sweeping,' or the brief pre-processing of cluster subhalos in groups. Each circle represents the projected location of a halo or subhalo; its radius corresponds to the object's $R_{200}$ value. Red circles represent the group, and green circles the group's subhalos identified at $z=0.2$. The grey dotted circle represents the cluster, and the grey dashed circles show the other cluster subhalos. \label{fig:fig2}}
  \end{center}  
\end{figure*}

Using AHF, we identified the group's and cluster's subhalos and subsubhalos, and we identified their progenitors and descendants at all available redshifts. Each panel of Figure~\ref{fig:fig2} shows the projected positions of the subhalos at a different redshift during the merger. All objects in this figure are represented as circles whose radii are equal to the subhalos' virial radii. The numbered subhalos are those identified as group members at $z = 0.2$.

Figure~\ref{fig:fig2a} shows the progenitors of the group's $z = 0.2$ subhalos at $z = 0.5$. Most $z = 0.2$ subhalos are \emph{outside} the group's virial radius at $z = 0.5$. The descendants of the $z = 0.5$ subhalos are, however, not identified as bound structures within the group at $z = 0.2$. This is most likely because these subhalos have been stripped to a mass below the halo finder's resolution limit ($2.45 \times 10^9  ~\mbox{M}_{\odot}$, corresponding to 40 particles) and the cosmological simulation's low spatial resolution ($48.8 h^{-1}$ kpc).

Figure~\ref{fig:fig2c} shows the descendants of the group's $z = 0.2$ subhalos at $z = 0$, after the group has begun its first pericentric passage. The radius of the bound group remnant has decreased as its outer weakly bound dark matter and subhalos have been gravitationally unbound by the cluster's potential. 

Taken together, the evolution of the resolved (and therefore most massive) subhalos through the merger suggests the following scenario. Halos that would have otherwise fallen into the cluster directly from the field are swept up by the merging group, where they may undergo some degree of pre-processing\footnote{Although it appears from Figure~\ref{fig:fig2a} that Subhalos 3 and 4 are closer to the cluster than the group, thereby seemingly violating the equivalence principle by becoming part of the group, this is a projection effect. When taking into account their three-dimensional positions, these subhalos are in fact closer to the group than the cluster.}. This can include removal of material due to tidal stripping within the group, removal of gas due to the ram pressure of the IGM, and even galaxy-galaxy interactions with the group's galaxies. The impact of these processes on swept-up subhalos should depend on the time spent within the group environment as well as the relative masses of the merging group 
and cluster.

\subsection{Bound versus unbound group material} 
\begin{figure*} 
  \begin{center}
    %\subfigure[Group remnant]
    {\includegraphics[width=6in]{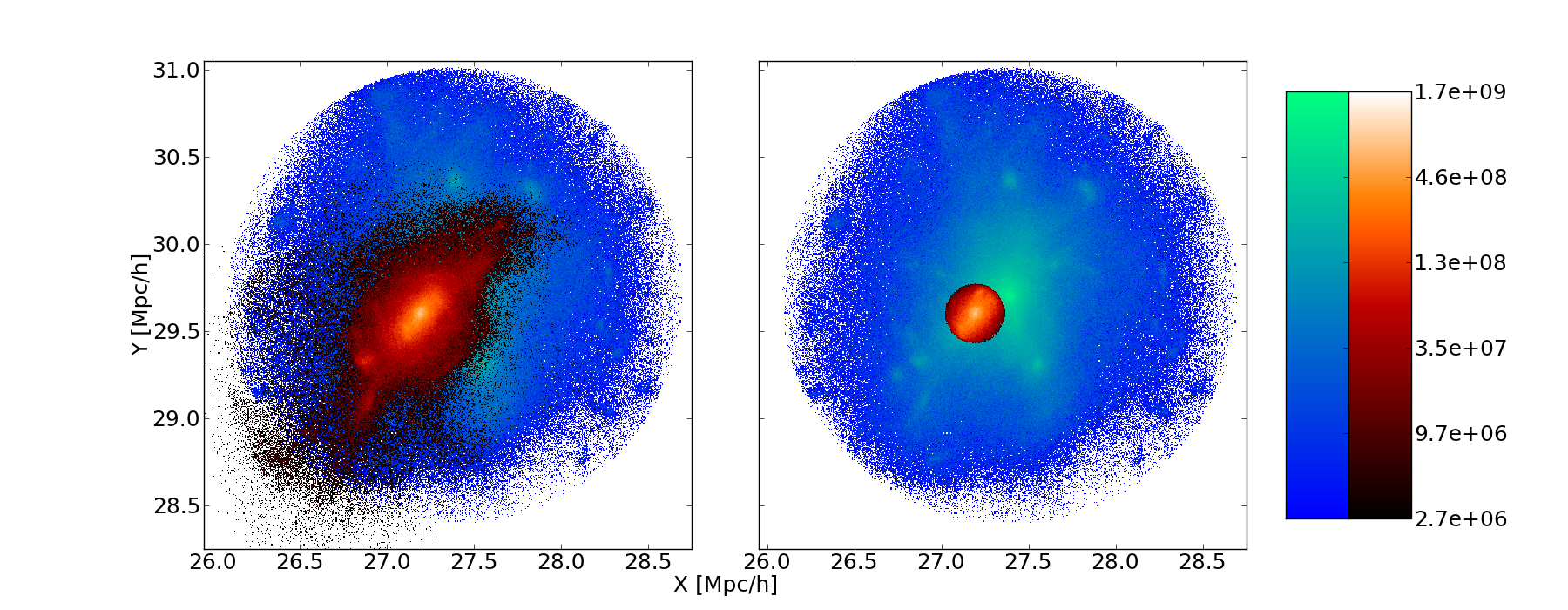}}%\label{fig:fig3a}}
    %\subfigure[Bound group remnant]
    %{\includegraphics[width=2.5in]{subrem2.png}\label{fig:fig3b}}
    \caption{Projected mass densities (in $ ~\mbox{M}_{\odot}$ kpc$^{-2}$) of the group and cluster particles at $z=0$. Blue/green colors represent cluster particles. Left: Red/orange colors indicate the density of all particles identified at $z=0.2$ as being bound to the group. Right: Red/orange colors indicate particles identified as being bound to the group at $z=0$. \label{fig:fig3}}
  \end{center}  
\end{figure*}

Figure~\ref{fig:fig3} shows color-coded maps of the projected density of the cluster and group particles (including their subhalos) after the merger. The merged group's projected density is overlaid on the cluster's density map. These maps distinguish between particles formerly bound to the merging group at $z = 0.2$ (Figure~\ref{fig:fig3}, left) and those still identified as part of the merged group at $z = 0$ (Figure~\ref{fig:fig3}, right).

We note here that the halo finder, AHF, identifies particles as being bound if their velocities are less than the local escape velocity, $v_{\rm esc}$, where $v_{\rm esc} = \sqrt{-2\phi_{\rm local}}$. $\phi_{\rm local}$ is the gravitational potential due to the subhalo's particles alone, computed using spherical averaging of the subhalo density distribution.

As the group's particles fall into the cluster's center during the merger, they experience a stronger cluster tidal field, and in response the group's potential becomes shallower. Those particles that are not stripped by the tidal field nevertheless become less well bound. Because the potential felt by the group's particles is changing with time, our definition of boundedness is not strictly correct. However, because the group responds to the cluster's tidal field by developing a shallower potential, it is reasonable to assume that particles that become unbound according to our criterion will remain unbound.

Although the group's former components have not been completely randomized in position within the cluster, the small size of the bound group remnant indicates that most of the outer material has been unbound from the group's potential and bound to the overall cluster potential. The mass of the bound group remnant is $1.26\times10^{13} ~\mbox{M}_{\odot}$, a factor of $\sim 2.5$ smaller than the group mass at $z = 0.2$ ($3.29 \times 10^{13}  ~\mbox{M}_{\odot}$). The density peaks in the former group's components, corresponding to the group's subhalos, further emphasize this point: these loosely bound, recently accreted group subhalos are quickly unbound from the group. 

\subsection{Group coherence in velocity space}
As the projected density maps show, when the group merges with the cluster, the positions of its components are not randomized within the cluster until at least the first pericentric passage of the group's core. Although gravitationally unbound, the components of the group can retain traces of their original infall velocity (and therefore the velocity of the main group remnant within the cluster) for some time. This is particularly important in the context of intra-group interactions within a cluster. 

\begin{figure*} 
  \begin{center}
    \subfigure[Radial velocity -- template]
    {\includegraphics[width=2.5in]{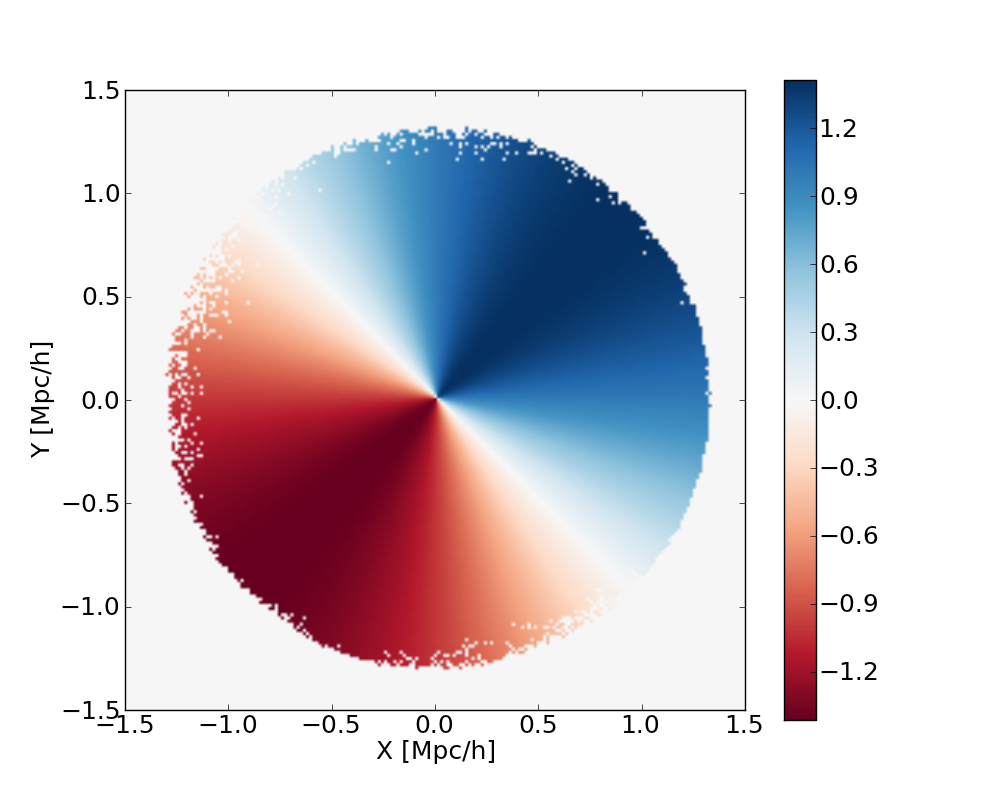}\label{fig:fig4a}}
    \subfigure[Radial velocity -- cluster at $z=0$]
    {\includegraphics[width=2.5in]{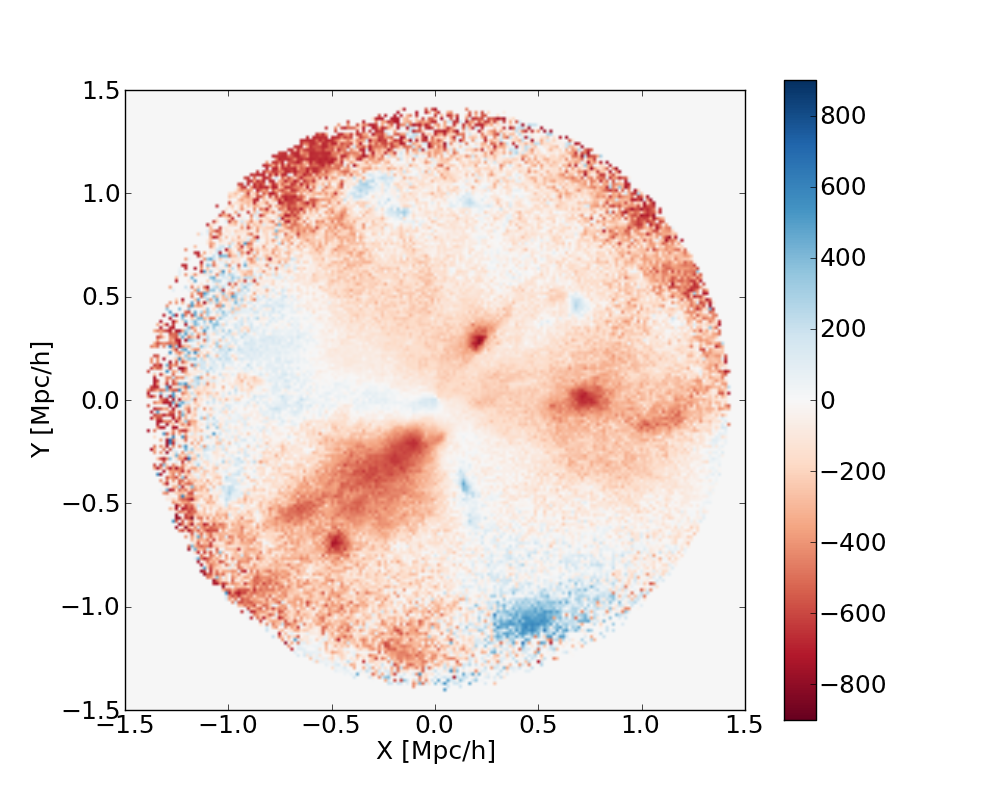}\label{fig:fig4b}}
    \\
    \subfigure[Radial velocity -- `old' cluster components]
    {\includegraphics[width=2.5in]{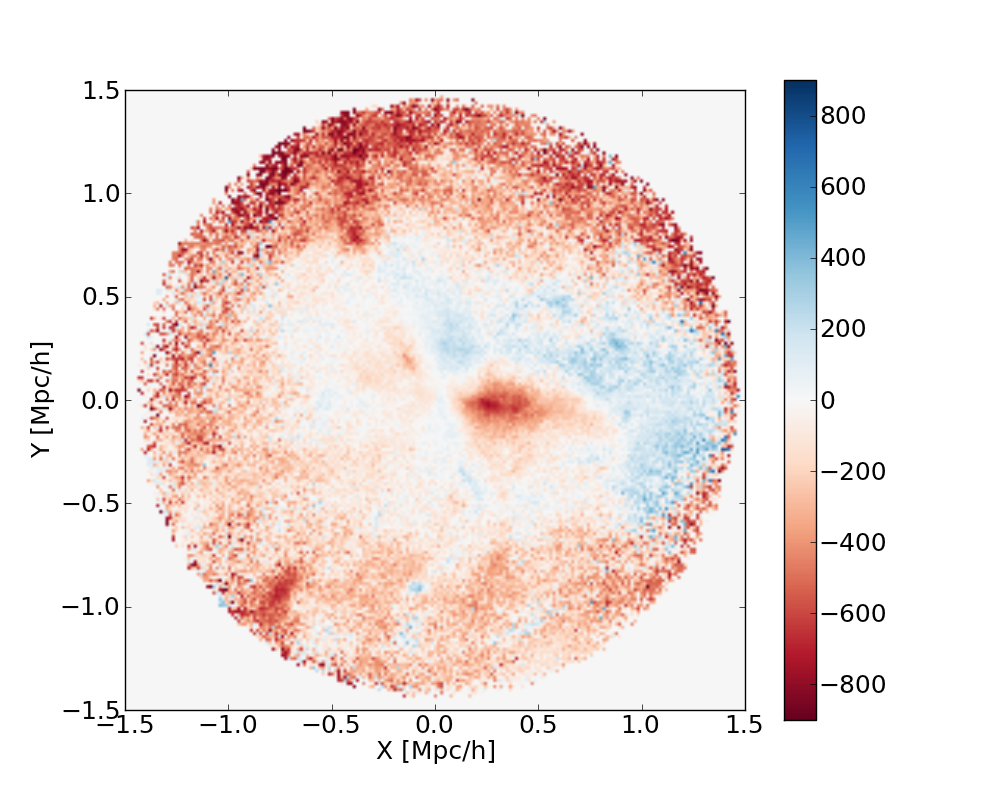}\label{fig:fig4c}}
    \subfigure[Radial velocity -- merging group]
    {\includegraphics[width=2.5in]{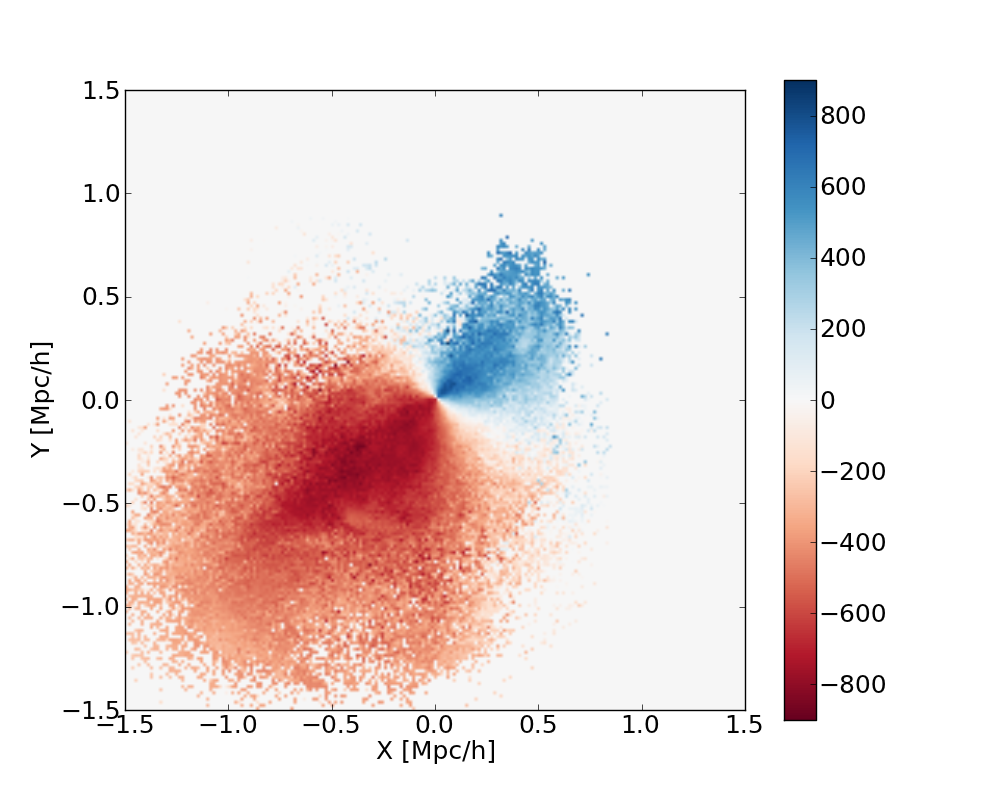}\label{fig:fig4d}}
    \caption{Top left: A template radial velocity map, where all the particles in the box have $v_{\rm x} = v_{\rm y} = 1$. Top right: $v_{\rm rad}$ map for all cluster compoents at $z = 0$. Bottom left: $v_{\rm rad}$ map of only those cluster components from $z = 0.2$ (`older' components). Bottom right: $v_{\rm rad}$ map of merged group's components. All velocities are in km s$^{-1}$ .\label{fig:fig4}}
  \end{center}  
\end{figure*}

\begin{figure*} 
  \begin{center}
    \subfigure[Tangential velocity -- template]
    {\includegraphics[width=2.75in]{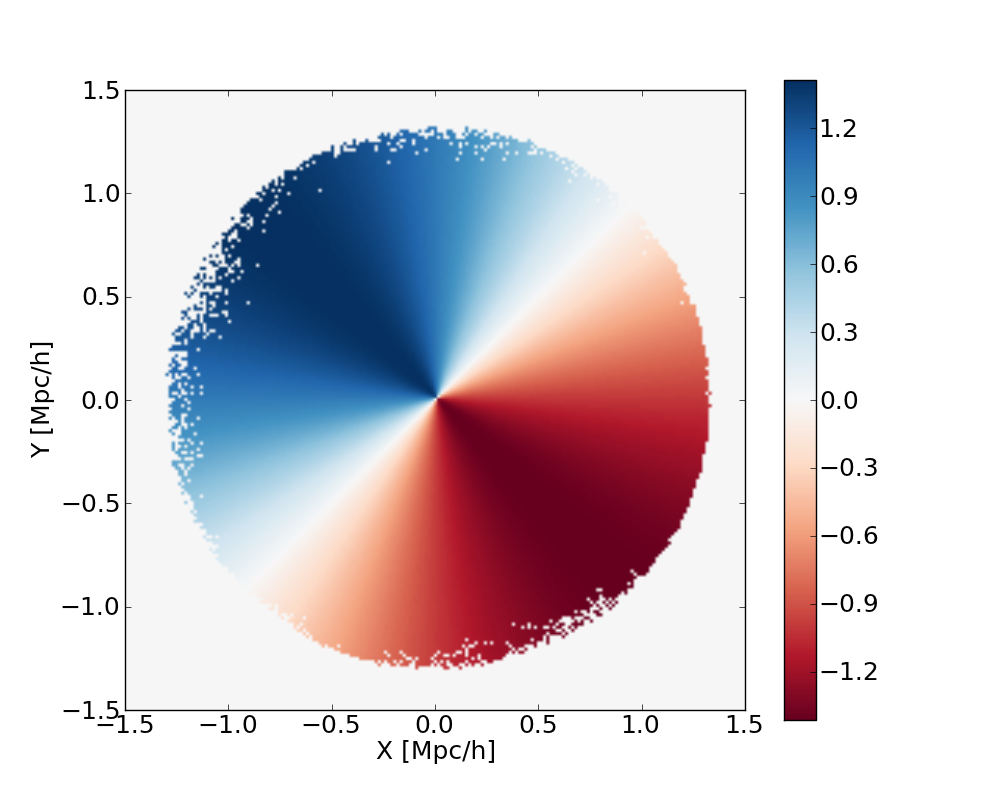}\label{fig:fig5a}}
    \subfigure[Tangential velocity -- merging group]
    {\includegraphics[width=2.75in]{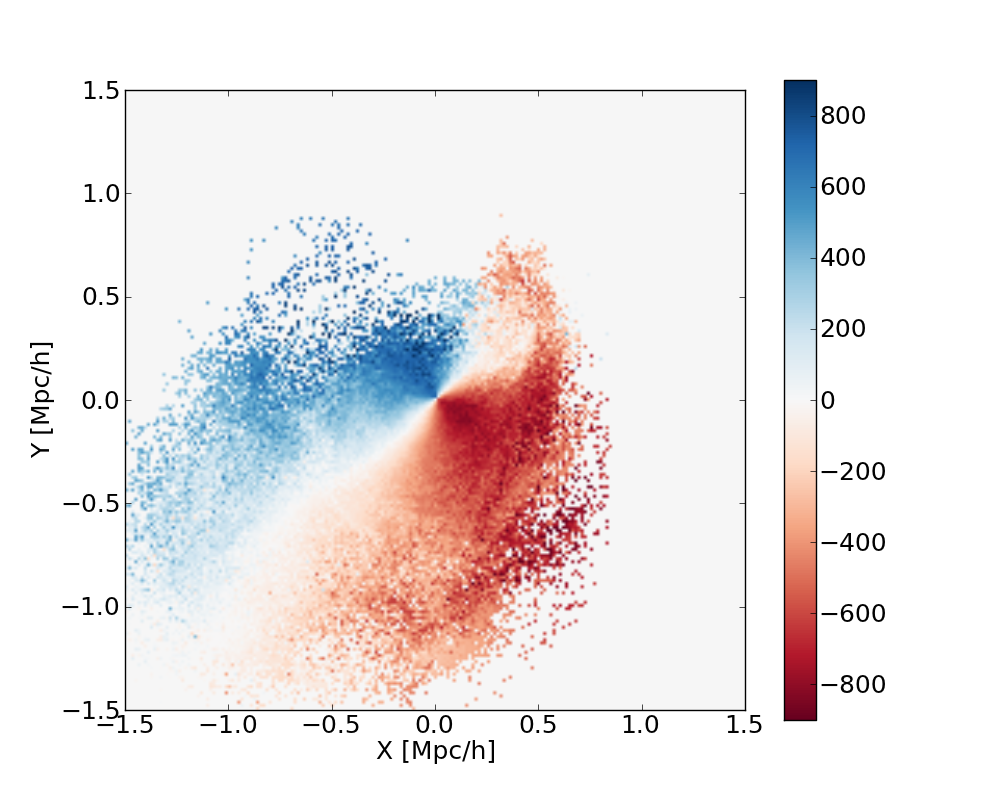}\label{fig:fig5b}}
    \caption{Left: Template tangential velocity map in which all particles have $v_{\rm x} = v_{\rm y} = 1$. Right: Tangential velocity map of merged group's components. All velocities are in km s$^{-1}$. \label{fig:fig5}}
  \end{center}  
\end{figure*}

To study the kinematic properties of the merged group within the cluster, we map the two-dimensional radial and tangential components of dark matter particle velocities' projections into the simulation volume's $xy$ plane. The radial velocity ($v_{rad}$) and tangential velocity ($v_{tan}$) are computed for each particle using
\begin{equation}
  v_{\rm rad} = v_{\rm x}\cos\phi + v_{\rm y}\sin\phi
\end{equation}  
\begin{equation}
  v_{\rm tan} = -v_{\rm x}\sin\phi + v_{\rm y}\cos\phi,
\end{equation}
where
\begin{equation}
  \phi = \tan^{-1}\left(\frac{y}{x}\right).
\end{equation}
$v_{\rm x}$ and $v_{\rm y}$ are the $x$ and $y$ components of the particle velocities. Figures~\ref{fig:fig4} and \ref{fig:fig5} are maps of the average radial and tangential velocity in each pixel of a $200 \times 200$ grid centered on the cluster's center at $z = 0$. To aid in interpretation, Figures~\ref{fig:fig4a} and \ref{fig:fig5a} show template maps in which all halo particles have been assigned $v_{\rm x} = v_{\rm y} = 1$. All velocities are in km s$^{-1}$.

Figure~\ref{fig:fig4b} shows the radial velocity map of all the cluster halo's particles at $z = 0$.  The largest feature in this map is the red clump falling into the cluster halo near its center, corresponding to the merging group (negative radial velocities, or redder regions, correspond to radial infall toward the center). We see other subhalos falling toward the cluster center as well, and the red regions near the edge of the halo correspond to material accreted by the cluster halo from the field. Figure~\ref{fig:fig4c} shows only those particles that were present in the cluster halo at $z = 0.2$. This map is unremarkable since it shows the velocity structure of particles that have been part of the cluster for at least $2.35$ Gyr, and thus have been virialized. Figure~\ref{fig:fig4d} shows particles that merged as part of the group. Here, we clearly see signs of the infalling group. A comparison with the template map shows that all the particles in this map have relatively uniform radial velocities. The bound remnant of the group occupies a much smaller region of the cluster than that encompassed by all former group components; this indicates that these particles, while not bound, are far from virialized and still have coherent velocities.

Figure~\ref{fig:fig5} shows tangential velocity maps for the merging group and cluster. Figure~\ref{fig:fig5b} is the $v_{\rm tan}$ map at $z=0$ of all former group components. Here too, the distribution of the tangential velocity components  is consistent with a group falling in toward the cluster center, retaining the group's infall velocity. The implications of the long timescale over which a merging group's components are coherent in velocity space are explored further in the following sections in the context of the idealized resimulation. 

\section{Results: Idealized Merger Resimulation}
\label{sec:results-ideal}
\subsection{Group and cluster stability}

\begin{figure*} 
  \begin{center}
    \subfigure[Dark matter]
    {\includegraphics[width=3.4in]{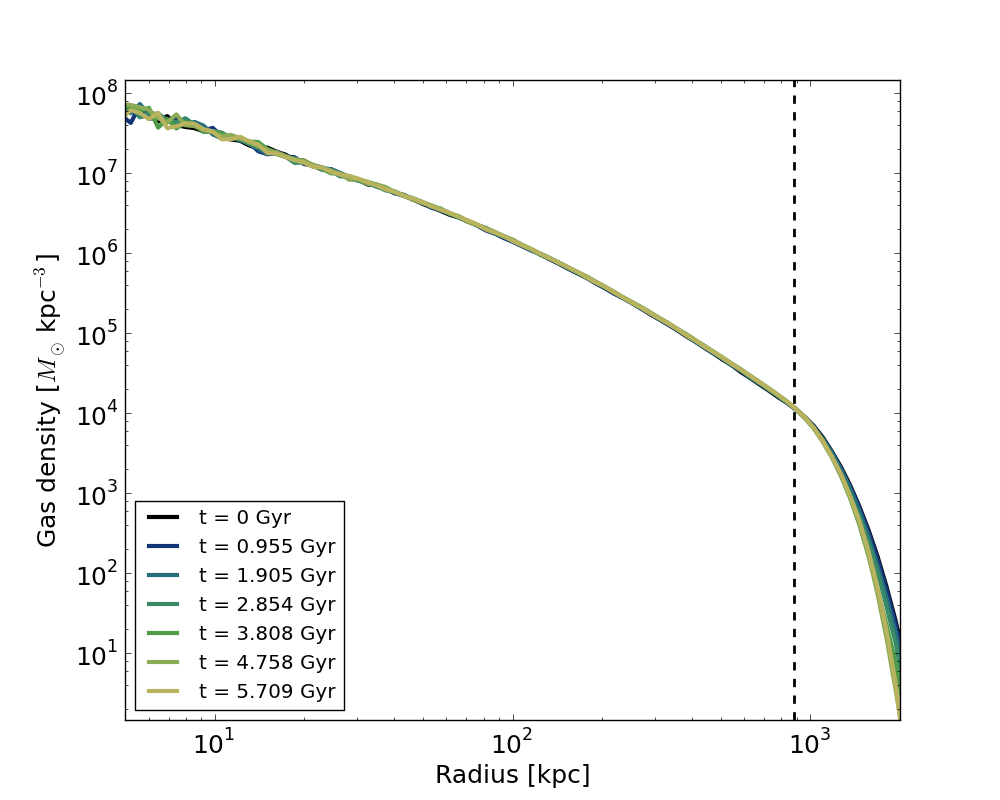}}
    \subfigure[Gas]
    {\includegraphics[width=3.4in]{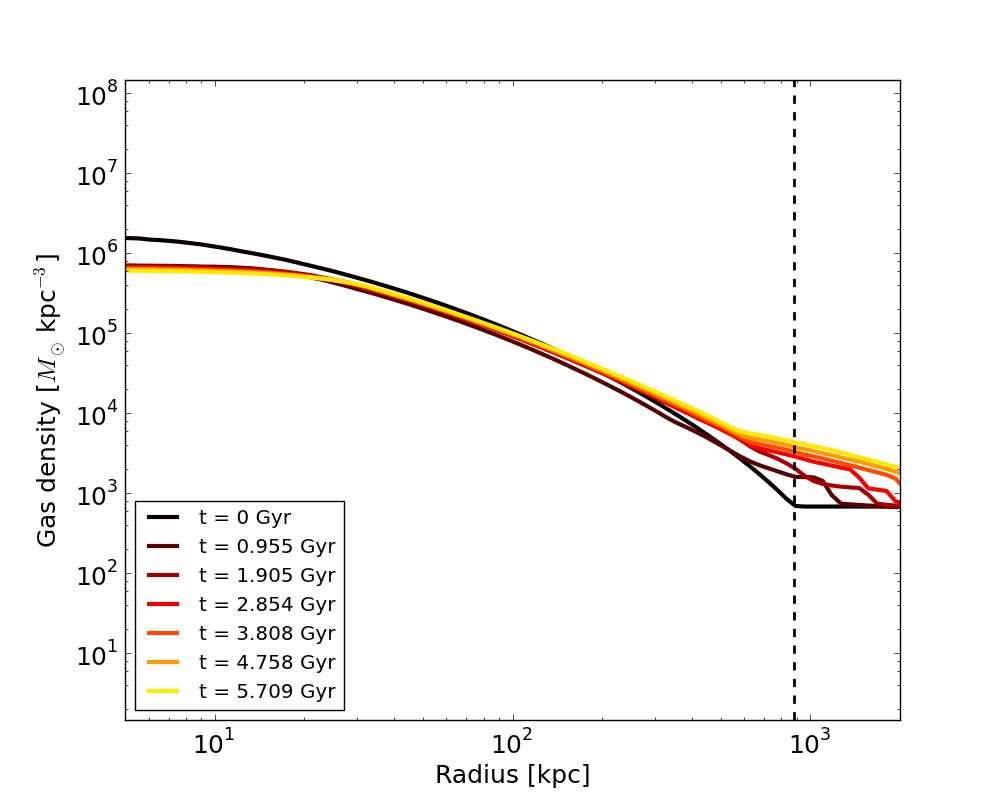}}
    \caption{Density profiles of the dark matter and gas components of the isolated cluster test. The dashed lines denote the virial radius of the cluster, $R_{200} = 880$ kpc.\label{stabprof}}
  \end{center}  
\end{figure*}

To test the stability of our idealized dark matter and adiabatic gas halos, we allowed the group and cluster halos to evolve in equilibrium for 6.34 Gyr. For $t_{\rm dyn} = \sqrt{3\pi/(32 G\overline{\rho})} \simeq 2.34$ Gyr, this equals $\sim 2.7$ dynamical timescales. These halos were refined with a maximum resolution of 7.6 kpc, and the total numbers of particles in the isolated group and cluster were the same as in the merging group and cluster respectively. Figure~\ref{stabprof} shows the evolution of the radial density profiles of dark matter and gas for the isolated cluster at six times over the course of the simulation. The isolated cluster is stable, as is the isolated group (not shown here). At the cluster's scale radius, $r_{\rm s}$, the mean density of dark matter fluctuates by a maximum of of $6.6\%$ and the mean gas density fluctuates by a maximum of $14\%$ during the simulation.  

\subsection{Orbit of the merging group}

\begin{figure*} 
  \begin{center}
    {\includegraphics[width=4in]{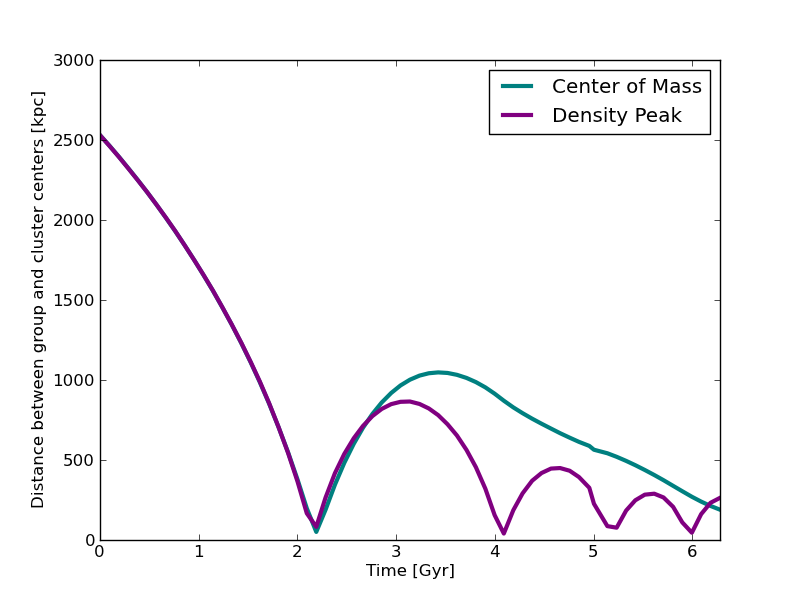}}
    \caption{Separation between the group and cluster centers through the merger, in kpc. The teal line is the separation between the two centers of mass, and the purple line is the separation between the density peaks. \label{fig:gc_sep}}
  \end{center}  
\end{figure*}

In our merger resimulation, we allow the group to fall into the cluster with an initial infall velocity vector (in km s$^{-1}$) $(v_{\rm x}, v_{\rm y}, v_{\rm z}) = (455.43, 500.76, 0)$. The group and cluster centers are initially (at simulation time $t = 0$~Gyr) separated by the vector (in Mpc) $(\Delta r_{\rm x}, \Delta r_{\rm y}, \Delta r_{\rm z}) = (1.84, 1.74, 0)$. These parameters are equal to the values from the cosmological simulation at $z = 0.2$, which corresponds to a lookback time of 2.35 Gyr. We see the evolution of the merging group's orbit in Figure~\ref{fig:gc_sep}, which shows the separation between the group and cluster centers as a function of time. The group makes its first pericentric passage at $t \simeq 2.2$ Gyr. As the group becomes tidally deformed by the cluster, its density peak ceases to coincide with its center of mass. The group's central dense core makes a second pericentric passage at $t \simeq 4$ Gyr, a third pericentric passage at $t \simeq 5.2$ Gyr, and a final pericentric passage at $t \simeq 6$ Gyr. The apocenters of the group's orbits are reached at $t \simeq $ 3 Gyr, 4.6 Gyr, and 5.7 Gyr. We see the decay of the group's orbit due to the combined effects of dynamical friction and the virialization of its components.

\begin{figure*} 
  \begin{center}
    \subfigure[{$t = 0$} Gyr]
    {\includegraphics[width=2.3in]{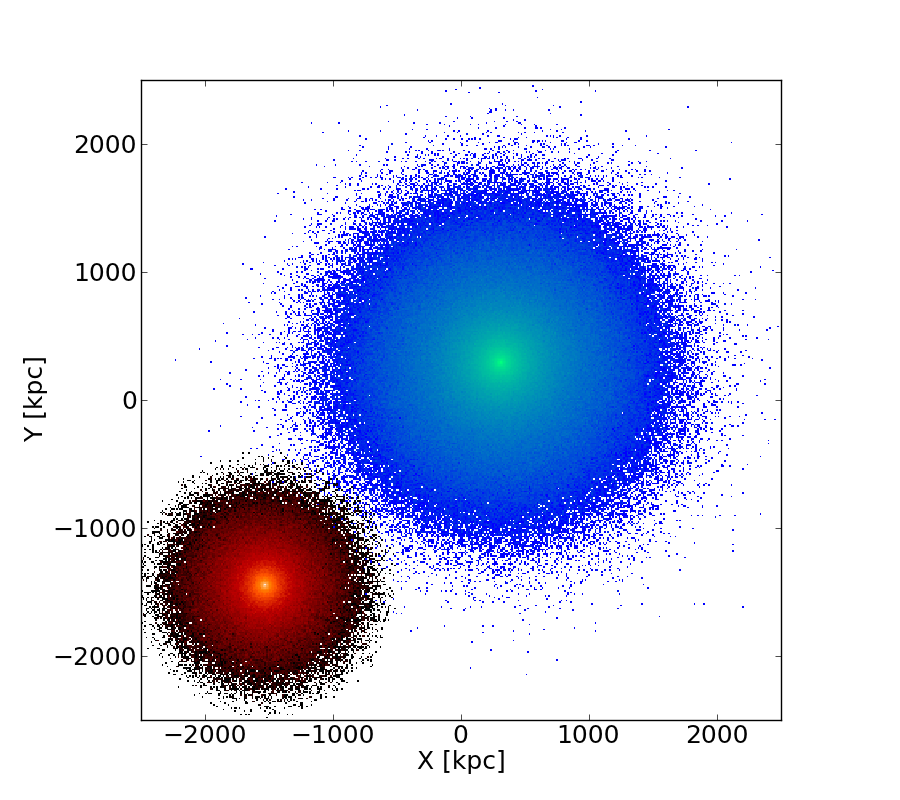}\label{fig:gcdma}}\hfill
    \subfigure[{$t = 1.526$} Gyr]
    {\includegraphics[width=2.3in]{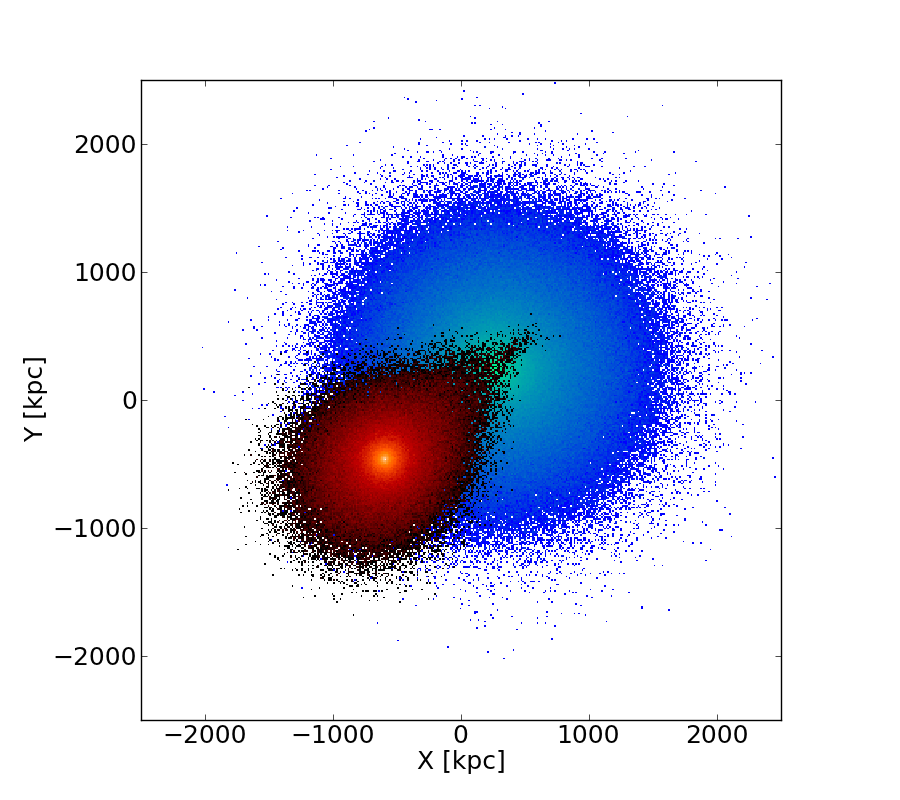}\label{fig:gcdmb}}\hfill
    \subfigure[{$t = 2.191$} Gyr]
    {\includegraphics[width=2.3in]{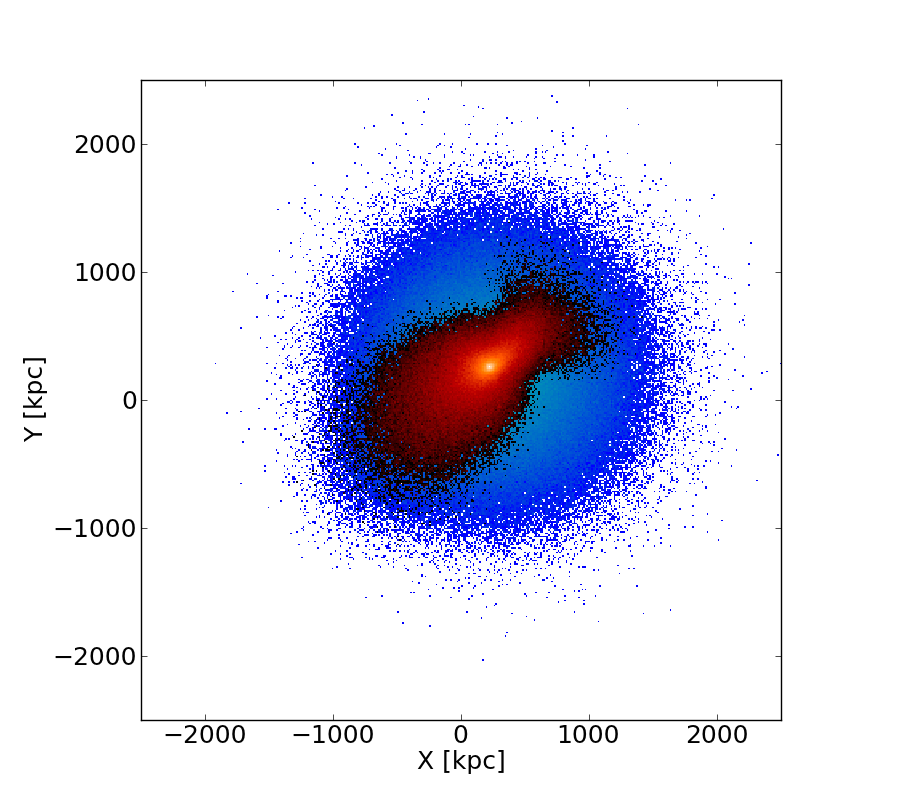}\label{fig:gcdmc}}
    \\
    \subfigure[{$t = 3.331$} Gyr]
    {\includegraphics[width=2.3in]{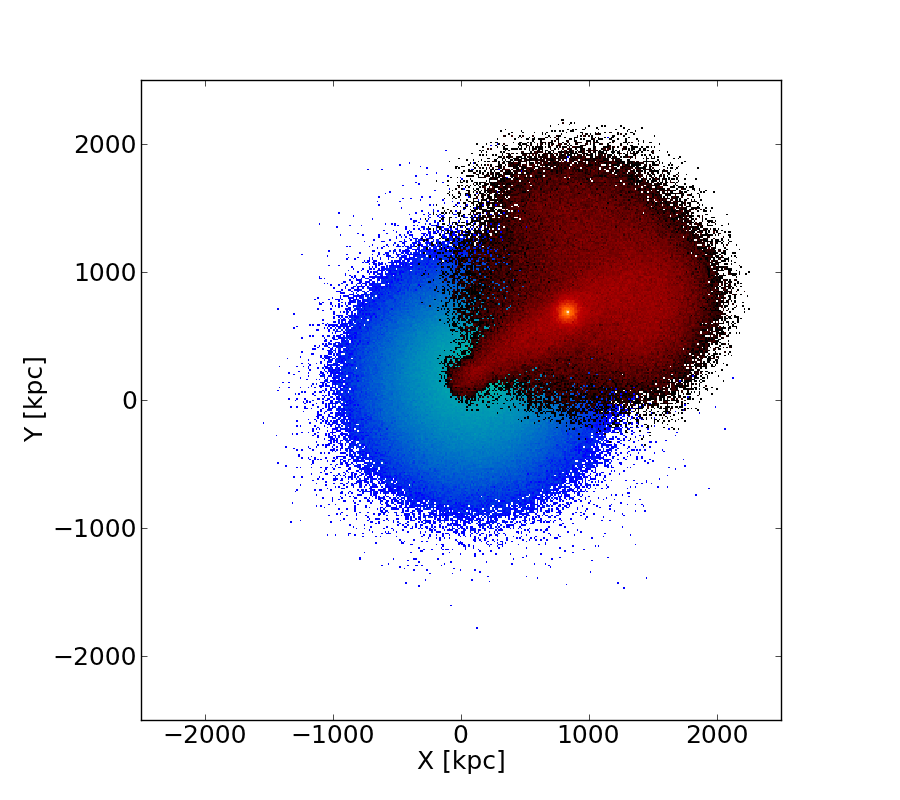}\label{fig:gcdmd}}\hfill
    \subfigure[{$t = 4.758$} Gyr]
    {\includegraphics[width=2.3in]{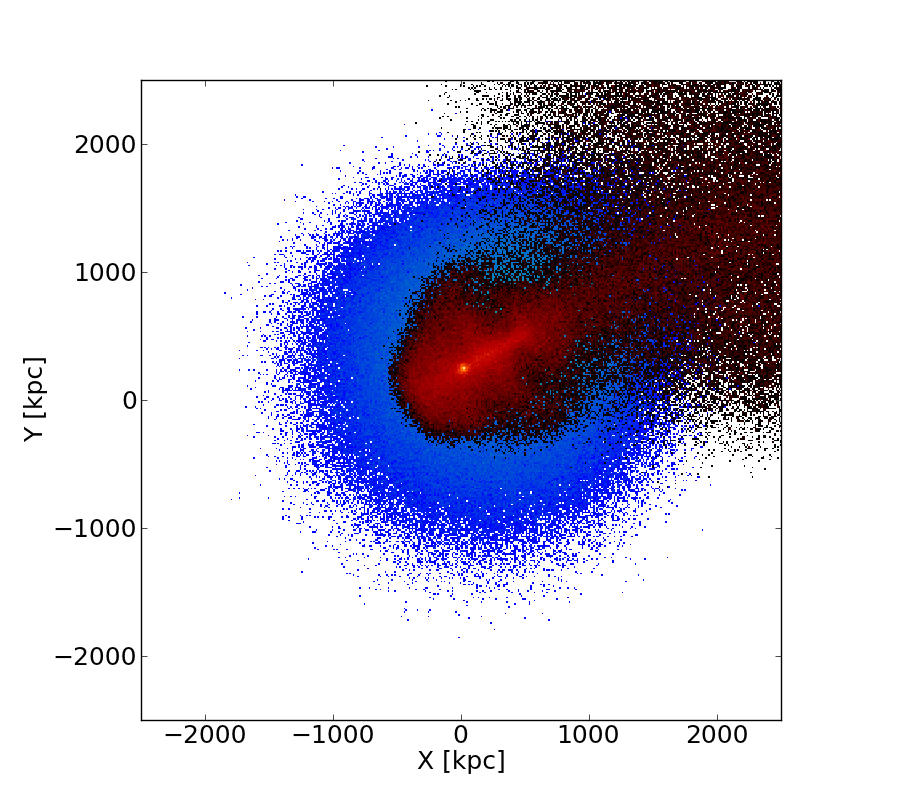}\label{fig:gcdme}}\hfill
    \subfigure[{$t = 5.709$} Gyr]
    {\includegraphics[width=2.3in]{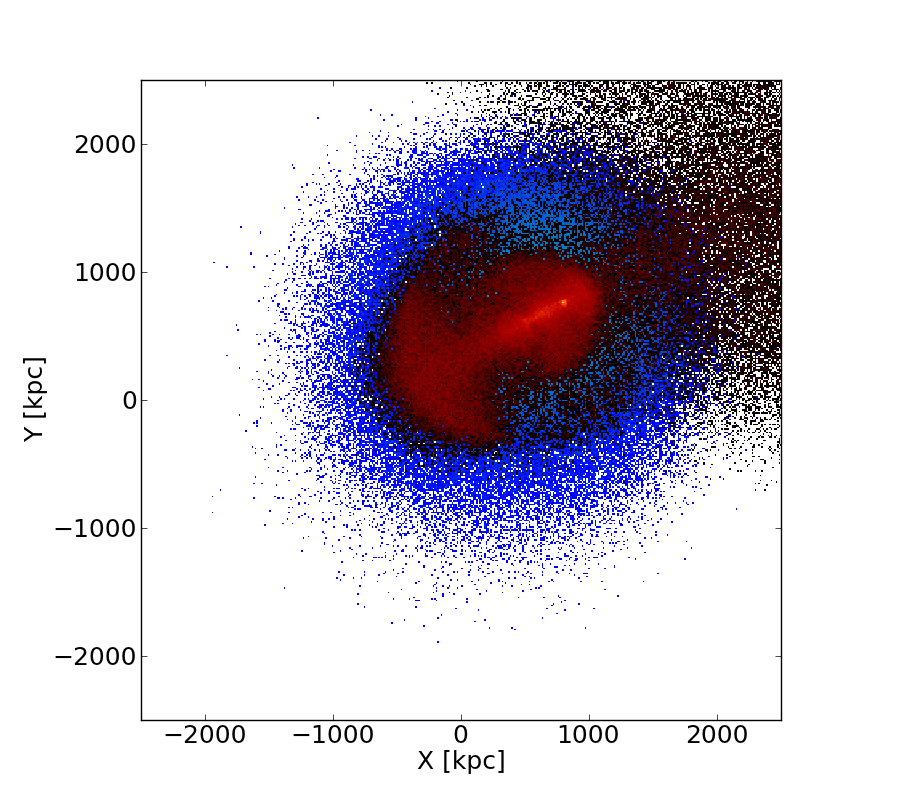}\label{fig:gcdmf}}
    \\
    %\subfigure[Surface mass density, $ ~\mbox{M}_{\odot}$ kpc$^{-2}$]
    {\includegraphics[width=6in]{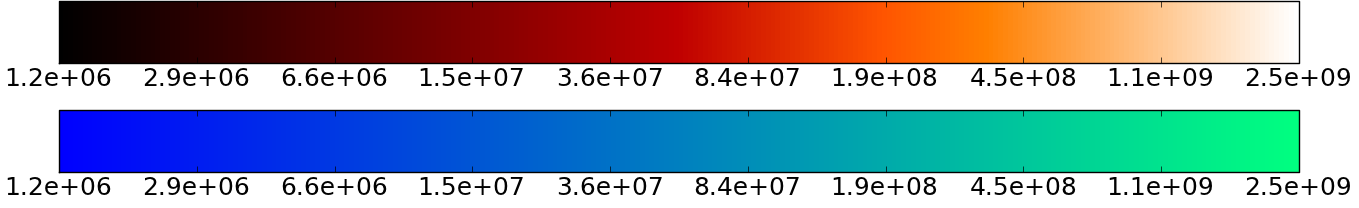}}
    \caption{The evolution of the group and cluster particle densities over the course of the merger.  These are plots of the mass density, projected along the $z$-axis, of the group and cluster dark matter (in units of $\mbox{M}_{\odot}$kpc$^{-2}$). 
    Colors correspond to the halos to which the particles originally belong. Group particles are in red-black, and cluster particles are in blue-green.\label{fig:figgcdm} 
    } 
  \end{center}  
\end{figure*}

Figure~\ref{fig:figgcdm} shows the evolution of the densities of the group and cluster particles projected onto the plane of the merger\footnote{The boundaries of the density maps in Figure~\ref{fig:figgcdm} (5 Mpc per side) do not encompass the entire simulation box (6.28 Mpc per side). Less than $1\%$ of the total mass is lost through the outflow boundaries.}. We allow the system to evolve for a total of 6.34 Gyr. The dense group core is distinctly visible up to $\sim$ 5 Gyr. Over the course of the merger, the group is tidally disrupted by the cluster. The group's components gain kinetic energy as they pass through the cluster's potential well, approach the apocenter of their orbits, and then fall back into the cluster. The group core orbits the cluster center with progressively smaller orbital amplitudes and shorter orbital periods under the influence of dynamical friction. The group's stripped components are randomized within the cluster and `forget' their original velocities and positions.

\subsection{Velocity coherence}
\label{sec:vcoh}
The results of the cosmological merger indicate that a merging group's components can retain traces of their original infall velocity long after becoming unbound. Here we quantify the timescale over which velocities remain coherent and the group's components virialize, and how the range of velocities of the group's components widens due to the group-cluster interaction. To do this, we define the pairwise normalized velocity difference, $v_{{\rm pd,}ij}$, between two particles $i$ and $j$ (where $i \ne j$) as
\begin{equation}
  v_{{\rm pd,}ij} = \frac{|\mathbf{v}_i - \mathbf{v}_j|}{\sigma_v}.\label{eqn:vpd}
\end{equation}
Here $\mathbf{v}_i$ and $\mathbf{v}_j$ are the velocities of the two particles, and $\sigma_v$ is the velocity dispersion of the merging group and cluster system taken as a whole. $\sigma_v$ is estimated for a system of $N$ particles with velocities $\mathbf{v}_i$ and an average velocity $\mathbf{\bar{v}}$ using
\begin{equation}
\sigma_v = \sqrt{\frac{\sum_{i=1}^N(\mathbf{v}_i - \mathbf{\bar{v}})^2}{N}}.
\end{equation}
At each timestep, we evaluate $v_{\rm pd}$ for all pairs in a group of 1000 randomly selected particles chosen from two sets, group particles only ($v_{\rm pd,gg}$) and cluster particles only ($v_{\rm pd,cc}$). These particles are identified as `group' or `cluster' particles based on their initial locations (i.e., at $t = 0$) within the group or cluster respectively. We also evaluate pairwise velocity differences $v_{\rm pd,gc}$ for 1000 particles chosen from the group and 1000 particles chosen from the cluster. We note here that this calculation does not account for the velocity bias $b \equiv \sigma_{v{\rm ,sub}}/\sigma_{v{\rm ,DM}}$ of actual group or cluster subhalos with respect to the dark matter particles' velocity dispersion. Simulations (\citealt{Colin00}, \citealt{Diemand04}) indicate a positive velocity bias in massive clusters, from $b \sim 1.1$ in the outer regions of clusters of up to $b \sim 1.3$ in the centers of clusters. We discuss the implications of a velocity bias in \S\ref{sec:discussion}.

\begin{figure*} 
  \begin{center}
    \includegraphics[scale=0.5]{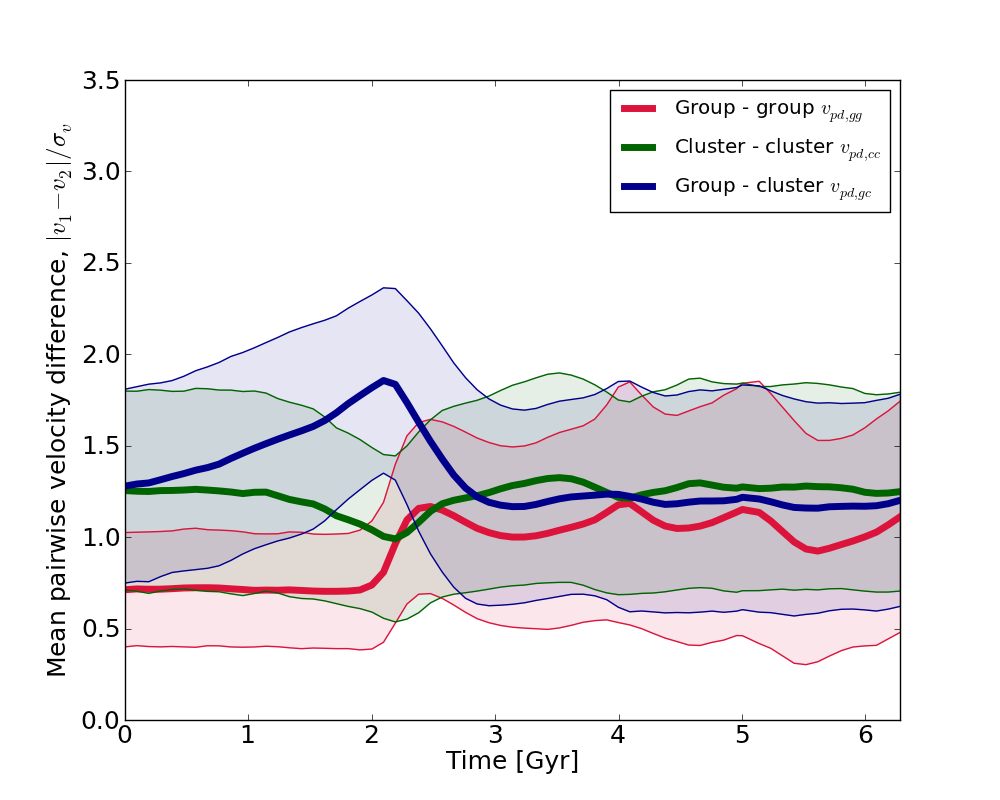}
    \caption{The evolution of $v_{\rm pd}$ for group and cluster particles. The thick lines show the mean value of $v_{\rm pd}$ for the given set of particles while the shaded regions represent the $1\sigma$ limits of the distribution.\label{fig:figvc1}}
  \end{center}    
\end{figure*}

The evolution of the distribution of pairwise velocity differences for all group and cluster particles is shown in Figure~\ref{fig:figvc1}. The normalization factor $\sigma_v$ appearing in Equation~\ref{eqn:vpd} is computed at each timestep for the entire group-cluster system. It increases from $\sim 800$~km~s$^{-1}$ at $t=0$ to $\sim 1250$~km~s$^{-1}$ at 2~Gyr before settling down to values around 1000~km~s$^{-1}$. Similarly, $\sigma_v$, when calculated only for the cluster particles, is initially $\sim 900$~km~s$^{-1}$, and varies by a maximum of $\sim 200$~km~s$^{-1}$. The group particles' $\sigma_v$, however, is initially $\sim 500$~km~s$^{-1}$, but increases to $\sim 1000$~km~s$^{-1}$ at each pericentric passage and at late times approaches the velocity dispersion of the system as a whole. 

At the beginning of the merger ($t \simeq 1.2$ Gyr), the mean value of $v_{\rm pd,gg}$ is less than 1, while that of $v_{\rm pd,cc}$ is greater than 1; this is because the cluster is `hotter' (has a greater velocity dispersion) than the group. The group components' large infall velocities, i.e., the large average relative velocity between group and cluster particles, imply that $v_{\rm pd,gc}$ is larger than $v_{\rm pd,cc}$ and $v_{\rm pd,gg}$ at the beginning of the merger. The group makes its first pericentric passage at $\sim 2.1 - 2.3$ Gyr, and the mean of $v_{\rm pd,gc}$ increases to its maximum value at that time. As it decreases from this peak, the mean and spread of $v_{\rm pd,gg}$ sharply increase, approaching the corresponding values for the distribution of $v_{\rm pd,cc}$. By $t \sim 3$~Gyr the three distributions become nearly constant and very similar, although $v_{\rm pd,gg}$ continues to oscillate as the group remnant makes successive pericentric passages. The virialization timescale depends on how virialization is defined, but visually it is at least $\sim 1$~Gyr. We note here that an accurate measure of virialization relies on $\sigma_{v}$ representing the velocities of already virialized particles. Since the particles in this system are not virialized during the merger, our estimate of the rate of `virialization' is in fact a measure of the rate at which the system's particles approach a steady state velocity distribution. 

\begin{figure*} 
  \begin{center}
    \subfigure[Group core, cluster core]
    {\includegraphics[width=3.4in]{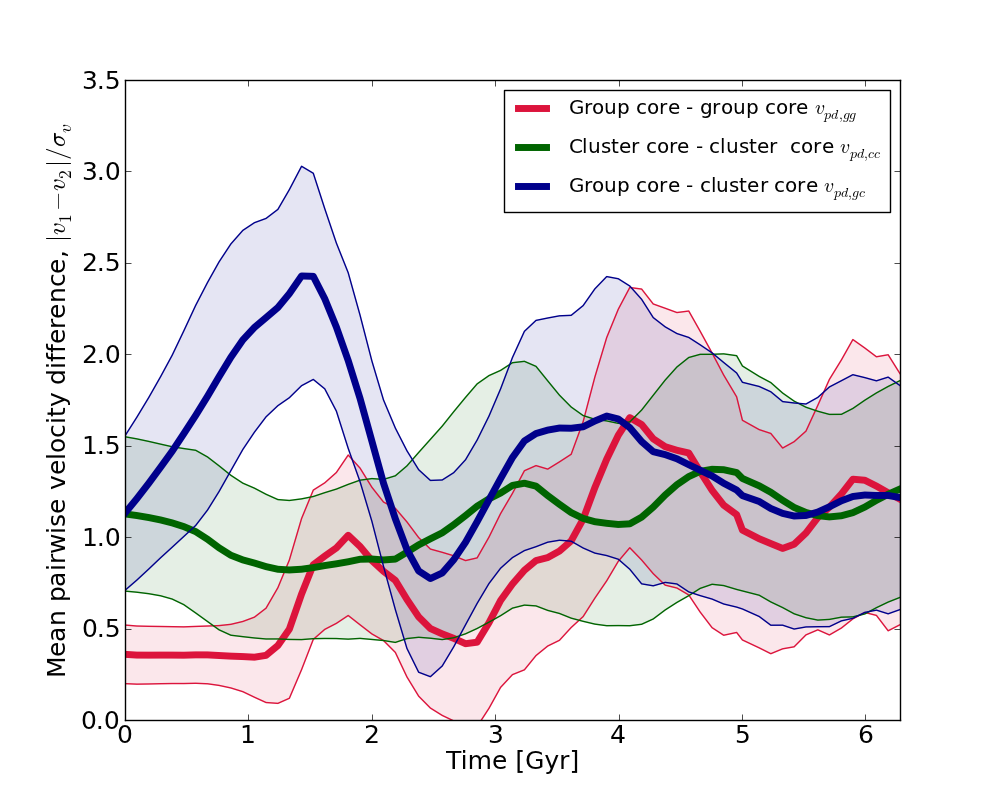}\label{fig:vcgccc}}
    \subfigure[Group core, cluster outskirts]
    {\includegraphics[width=3.4in]{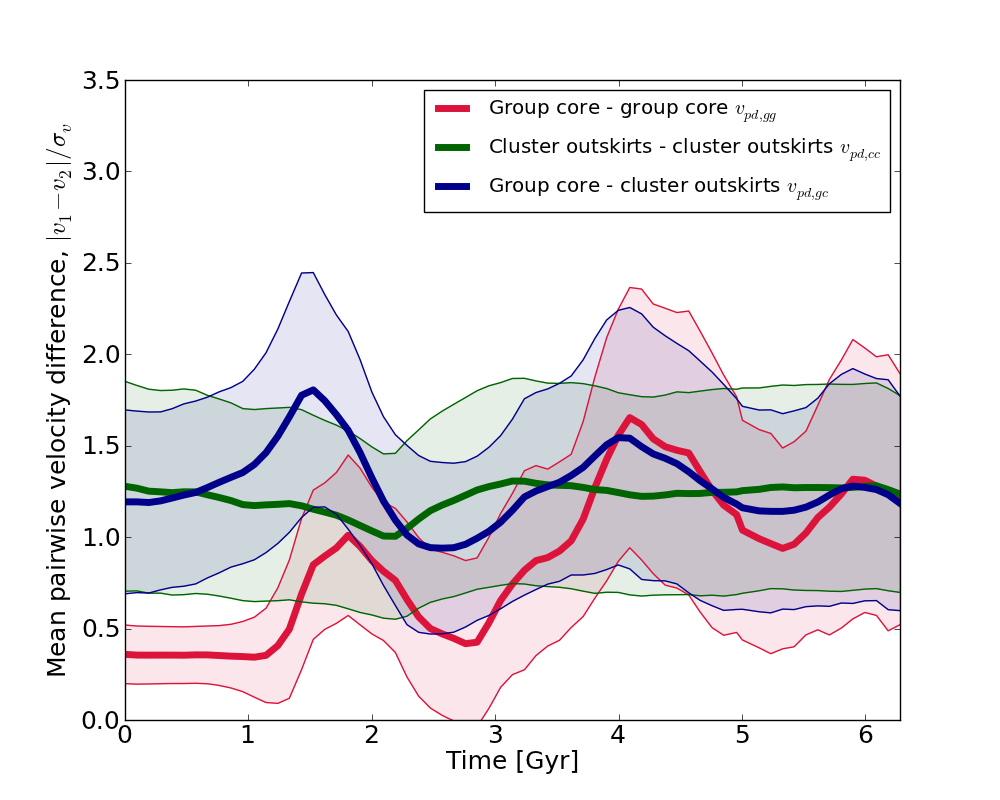}\label{fig:vcgcch}}
    \\
    \subfigure[Group outskirts, cluster core]
    {\includegraphics[width=3.4in]{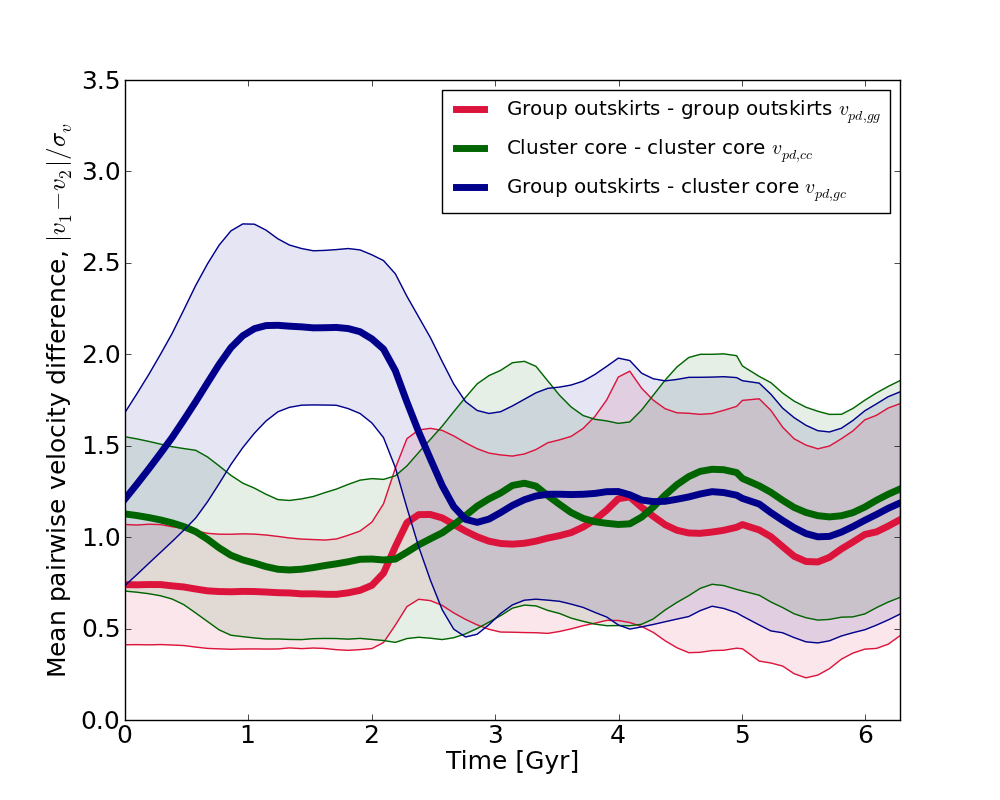}\label{fig:vcghcc}}
    \subfigure[Group outskirts, cluster outskirts]
    {\includegraphics[width=3.4in]{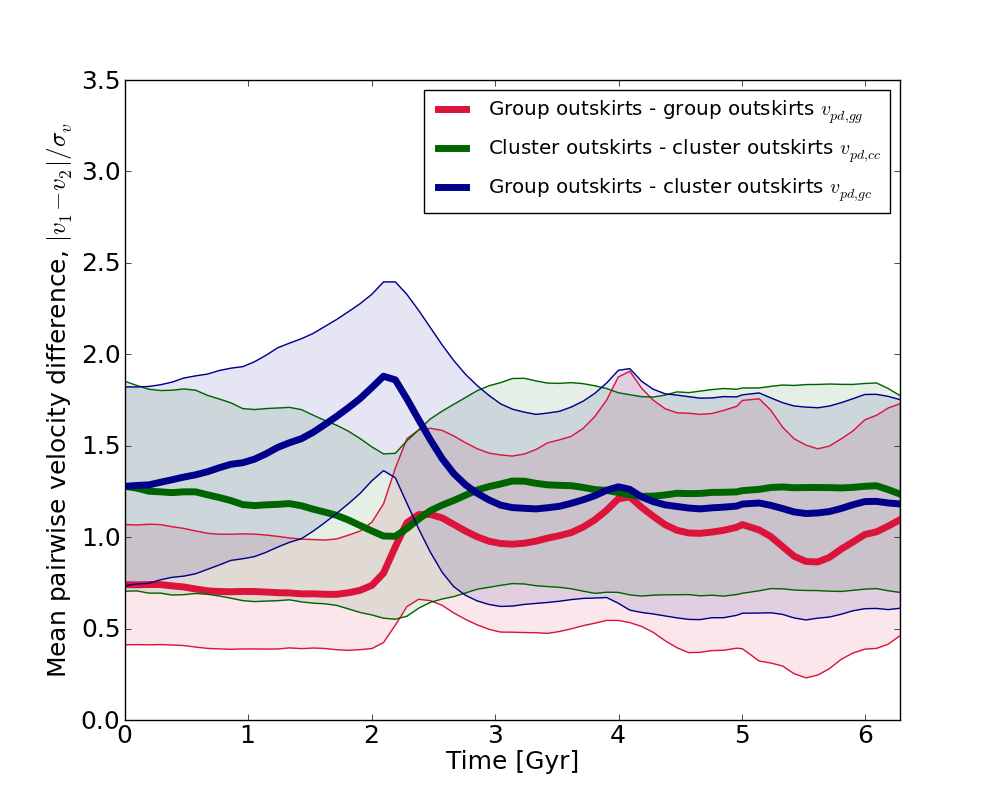}\label{fig:vcghch}}
    \caption{Mean and standard deviation in $v_{\rm pd}$ for group and cluster particles, distinguished by initial host-centric distance. `Core' particles started at $r < r_{\rm s}$, while  particles in the outskirts started at $ r < r_{\rm s}$. Thick lines show mean values of $v_{\rm pd}$, and shaded regions depict $1\sigma$ limits in the distribution of $v_{\rm pd}$. \label{fig:figvc2}} 
  \end{center}  
\end{figure*}

Figure~\ref{fig:figvc2} decomposes the evolution of $v_{\rm pd}$ for four separate classes of particles: group core, group outskirts, cluster core, and cluster outskirts. Here, `core' particles are those whose initial host-centric radii, $r$, are less than the host's scale radius, $r_{\rm s}$, while particles in the outskirts are those for which $r$ is greater than $r_{\rm s}$. Figure~\ref{fig:vcgccc} shows the evolution of $v_{\rm pd}$ for group and cluster core particles. At $t \sim 1.2$ Gyr, the relative velocities of group and cluster core particles with respect to other group and cluster core particles respectively are close to the overall velocity dispersion of all the particles in the merging group-cluster system. However, the relative velocities of the group and cluster core particles with respect to \emph{each other} are $2 - 2.5$ times greater than the overall velocity dispersion. In contrast, the relative velocities at this time between the particles that were initially in the group's and cluster's outskirts (Figure~\ref{fig:vcghch}) are only $1.5$ times the overall velocity dispersion. This reinforces the idea that the velocities of the group's less strongly bound outer particles approach the overall systematic velocity dispersion earlier than the strongly bound core. Additionally, the overall spread in the the values of $v_{\rm pd,gc}$ for particles in the outskirts approaches steady state before that of the core particles. 

At $t \sim 4$ Gyr, when the stripped group core makes its second pericentric passage, at least some of the group core's components receive a velocity boost, leading to an increase in the mean pairwise velocity difference (to $1.5 \times$ the velocity dispersion) between these particles and cluster particles as well as between group core particles themselves (as seen in Figure~\ref{fig:vcgccc} and Figure~\ref{fig:vcgcch}). The group's outer particles, on the other hand, do not have $v_{\rm pd,gc}$ values as large as and as variable as the core particles', consistent with a scenario where they are stripped off and their velocities are randomized before those of the core particles. The amplitude of the oscillations of the group core particles' $v_{\rm pd,gg}$ over the course of the group's orbit is larger that of the group's outer particles, implying that the group core remains coherent for a longer time and therefore also receives larger overall velocity boosts at each pericentric passage. 

The merger also affects the distribution of cluster particles' velocities. The spread in cluster components' velocities (Figure~\ref{fig:figvc1}) increases, and this increase is seen in both the cluster's core and outskirts (Figure~\ref{fig:figvc2}). The mean $v_{\rm pd,cc}$ of the cluster's components decreases slightly (in contrast to that of the group's components) at the beginning of the merger. This decrease can be attributed to an increase in the overall velocity dispersion of the system due to the merger, and therefore a decrease in $v_{\rm pd,cc}$ (which is normalized to the system's overall velocity dispersion).

The thick blue lines in all four figures of Figure~\ref{fig:figvc2} show the process of virialization for the group core and halo particles. The pairwise velocity difference between group core and cluster particles oscillates with a higher amplitude compared to that between group halo and cluster particles, a further confirmation of the fact that the bound group core is virialized long after the outskirts. 

\subsection{Merger and collision timescales}
In this section we calculate and compare the interaction timescales and merger rates of galaxies in the isolated group, isolated cluster, and merging group and cluster. For this calculation, we trace the orbits of randomly selected dark matter particles and identify these orbits as proxies for actual galaxy orbits within a host, since we do not directly include galaxy-sized subhalos in our idealized merger. We repeat these calculations for 100 random realizations of galaxy initial positions and velocities within the group and cluster, then average our results over these random realizations. The collision timescale for a galaxy is given by
\begin{equation}\label{eqn:tcoll}
  t_{\rm coll} = \frac{1}{n_{\rm gal}\sigma_{\rm cs}v_{\rm gal}}.
\end{equation}
Here $n_{\rm gal}$ is the local number density of galaxies at a galaxy's position, $\sigma_{\rm cs} = \pi r_{\rm gal}^2$ is the galaxy cross-section (we assume a galactic radius $r_{\rm gal} = 100$ kpc for all the galaxies in our calculation of $\sigma_{\rm cs}$), and $v_{\rm gal}$ is the galaxy's peculiar velocity with respect to the halo in which it originated. In this calculation, the assumed galactic radius is in general an overestimate for the radius of a galaxy in a cluster; hence our calculation will underestimate $t_{\rm coll}$. In \S\ref{sec:tidaltrunc}, we show that the tidal truncation of galaxies within group and cluster halos reduces $r_{\rm gal}$ to a few tens of kpc. The galaxy number density $n_{\rm gal}$ in Equation~\ref{eqn:tcoll} is computed at each galaxy particle's position by CIC mapping galaxy particles to the AMR mesh and then inverse mapping the resulting mesh density to the galaxy particle's position.

We use the conditional luminosity function (CLF) of \citet{Yang08} to estimate the number of galaxies in the group and cluster, given the masses of the group and cluster halos. Based on this CLF, we assume that the group ($M_{200} = 3.2 \times 10^{13}  ~\mbox{M}_{\odot}$) and cluster ($M_{200} = 1.2 \times 10^{14}  ~\mbox{M}_{\odot}$) have 26 and 152 galaxies more massive than $10^9  ~\mbox{M}_{\odot}$ respectively (assuming a mass-to-light ratio of 10~$\mbox{M}_{\odot}~\mbox{L}_{\odot}^{-1}$). For each group and cluster galaxy, we assign a position and velocity corresponding to that of a group or cluster particle. We can thus track an ensemble of realizations of galaxy orbits over the course of the simulation. 

\begin{figure*} 
  \begin{center}
    \subfigure[Collision timescales]
    {\includegraphics[width=3.4in]{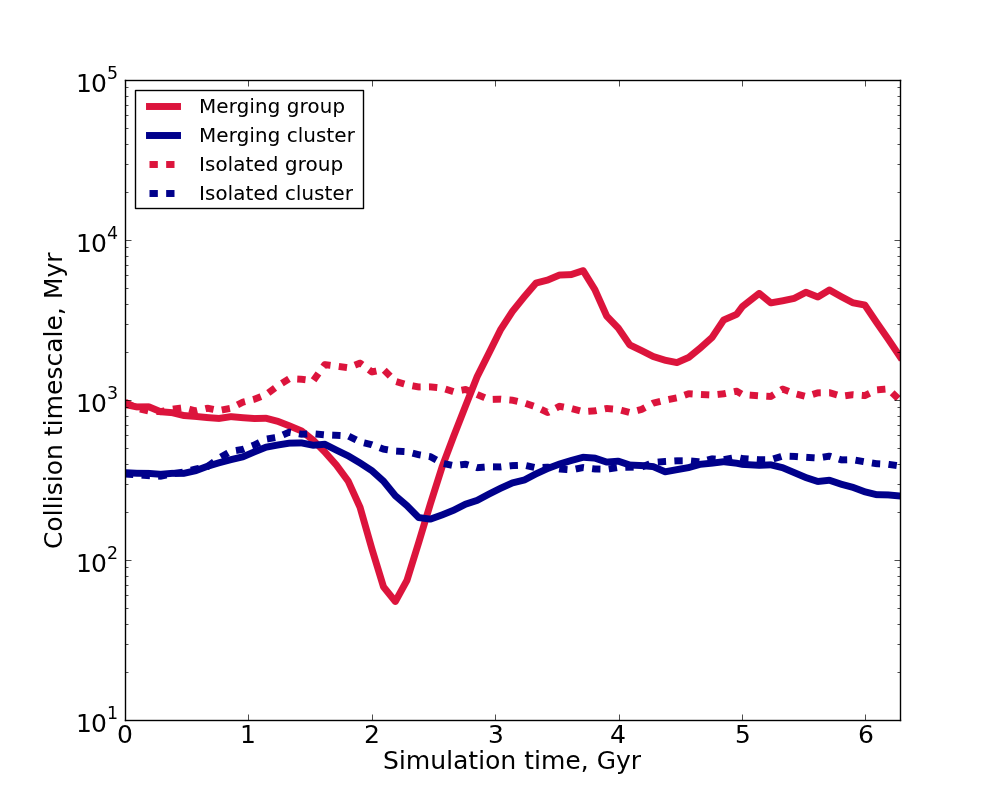}\label{fig:tcoll_gal}}
    \subfigure[Merger timescales]
    {\includegraphics[width=3.4in]{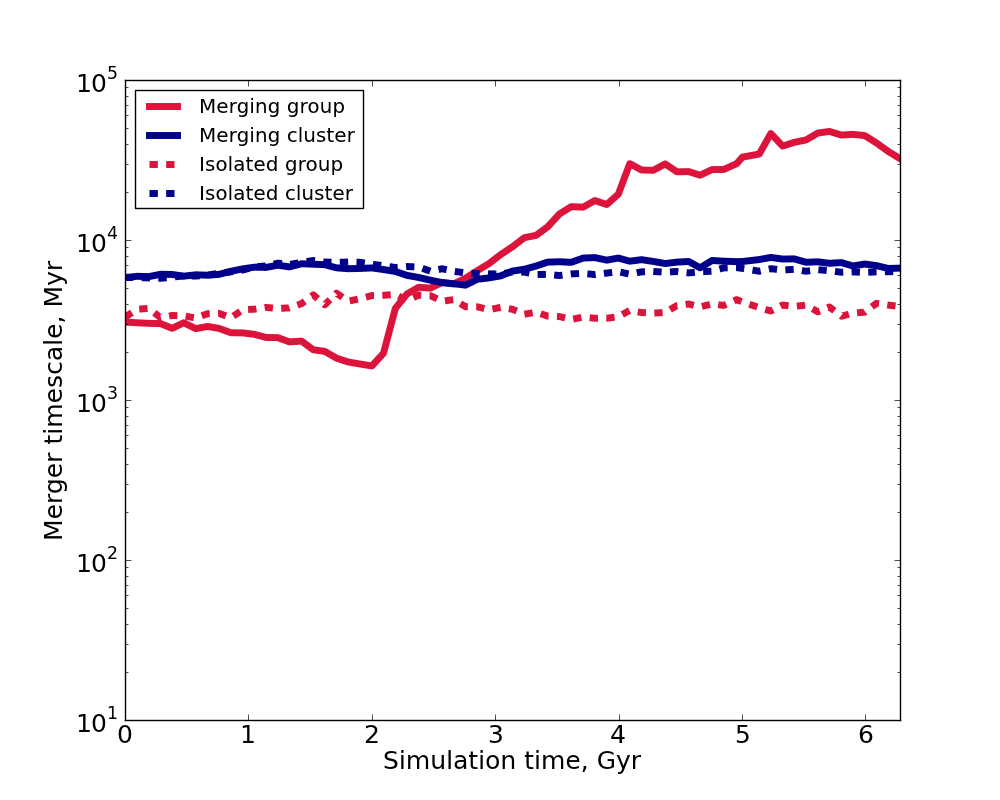}\label{fig:tmerg_gal}}
    \caption{Collision and merger timescales, in Myr, for galaxy particles in the merging group and cluster and isolated group and cluster. \label{fig:tcoll_tmerg_gal}}
  \end{center}  
\end{figure*}

The left panel of Figure~\ref{fig:tcoll_tmerg_gal} shows the time evolution of the average galaxy collision timescale. We see a drop in collision timescales for both the merging group and cluster at the pericentric passage, with the group showing a much larger decrease. The group reaches its minimum collision time $\sim 300$~Myr before the cluster. The collision timescale for the merging group increases from $\sim 50$ Myr at the group's pericentric passage to almost 8~Gyr at the group's apocentric passage, then oscillates between 2 and 5~Gyr. The cluster's $t_{\rm coll}$, on the other hand, remains relatively stable at $\sim 300$ Myr throughout the merger. The isolated group and cluster show much smaller-amplitude oscillations due to equilibration in the control runs.

To calculate the merger timescale, $t_{\rm merg}$, we use equation~\ref{eqn:tcoll}, but in place of $n_{\rm gal}$ we use the number density of merging galaxies, $n_{\rm merg}$, which only includes those galaxies with speeds relative to a given galaxy $v_{\rm rel}$ less than $3\sigma_{\rm gal}$. Here $\sigma_{\rm gal}$ is the internal velocity dispersion of a galaxy, for which we assume a uniform value $\sigma_{\rm gal} = 200$ km s$^{-1}$. The time evolution of $t_{\rm merg}$ is shown in the right panel of Figure~\ref{fig:tcoll_tmerg_gal}. We see that the group's galaxies have shorter merger timescales compared to the cluster before the merger ($t_{\rm merg} \simeq 3$~Gyr for the group and $\sim 6$~Gyr for the cluster). The average merger timescale of the group's galaxies decreases during the group's first pericentric passage. After the initial pericentric passage it steadily increases to almost $40$ Gyr as the group's components become distributed throughout the cluster. As with the collision timescale, the cluster galaxies' merger timescale remains relatively stable throughout the merger. In \S~\ref{sec:ggint} we discuss the reasons why the group's merger timescale does not approach the same 
value as the cluster's.

An important caveat to our calculations here is that galaxies that are less massive than those in our assumed-uniform population will have smaller radii and lower internal velocity dispersions \footnote{$r_{\rm gal} \propto M_{\rm gal}^{1/3}$ and $\sigma^2_{\rm gal} \propto M_{\rm gal}/r_{\rm gal} \sim M_{\rm gal}^{2/3}$.}. Consequently, the collision and merger timescales of lower-mass galaxies will be higher \footnote{$t_{\rm coll} \propto r_{\rm gal}^{-2} \sim M_{\rm gal}^{-2/3}$, in addition to fewer galaxies meeting the $v_{\rm rel} < 3\sigma_{\rm gal}$ merger criterion.}.

\subsection{Ram pressure and strangulation}
\label{sec:pram}
A galaxy containing diffuse gas and moving through a diffuse gaseous medium like the IGM or ICM experiences ram pressure which can strip it of its gas (\citealt{Gunn72}). The ram pressure experienced by a galaxy moving through a fluid medium is given approximately by
\begin{equation}
  P_{\rm ram} = \rho_{\rm gas}v_{\rm gal, ICM}^2 .
\end{equation}
Here $\rho_{\rm gas}$ is the density of the ambient ICM/IGM gas and $v_{\rm gal, ICM}$ is the velocity of the galaxy with respect to the surrounding gas. Stripping is effective when the ram pressure on a galaxy is greater than the gravitational restoring force on the galaxy's bound gas.

\begin{figure*} 
  \begin{center}
    \subfigure[Merging group and cluster]
    {\includegraphics[width=3.4in]{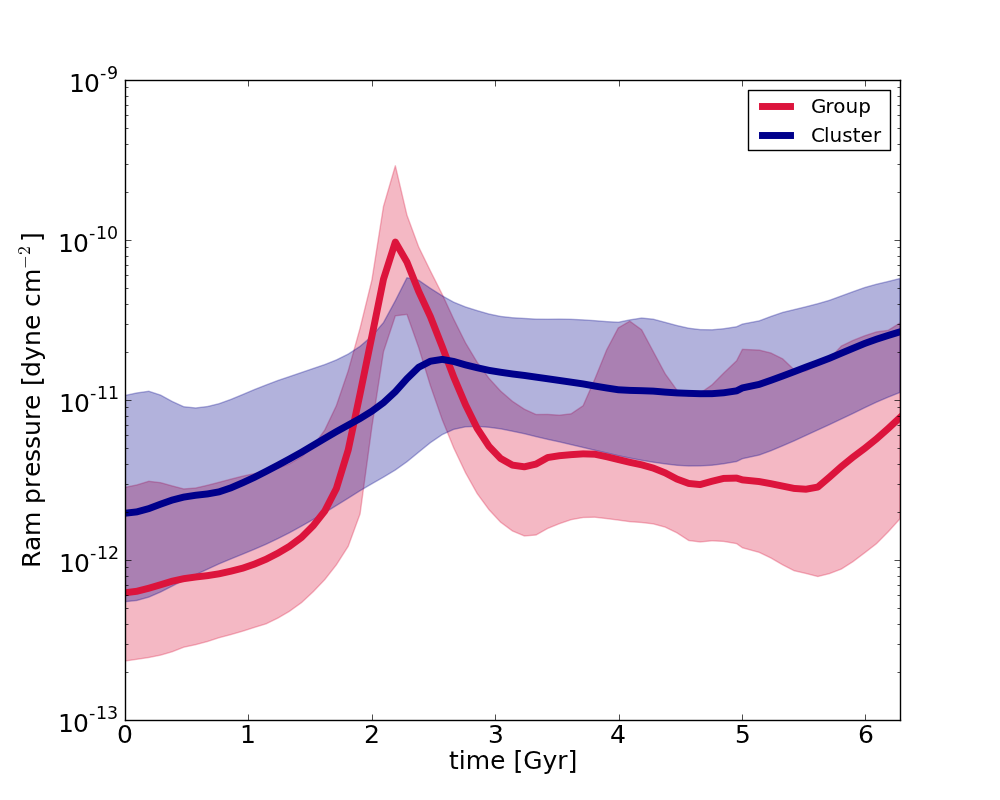}\label{fig:pram_merg}}
    \subfigure[Isolated group ancd cluster]
    {\includegraphics[width=3.4in]{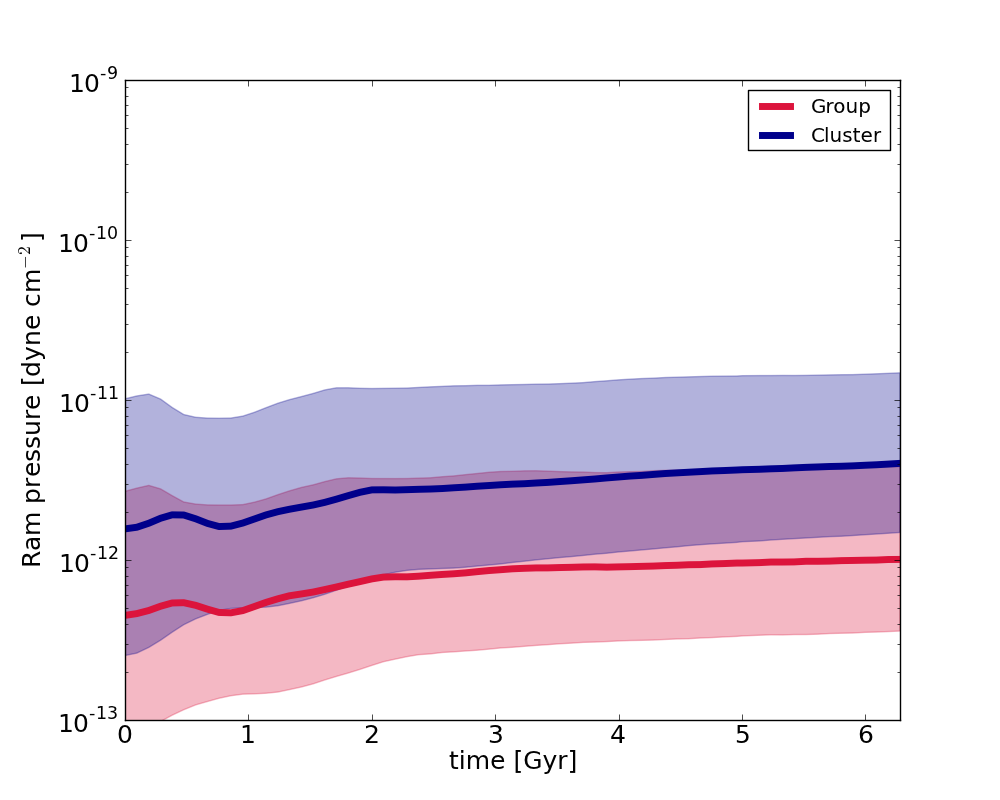}\label{fig:pram_iso}}
    \caption{Ram pressure on the merging group and cluster (left) and isolated group and cluster (right). The thick lines show the median values of the ram pressure and the shaded region shows the range of $P_{\rm ram}$ values between the 25th and 75th percentiles of the distribution. \label{fig:pram_merg_iso} }
  \end{center}  
\end{figure*}

Figure~\ref{fig:pram_merg_iso} shows the ram pressure on the collisionless particles (to which we later attach galaxy models) of the merging group and cluster and isolated group and cluster. Here too, we see the effect of the group's pericentric passage: at $t \simeq 2.3$ Gyr, there is a significant increase in the ram pressure on the group's particles. Additionally, we see that the merger leads to an overall increase in the ram pressure on the cluster's particles. Ram pressures of $10^{-11}$ dyne cm$^{-2}$ can strip a typical disk galaxy of its gas within a few million years (\citealt{Gunn72}, \citealt{Bruggen08}). From Figure~\ref{fig:pram_merg_iso}, we see that while the cluster's galaxy particles can be subject to ram pressures of $10^{-11}$ dyne cm$^{-2}$ even before the merger, the group's particles experience ram pressures of this extent only during and after the merger. 

\begin{figure*} 
  \begin{center}
    \subfigure[Group core and halo]
    {\includegraphics[width=3.4in]{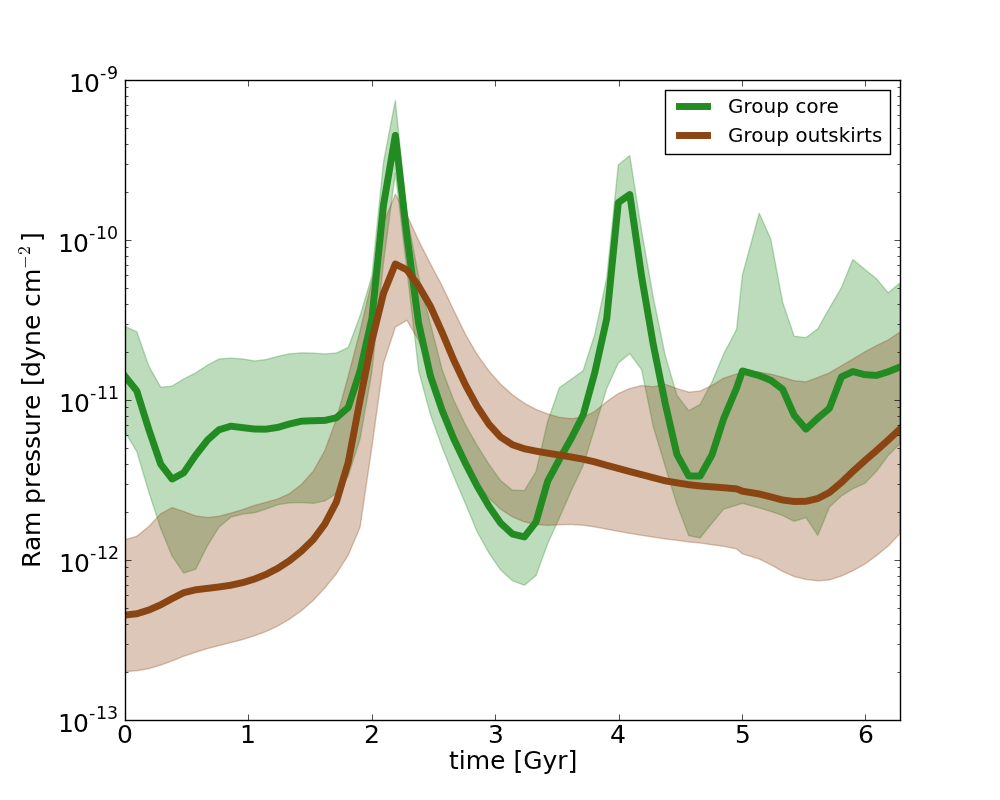}\label{fig:pram_mgroup}}
    \subfigure[Cluster core and halo]
    {\includegraphics[width=3.4in]{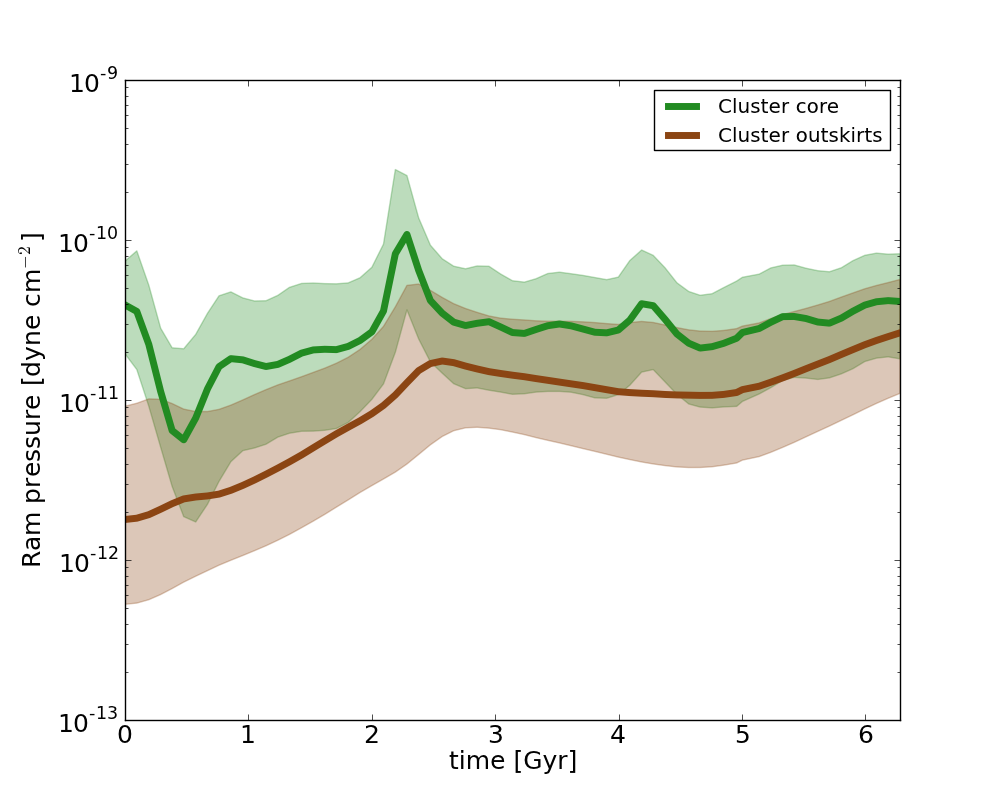}\label{fig:pram_mcluster}}
    \caption{Ram pressure on the core ($r < r_{\rm s}$) and outskirts ($r > r_{\rm s}$) of the group and cluster. As in the previous figure, the thick lines show the median values of the ram pressure, and the shaded regions show the range of $P_{\rm ram}$ values between the 25th and 75th percentiles of the distribution. \label{fig:pram_core_halo}}
  \end{center}  
\end{figure*}

We calculate the ram pressures on the merging group and cluster's core ($r < r_s$) and outskirts ($r > r_s$) particles to see where the boost in ram pressure due to the merger is most effective. Figure~\ref{fig:pram_core_halo} shows the evolution of these quantities. We see that while particles in both the group's and cluster's cores are subject to higher overall ram pressures, these regions do not experience a significant change in ram pressure due to the merger until the group's initial pericentric passage. The outer particles, on the other hand, are initially subject to much lower values of ram pressure, but $P_{\rm ram}$ increases rapidly as the group falls in. At the pericentric passage, the ram pressure on the group's outskirts is comparable to that on the group's core. The infalling group also boosts the ram pressure on the cluster's outer particles, even at the beginning of the merger. 

\begin{figure*} 
  \begin{center}
    \subfigure[Gas density]
    {\includegraphics[width=3.4in]{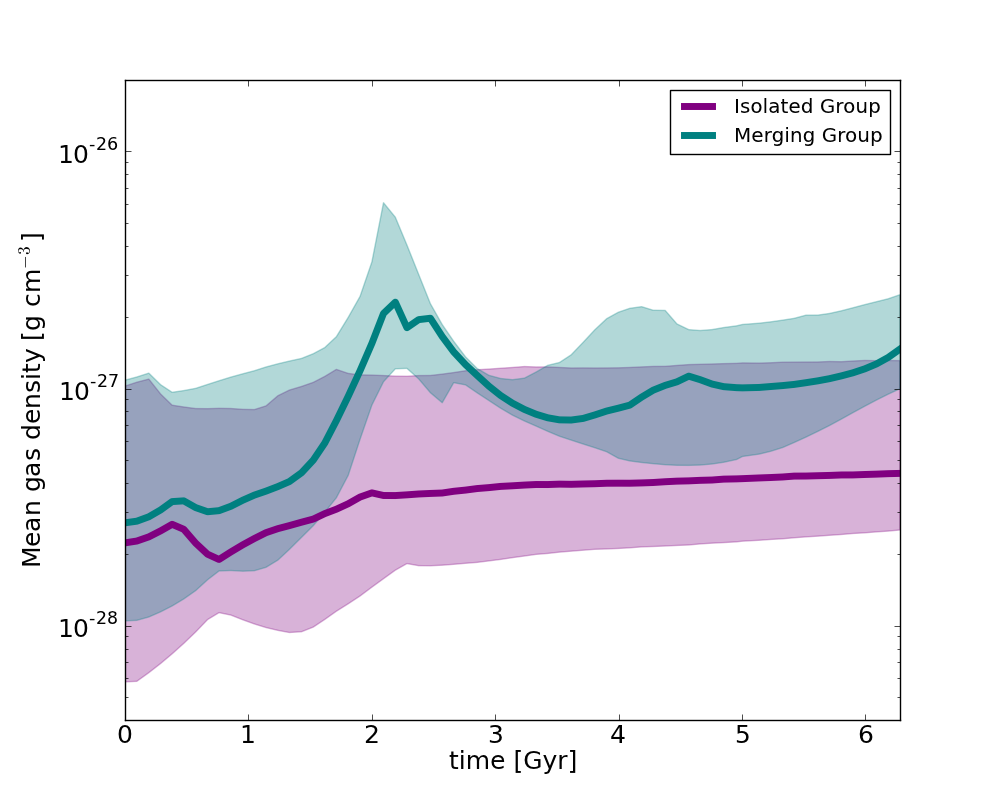}\label{fig:rho_pram_group}}
    \subfigure[Velocity$^2$ w.r.t. gas]
    {\includegraphics[width=3.4in]{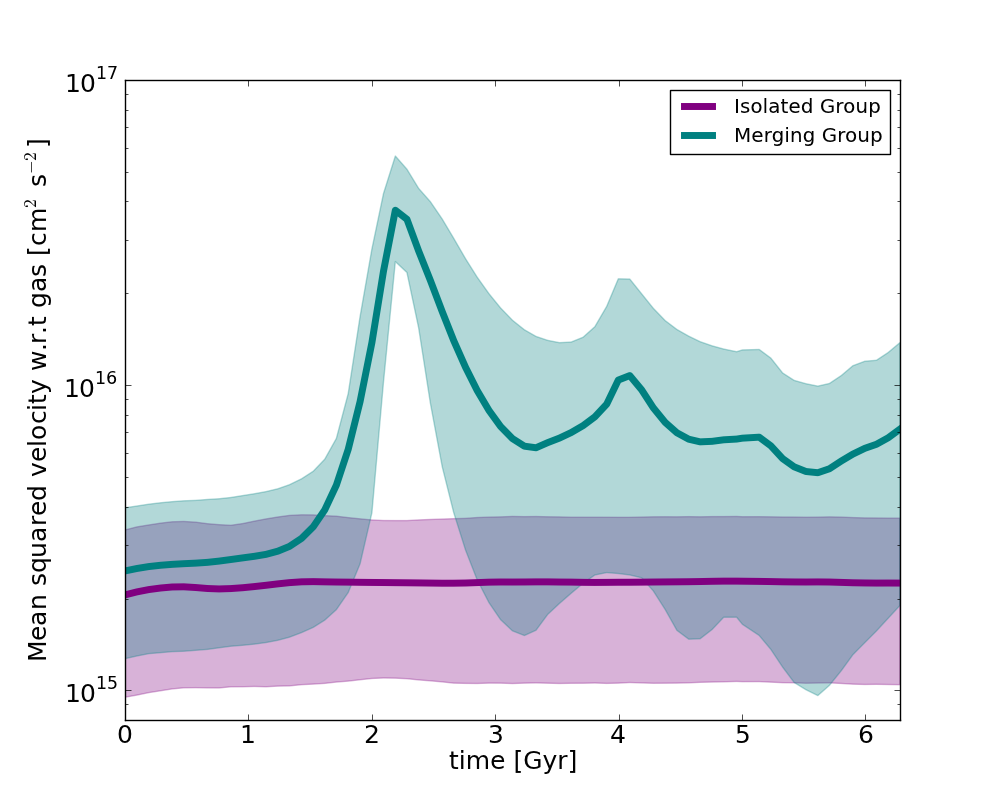}\label{fig:vel_pram_group}}
     \caption{Left: The average gas density mapped at the positions of the group's collisionless particles. The thick lines represent the average gas density, and the shaded regions represent the spread in densities between the 25th and 75th percentiles of the distribution. Right: The evolution in the average squared velocity of the group's particles with respect to the gas. The thick lines are the median relative velocities, and the shaded regions show the limits corresponding to the 25th and 75th percentiles of the distribution. The purple curves correspond to the isolated group, and the teal curves correspond to the merging group.  \label{fig:rho_vel_pram_group}}
  \end{center}  
\end{figure*}

\begin{figure*} 
  \begin{center}
    \subfigure[Gas density]
    {\includegraphics[width=3.4in]{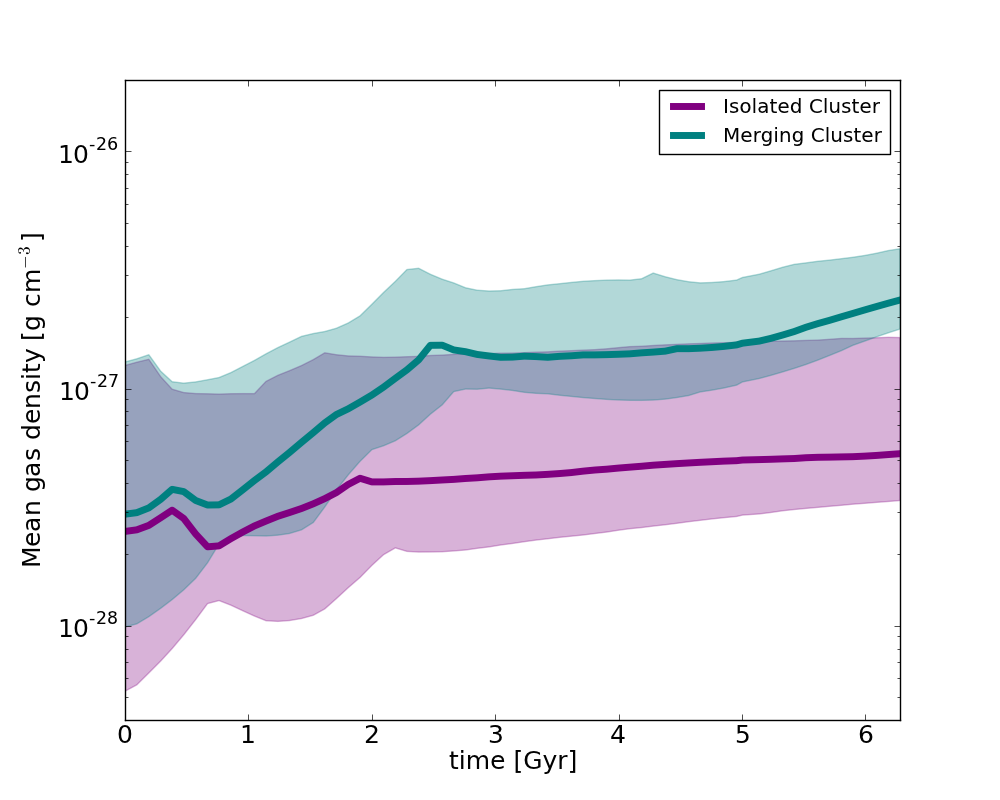}\label{fig:rho_pram_cluster}}
    \subfigure[Velocity$^2$ w.r.t. gas]
    {\includegraphics[width=3.4in]{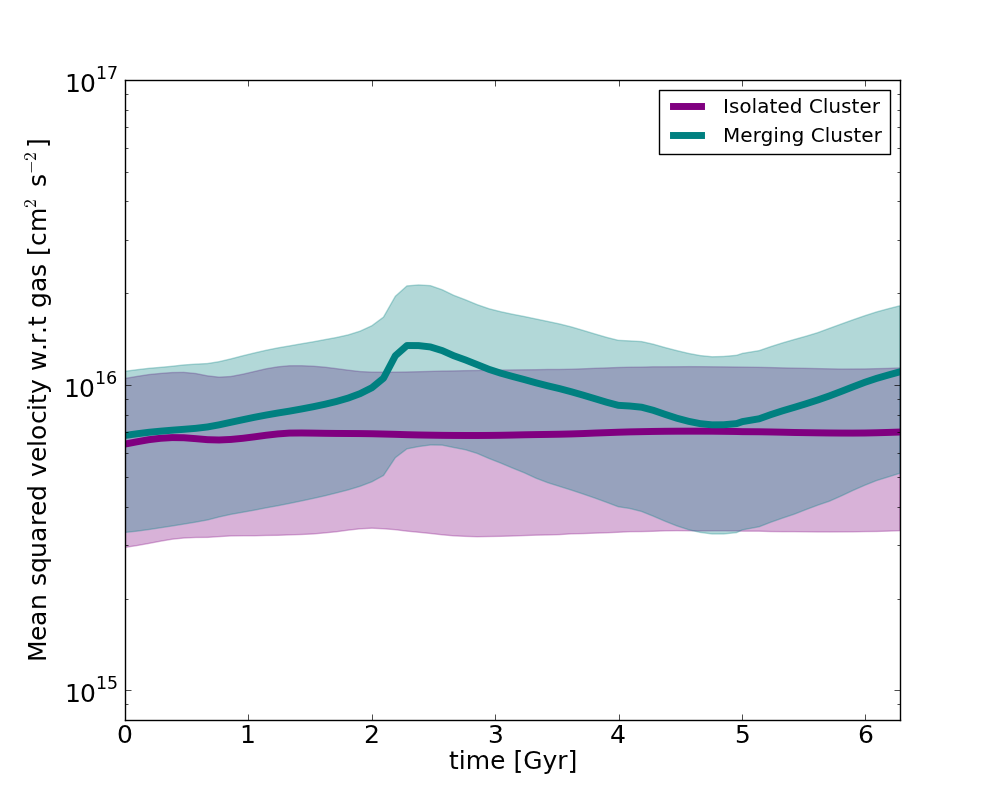}\label{fig:vel_pram_cluster}}
     \caption{Left: The average gas density mapped at the positions of the cluster's collisionless particles. Right: The average velocity squared of the cluster's collisionless components with respect to the gas. The lines and colors have the same meanings as those in Figure~\ref{fig:rho_vel_pram_group} \label{fig:rho_vel_pram_cluster}}
  \end{center}  
\end{figure*}

The sharp increase in ram pressure on the group and cluster during initial infall is a consequence of an increase in both the average gas density (Figures~\ref{fig:rho_pram_group} and \ref{fig:rho_pram_cluster}) and the velocity of the particles with respect to the gas  (Figures~\ref{fig:vel_pram_group} and \ref{fig:vel_pram_cluster}). Compression due to the merger shock increases the gas density encountered by the particles, and the shock also sets the gas into motion with respect to the average rest frames of the group and cluster's particles. As a result, $P_{\rm ram}$ increases temporarily by a factor of $\sim 100$. As with the relative velocity of the collisionless components, the spread in the relative velocity of the gas also increases due to the merger. At later times, the average velocity of the cluster particles with respect to the gas briefly decreases and then increases as the gas sloshes relative to the cluster's particles. The median gas density also increases compared to the density before the merger, following the deepening of the cluster's gravitational potential well after the merger. Thus, there is an overall increase in the cluster's ram pressure.

The ram pressure acting on a galaxy can remove its hot gaseous halo (\citealt{Larson80}, \citealt{McCarthy08}). This removal of gas that can eventually fuel star formation is sometimes referred to as `strangulation' or `starvation'. To quantify the contribution of ram pressure stripping-driven strangulation towards pre-processing, we compare the gravitational restoring force within a model galaxy to the ram pressure acting on it at a given time. Our two-component (dark matter + gas) model galaxies are spherically symmetric and have total density profiles corresponding to an NFW profile, with initial parameters $R_{200} = 100$ kpc, $M_{200} = 1.7 \times 10^{11} ~\mbox{M}_{\odot}$, and $c_{200} = 10$. We assume that the density profile of the hot halo gas is described by that of a singular isothermal sphere, $\rho_{\rm gas}(r) = \rho_0 r_0^2 / r^2$, and the total gas mass is 10\% of the total mass ($M_{\rm gas} = 1.7 \times 10^{10} ~\mbox{M}_{\odot}$). Following \citet{Gunn72} and \citet{McCarthy08}, we use $P_{\rm ram} > F_{\rm rest}/A$ as the condition for ram pressure stripping. $F_{\rm rest}/A$ is the gravitational restoring force per unit surface area on the gas, given by
\begin{equation}
  F_{\rm rest}(r)/A = \Sigma_{\rm gas}(r)a_{\rm max}(r).
\end{equation} 
$\Sigma_{\rm gas}(r)$ is the projected surface density of gas at a radius $r$ and can be calculated using
\begin{equation}
  \Sigma_{\rm gas}(r) = \int_{-\infty}^{\infty} \rho_{\rm gas}(\sqrt{r^2+z^2}) dz = \pi r \rho_{\rm gas}(r).
\end{equation}
The maximum of the acceleration due to gravity in the direction of motion of the galaxy with respect to the gas at a radius $r$ is $a_{\rm max}(r) = GM_{\rm tot}(r)/2r^2$. 

We use the positions and velocities of 26 and 152 randomly selected group and cluster particles respectively as those of galaxy particles within the group and cluster. We then compare the ram pressure on these particles through the course of the simulation to the maximum internal galactic gravitational restoring force per unit surface area. At a given timestep, we can calculate a radius $r_{\rm ram}$ (the `stripping radius') at which $P_{\rm ram} \geq F_{\rm rest}/A$. We do not allow the density profiles of our model galaxies to adjust in response to the stripping of gas. Consequently, even if the ram pressure on a model galaxy particle decreases at a later time in the simulation, $r_{\rm ram}$ cannot increase. We assume that all the hot gas outside $r_{\rm ram}$ is lost due to stripping instantaneously when the above condition is met. We repeat this calculation for an ensemble of model galaxies and average our calculated values of $r_{\rm ram}$ in different radial bins over 50 random realizations of galaxy particles' positions and velocities within the merging and isolated group and cluster.  

\begin{figure*} 
  \begin{center}
    \subfigure[Merging group particles]
    {\includegraphics[width=3.4in]{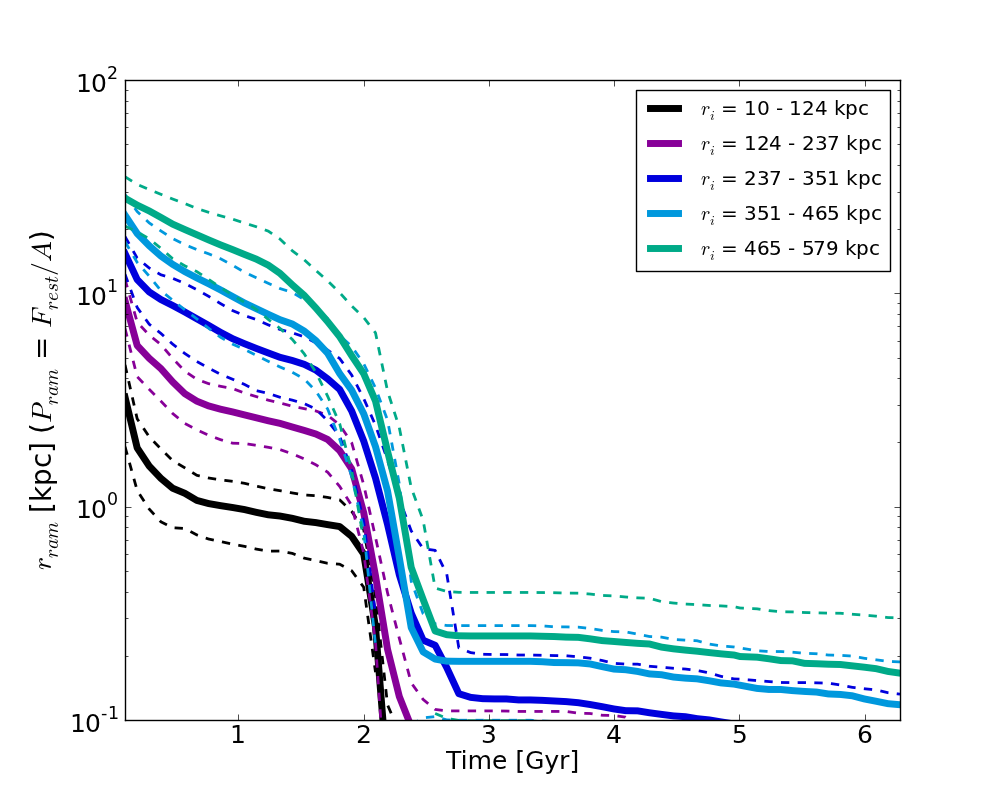}\label{fig:req_mgroup}}
    \subfigure[Merging cluster particles]
    {\includegraphics[width=3.4in]{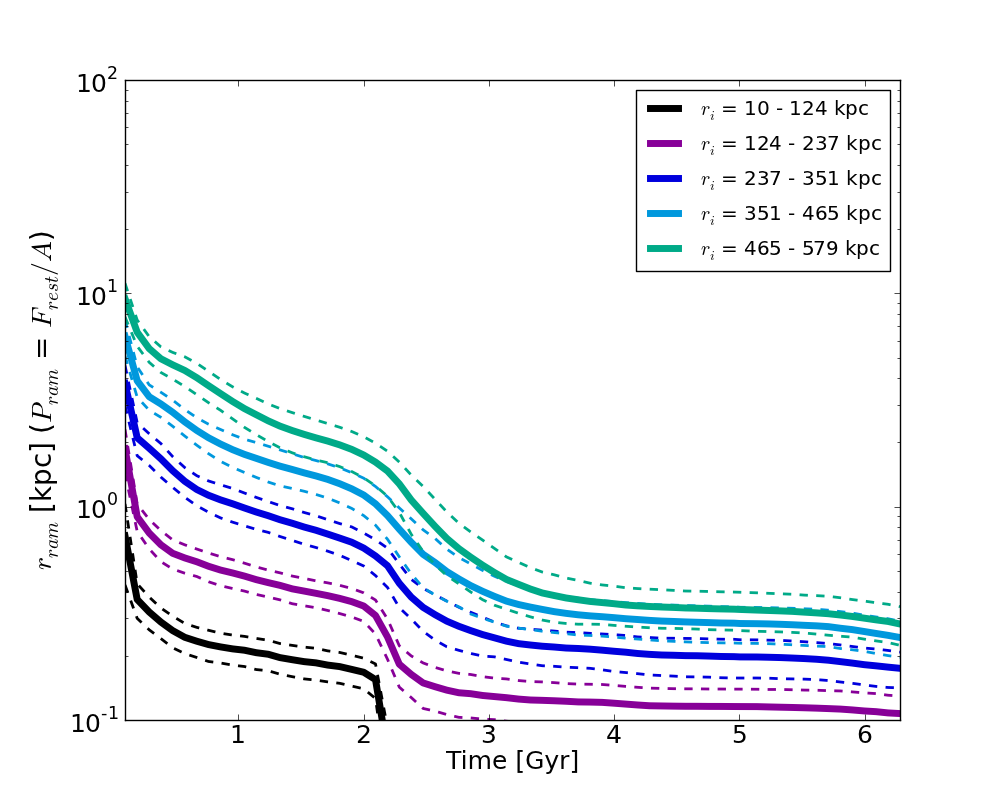}\label{fig:req_mcluster}}
    \\
    \subfigure[Isolated group particles]
    {\includegraphics[width=3.4in]{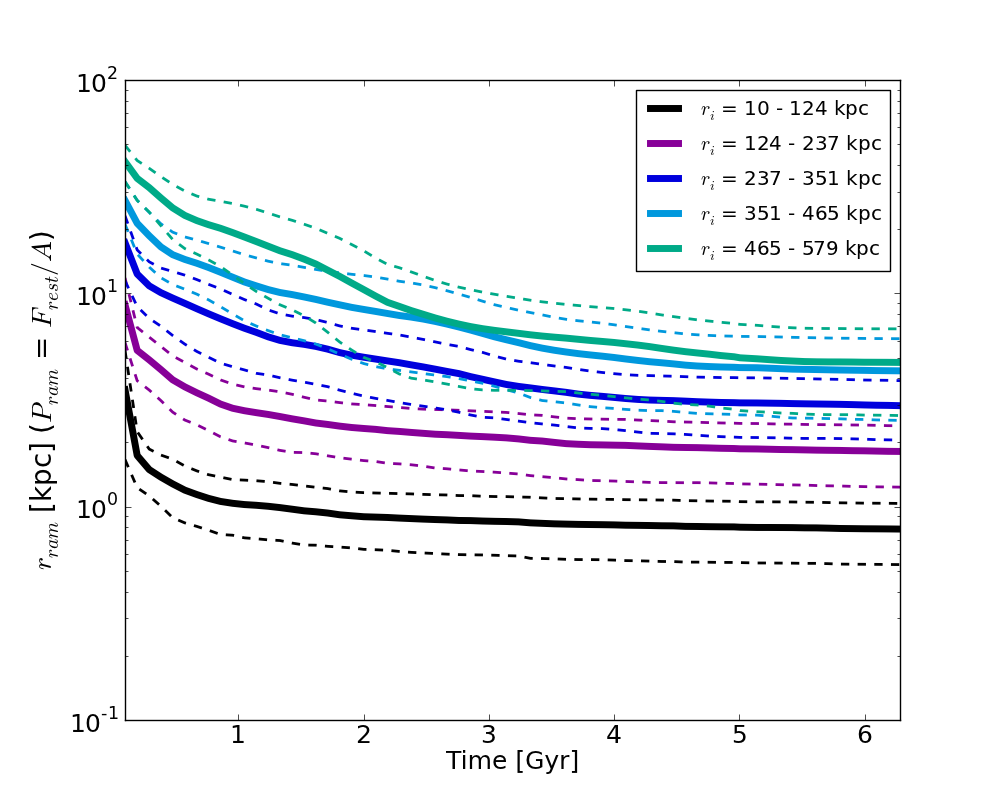}\label{fig:req_igroup}}
    \subfigure[Isolated cluster particles]
    {\includegraphics[width=3.4in]{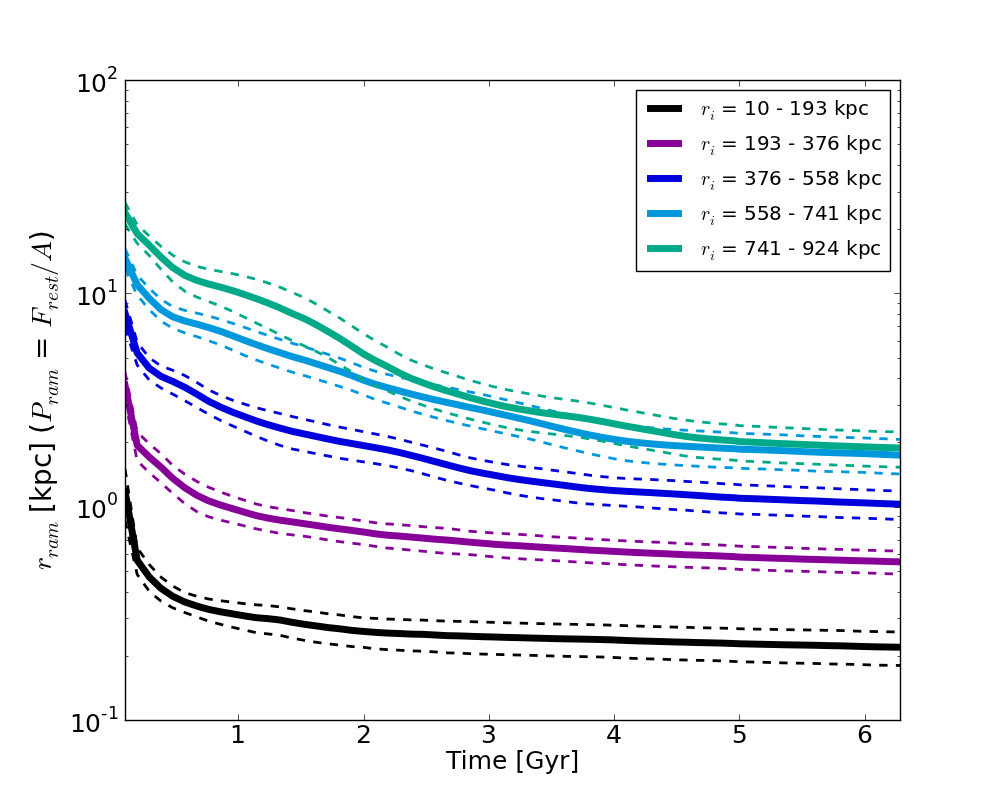}\label{fig:req_icluster}}
    \caption{Evolution of the minimum radius ($r_{\rm ram}$) where ram pressure exceeds gravitational restoring force per unit surface area ($P_{\rm ram} \geq F_{\rm rest}/A$) for galaxy particles in the group and cluster. Galaxy particles are binned in five radial bins according to their initial halo-centric radius, $r_i$. The cluster has a larger virial radius; therefore its galaxy particles have a larger range of $r_i$ values. The dashed lines correspond to the $1\sigma$ limits in the distribution of $r_{\rm ram}$ in each radial bin. \label{fig:req_migroup_micluster}}
  \end{center}  
\end{figure*}

Figure~\ref{fig:req_migroup_micluster} shows the evolution of $r_{\rm ram}$ for galaxy particles in the merging group and cluster (top) and isolated group and cluster (bottom). The effect of the merger and the impact of the group's first pericentric passage are evident when comparing the $r_{\rm ram}$ values of the merging and isolated group's model galaxies (Figs.~\ref{fig:req_mgroup} and \ref{fig:req_igroup}). At the beginning of the simulation, before the onset of the merger, the values of $r_{\rm ram}$ for the group and cluster galaxy particles are comparable. At $t \simeq 2$ Gyr, during the group's first pericentric passage, the ram pressure on the merging group's galaxy particles is strong enough to remove almost all the hot halo gas bound to a galaxy. The isolated group's galaxy particles, on the other hand, do not show as dramatic an evolution in $r_{\rm ram}$. Additionally, we see that particles that are initially closer to the halo center are stripped of their gas before those with larger halo-centric radii. We bin particles in five radial bins, up to the virial radius, using their initial halo-centric radii $r_i$. Of the merging group's galaxy particles, those in the first three radial bins ($r_i < 350$ kpc) are seen to be stripped to a few kpc before $\sim$ 1 Gyr, i.e., before the group falls into the cluster. The merger also enhances the ram pressure on the cluster's particles, as seen on comparing Figures~\ref{fig:req_mcluster} and \ref{fig:req_icluster}. The merging cluster's particles lose their gas before the isolated cluster's particles, and this loss predominantly happens after the group's pericentric passage, an event that corresponds to an overall increase in the cluster's ram pressure (seen in Figure~\ref{fig:pram_merg}). 

\subsection{Tidal stripping and truncation}
\label{sec:tidaltrunc}

One can define a tidal radius in a fashion analogous to our definition of the ram pressure stripping radius, $r_{\rm ram}$. The tidal truncation radius (or alternatively  the tidal radius), $r_{\rm tid}$, of a galaxy within a massive group or cluster halo is the galaxy-centric radius at which the tidal force due to the group or cluster halo balances the galaxy's gravitational force. $r_{\rm tid}$ can be estimated for a given galaxy as the galaxy-centric radius at which the density of the background halo at the galaxy's position exceeds the galaxy's density (\citealt{Gnedin03b}): $\rho_{\rm halo}(\mathbf{x}_{\rm gal}) \geq \rho_{\rm gal}(r_{\rm tid})$. We assign model galaxies to the positions and velocities of randomly selected particles, as in the previous section, and then calculate the evolution of $r_{\rm tid}$. The model galaxies initially have a total density profile corresponding to an NFW profile with the same parameters as in the previous section. The density of the background halo at the galaxy's position is $\rho_{\rm halo}(\mathbf{x}_{\rm gal}) = \rho_{\rm DM}(\mathbf{x}_{\rm gal}) + \rho_{\rm gas}(\mathbf{x}_{\rm gal})$. The dark matter density, $\rho_{\rm DM}$, is calculated from the positions of dark matter particles using a cloud-in-cell interpolation technique. At each timestep, we tabulate the values of $r_{\rm tid}$ for each of the model galaxies. If the background halo density increases at a later time, we tabulate the corresponding new value of $r_{\rm tid}$. Thus, as with $r_{\rm ram}$, $r_{\rm tid}$ cannot increase at a later time in the simulation, since we assume that all of a galaxy's components outside $r_{\rm tid}$ are instantaneously stripped when $\rho_{\rm halo}(\mathbf{x}_{gal}) \geq \rho_{\rm gal}(r_{\rm tid})$. 

\begin{figure*} 
  \begin{center}
    \subfigure[Merging group particles]
    {\includegraphics[width=3.4in]{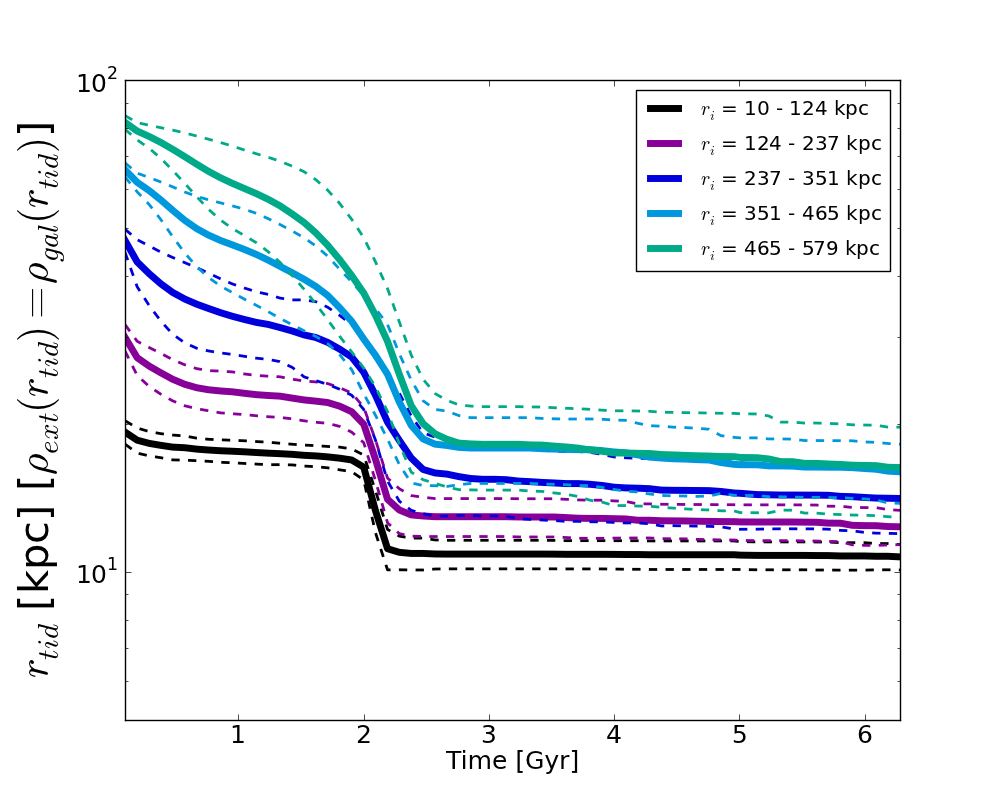}\label{fig:rtid_mgroup}}
    \subfigure[Merging cluster particles]
    {\includegraphics[width=3.4in]{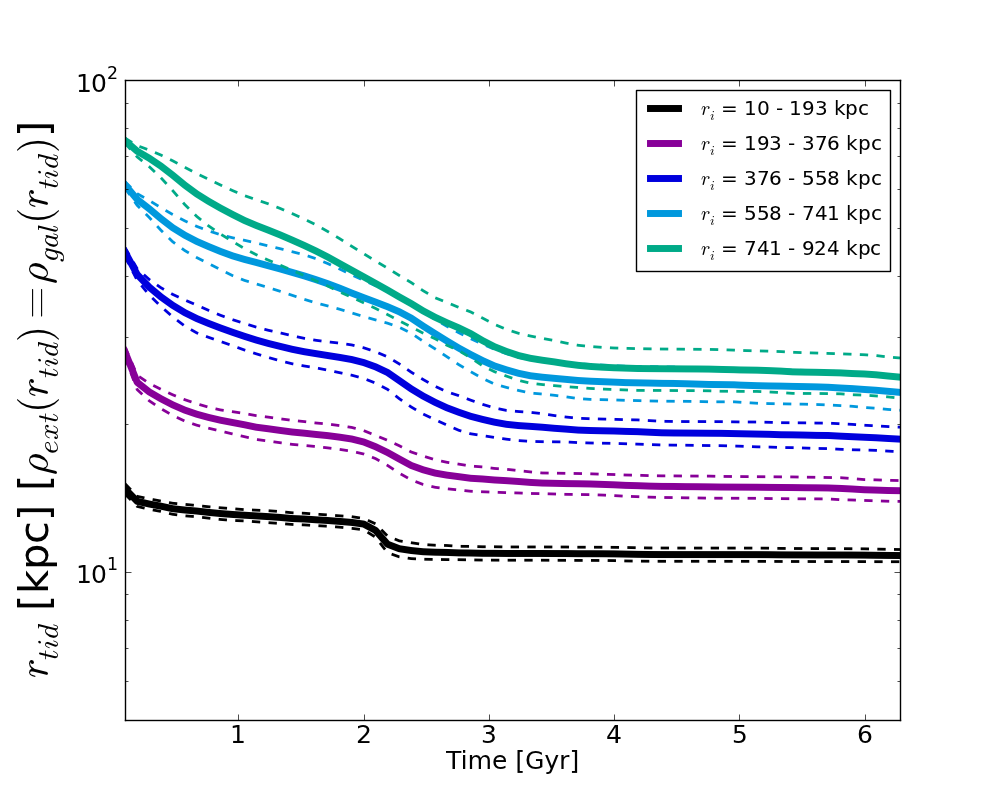}\label{fig:rtid_mcluster}}
    \\
    \subfigure[Isolated group particles]
    {\includegraphics[width=3.4in]{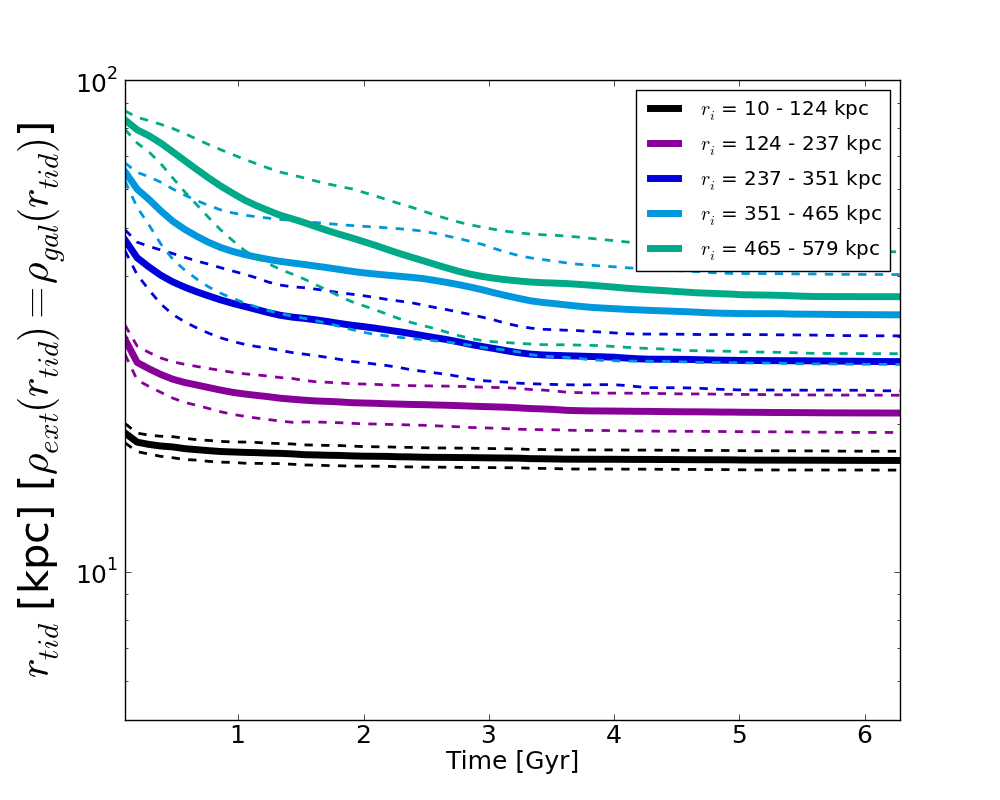}\label{fig:rtid_igroup}}
    \subfigure[Isolated cluster particles]
    {\includegraphics[width=3.4in]{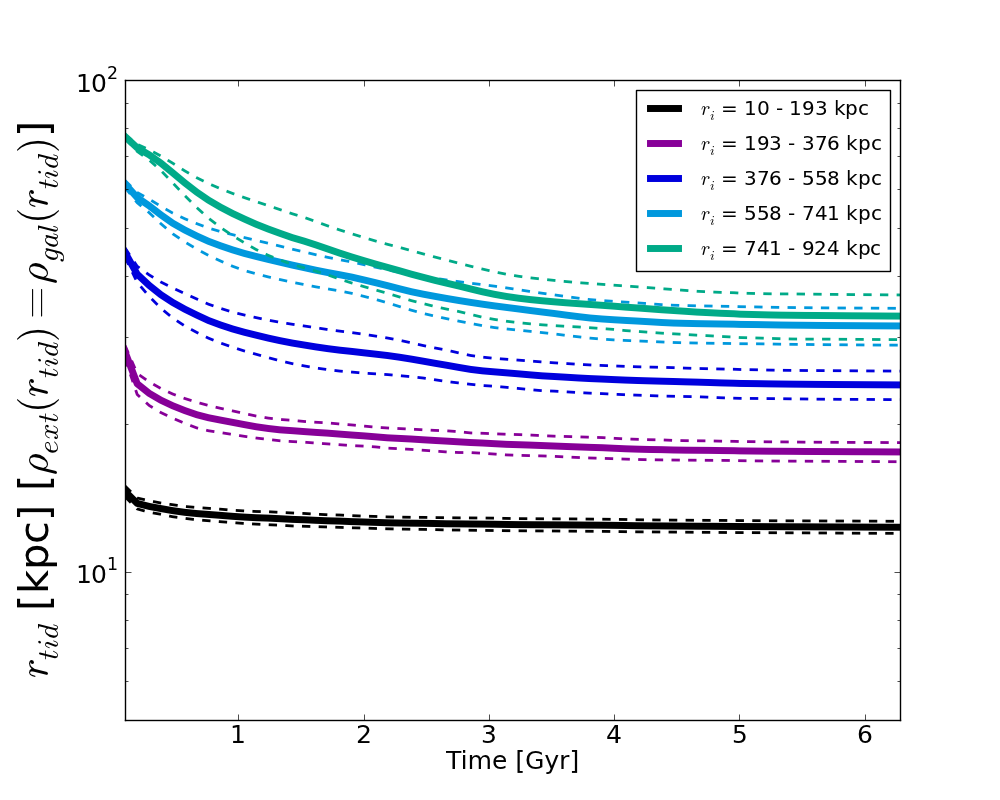}\label{fig:rtid_icluster}}
    \caption{Evolution of the tidal radius ($r_{\rm tid}$) for galaxy particles in the merging and isolated group and cluster. As in the plots of $r_{\rm ram}$, galaxy particles are binned by their initial halo-centric radius $r_i$. The dashed lines correspond to the $1\sigma$ limits in the distribution of $r_{\rm tid}$ in each radial bin \label{fig:rtid_migroup_micluster}}
  \end{center}  
\end{figure*}

Figure~\ref{fig:rtid_migroup_micluster} shows the evolution of $r_{\rm tid}$ for model galaxy particles (averaged over 50 ensembles of galaxies) in the merging (top) and isolated (bottom) group (right) and cluster (left). As in the evolution of $r_{\rm ram}$ in Figure~\ref{fig:req_migroup_micluster}, we see that the group's model galaxies are subject to a significant enhancement in the local density of the background halo, and therefore, decrease in tidal radius, at the group's pericentric passage (Figure~\ref{fig:rtid_mgroup}). The isolated group's model galaxies, on the other hand, do not show any significant decrease in their tidal radii (Figure~\ref{fig:rtid_igroup}). After about 4 Gyr, the tidal radii of the merging group's galaxies (with $r_i < 350 $ kpc) are $\sim 10$ kpc, while those of the isolated group are $\sim 20 - 30$ kpc. Additionally, as with $r_{\rm ram}$, the merging cluster's model galaxies (Figure~\ref{fig:rtid_mcluster}) are subject to greater tidal stripping around the time of the group's pericentric passage, a feature absent in the isolated cluster (Figure~\ref{fig:rtid_icluster}). At the end of the simulation, the lower limits of the tidal radii of the merging group and cluster particles are comparable, while the isolated group particles' tidal radii are larger than those of the isolated cluster. The merging group's particles also have smaller overall tidal radii than the merging cluster particles after $\sim 3$ Gyr. 

We note here that our assumption that $r_{\rm tid}$ is nonincreasing is not strictly accurate, since a galaxy can recapture its tidally stripped material once it moves beyond its orbital pericenter. However, if we allow for such an increase in our calculation, $r_{\rm tid}$ can increase to values larger than $r_{\rm tid}$ at $t = 0$, particularly during a galaxy's apocentric passages. Our simple model of proxy galaxy particles cannot properly account for this tidal recapture. We see in Figures~\ref{fig:rtid_mgroup} and~\ref{fig:rtid_mcluster} that most of the decrease in $r_{\rm tid}$ happens during the group's pericentric passage, when the potential on the group's (and some of the cluster's) galaxies changes rapidly. Hence tidally stripped material may not remain close to the galaxy from which it was removed. Repeated pericentric passages of a galaxy or a group within a cluster should make any recapture a temporary phenomenon.

\subsection{Tidal distortion of the merging group}
In our idealized resimulation, we assume that the group and cluster merge as spherically symmetric halos in equilibrium. However, as seen in Figure~\ref{fig:fig1}, the group is stretched out along the direction of infall before it falls into the cluster (when the group and cluster centers are separated by at least 4 Mpc). This could be a consequence of the group falling in along a cosmological filament or the effect of the cluster's tidal field, or a combination of both. We investigate the second possibility in the idealized resimulation by increasing the initial separation of the merging group and cluster to 10 Mpc. We can therefore study the effect of tidal distortion on the group's components before they are stripped away by the cluster. 

We use the angular variation in the group's radial density profile to quantify its tidal distortion. To do so, we bin the group's particles into a uniform $100^3$ grid of side 2 Mpc. We then calculate the mean number of particles ($\mu_{\rm part}$) and the standard deviation in the number of particles in a grid cell ($\sigma_{part}$) as a function of the cell's halo-centric radius. We use the coefficient of variation, $c_v$, as a measure of the distorted density profile, where $c_v = {\sigma_{\rm part}}/{\mu_{\rm part}}$. For a given radius, as the tidal force increases, the density increases along the direction of the tidal gravitational force and decreases in the directions normal to the force (\citealt{Chandrasekhar33}): the larger the tidal force, the larger the difference in densities and the deviation from spherical symmetry. As the deviation from spherical symmetry increases, the density distortion measured by $c_v$ increases. 

\begin{figure*}
  \begin{center}
    {\includegraphics[width=5in]{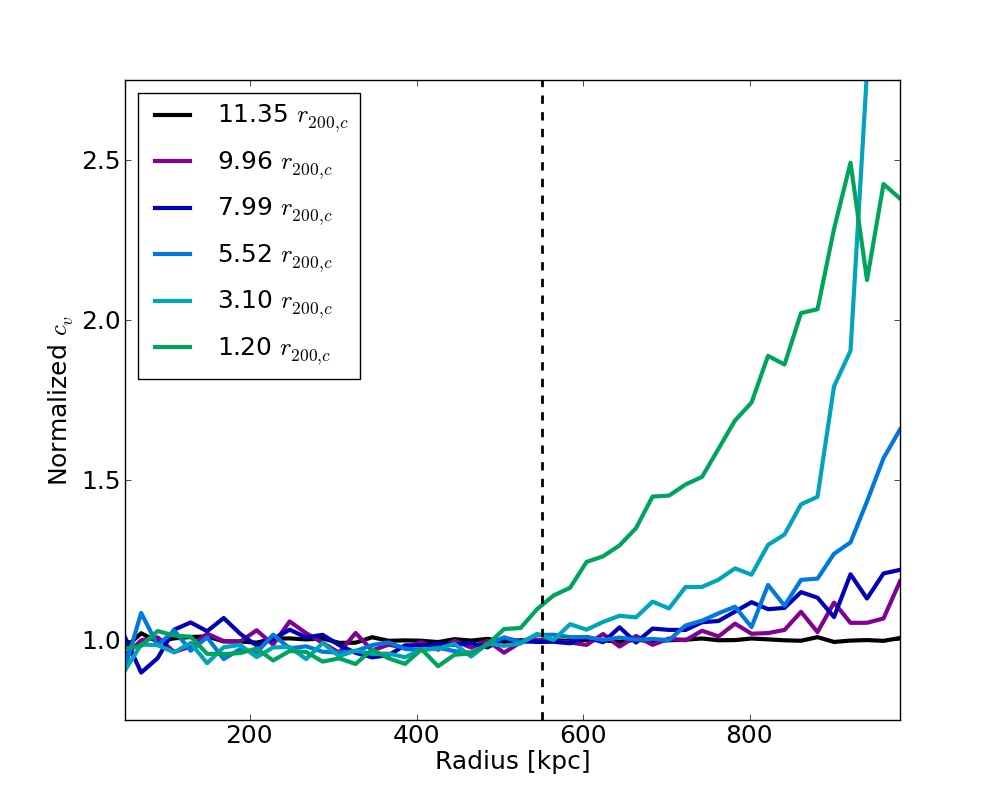}}
    \caption{Coefficient of variation, $c_v = {\sigma_{\rm part}}/{\mu_{\rm part}}$, for the merging group. Line colors correspond to the separation between the group and cluster centers at a given time in units of $R_{200,c}$, the cluster's virial radius (880 kpc). Here $c_v$ is normalized to the corresponding value in an isolated group at the same simulation time. The black dashed line shows the location of the group's virial radius ($R_{200,g} = 551$ kpc). \label{fig:cv_groupdens_norm}}
  \end{center}  
\end{figure*}

Figure~\ref{fig:cv_groupdens_norm} shows $c_v$ as a function of radius for the merging group. Here $c_v$ is normalized to the corresponding value calculated for an isolated group at the same simulation time. Thus, we can minimize the effects of particle shot noise in this calculation and account for the group halo's `breathing' due to initialization errors. The colors of the curves in this plot correspond to the separation between the group and cluster centers, in units of the cluster's virial radius, at different points in time. From this figure, we see that $c_v$, and therefore the tidal distortion of the group, increases with decreasing group-cluster separation. $c_v$ increases significantly starting at a separation of $\sim$ 4.6 $R_{200,c}$ and continues to increase until the group reaches the cluster. However, this effect is not significant within the group's virial radius (550 kpc). Inside this region the group's self-gravity is stronger than the cluster's tidal field. 

We also use power ratios as a second independent estimate of the distorted morphology of the merging group. These have been in used in cluster X-ray studies (\citealt{Buote95}, \citealt{Yang09}) to quantify the morphologies of cluster surface brightness maps. Power ratios are based on the multipole expansion of the two-dimensional gravitational potential, $\Psi(R, \phi)$, which satisfies
\begin{equation}
  \nabla^2\Psi(R, \phi) = 4\pi G \Sigma(R, \phi).
\end{equation}
Here, $\Sigma(R, \phi)$ is the density of the group halo projected along the $z$ axis, $R$ is the halo-centric radius in the $xy$ plane, and $\phi$ is the azimuthal angle. The multipole expansion of $\Psi(R, \phi)$ is
\begin{equation}
  \Psi(R, \phi) = -2G \left [ a_0 \ln \left(\frac{1}{R}\right) + \sum_{m=1}^{\infty} \frac{1}{m R^m} (a_m \cos m\phi + b_m \sin m\phi) \right ].
\end{equation}
$a_m$ and $b_m$ are the moments, given by
\begin{equation}
  a_m (R) = \int_{R' \leq R} \Sigma(x')(R')^m \cos{m\phi'} d^2 x'
\end{equation}
\begin{equation}
  b_m (R) = \int_{R' \leq R} \Sigma(x')(R')^m \sin{m\phi'} d^2 x'.
\end{equation} 
The power ratios are then defined as $P_m / P_0$, where
\begin{equation}
  P_0 (R) = \left(a_0 \ln R\right)^2
\end{equation}
\begin{equation}
  P_m (R) = \frac{1}{2m^2R^{2m}}\left(a_m^2 + b_m^2\right),\ m > 0.
\end{equation}
Each power ratio $P_m / P_0$ is an estimate of the $m$th multipole moment of the surface density of the tidally stretched group. $P_1/P_0$ is a measure of the dipole power, or mirror asymmetry, and should not change by definition ($R = 0$ corresponds to the group's center). $P_2/P_0$ is a measure of the quadrupole power and increases with more elliptical morphologies; it should therefore increase monotonically for the tidally stretched group as the group moves closer to the cluster. $P_3/P_0$ is a measure of unequally sized bimodal structures and should not change significantly as long as the group is well outside the cluster. 

\begin{figure*} 
  \begin{center}
    {\includegraphics[width=5in]{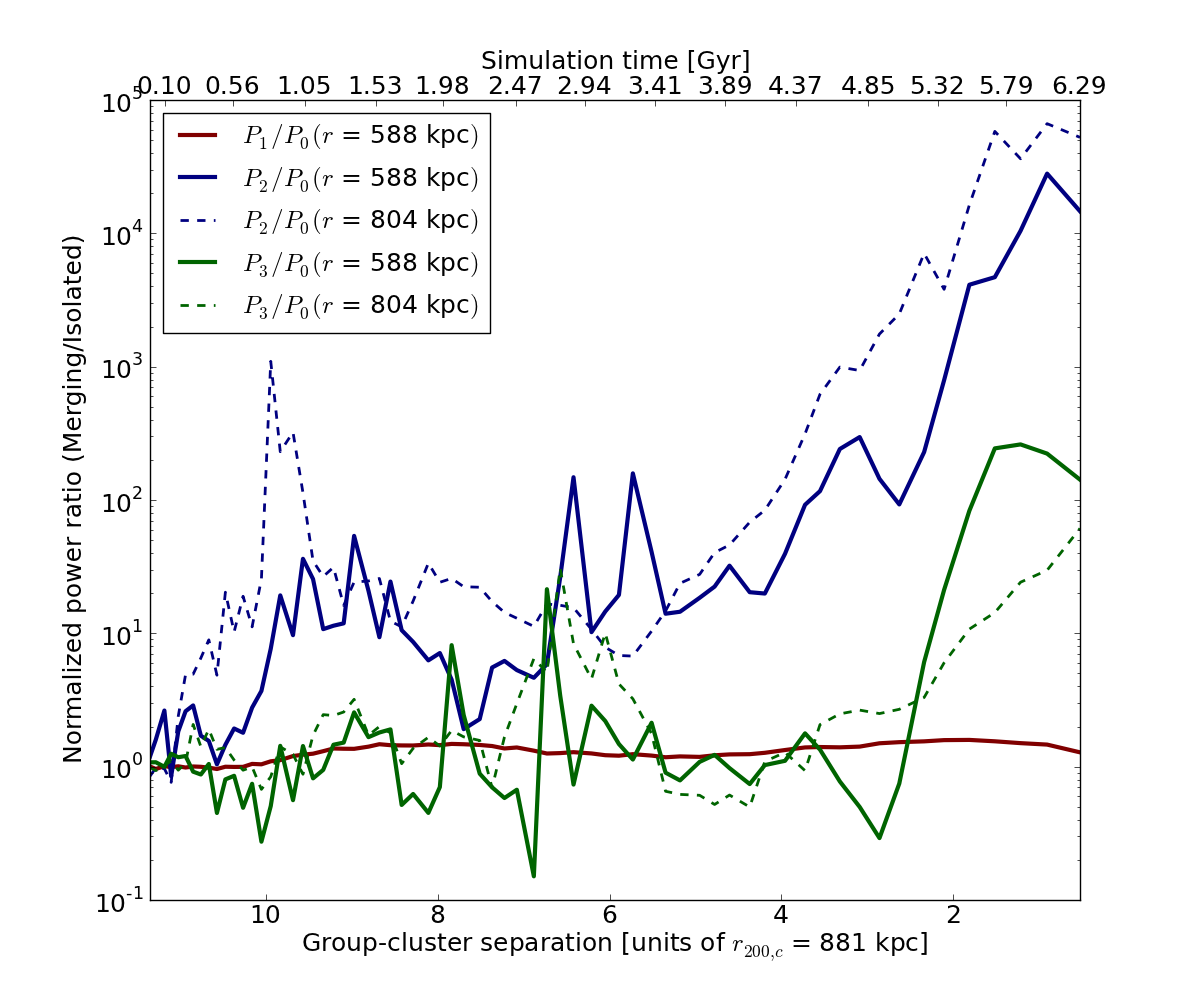}}
    \caption{Normalized power ratios $\left(\frac{P_m}{P_0}\right)\lvert_{\rm merging} / \left(\frac{P_m}{P_0}\right)\lvert_{\rm isolated}$ as functions of time for the merging group. Solid and dashed lines correspond to the two different values of group-centric radius within which the contribution of the power due the $m$th multipole moment is measured.\label{fig:power_ratio}}
  \end{center}  
\end{figure*}

Figure~\ref{fig:power_ratio} shows the evolution of $P_1/P_0$, $P_2/P_0$, and $P_3/P_0$ for the 
merging group (normalized to the corresponding values for the isolated group at each point in time). The solid lines correspond to the power ratios measured for the multipole moments within 580 kpc, just outside the group's virial radius, and the dashed lines are the power ratios well outside the virial radius at 804 kpc. $P_1/P_0$ shows little evolution, as expected. The normalized value of $P_2/P_0$ is expected to be the most sensitive of all power ratios for the elliptically distorted group, and this is indeed seen in Figure~\ref{fig:power_ratio}. When the group-cluster separation decreases to less than $5R_{200,c}$, $P_2/P_0$ begins to steadily increase. The morphology of the outer part of the group's halo is more distorted than the inner regions, and this property manifests itself in the higher values of $P_2/P_0$ at the larger halo-centric radius. $P_3/P_0$ does not change significantly, as expected given the lack of any substructure within the group, until the group and cluster are separated by less than $2R_{200,c}$. $P_3/P_0$ increases beyond this as the group and cluster halos start to overlap.   

\section{Discussion}
\label{sec:discussion}
\subsection{The importance of pre-processing}

\subsubsection{Strangulation and star formation}

In this section we relate our results to observed trends that indicate pre-processing in galaxies up to (and beyond) $2-3 R_{200}$ of clusters (as described in \S\ref{sec:intro}). The ratio between the average star formation rate (SFR) of galaxies just outside a cluster and the SFR of galaxies in the field is roughly given by
\begin{equation}
f_{\rm q} = 1 - f_{\rm group} (1 - f_{\rm qi})\ ,
\end{equation}
where $f_{\rm group}$ is the fraction of galaxies that fall into clusters as members of groups and $f_{\rm qi}$ is the ratio between the average SFRs of group and field galaxies. Here we assume that only the group environment acts to quench star formation outside a cluster's virial radius and that galaxies falling into clusters directly from the field are unquenched. We also assume that all galaxies in the region just outside a cluster's virial radius eventually make their way into the cluster. If we further assume that the efficiency with which stars form from cold gas is unaffected by a galaxy's interactions with its environment, we can take $f_{\rm qi} \approx f_{\rm cold}$, the average fraction of a group galaxy's gas that is able to cool without being stripped away.

As noted in \S\ref{sec:intro}, the fraction $f_{\rm group}$ has been estimated by several authors using $N$-body simulations (\citealt{Berrier09}, \citealt{White10}, \citealt{McGee09}, \citealt{DeLucia12}), yielding values $\sim 30 - 50\%$. Thus if preprocessing were highly efficient, we would expect $f_{\rm q} \sim 0.5 - 0.7$. We may therefore take 0.5 as a lower limit on $f_{\rm q}$.

We can arrive at an upper limit for $f_{\rm q}$ by examining the efficiency which which strangulation in a group removes gas that would otherwise have cooled and formed stars during the time it takes for a group to fall from $\sim 3 R_{200}$ into a cluster. In our merger simulation, this interval is $t_{\rm ff} \sim 2$~Gyr. Our model galaxies in \S\ref{sec:pram} are isothermal, so as a galaxy evolves in time its cooling radius $r_{\rm cool}$ (defined via $t_{\rm cool}(r_{\rm cool}) = t$) increases because the gas is centrally concentrated. Assuming radiative cooling due to bremsstrahlung emission, the local cooling time at a radius $r$ about a galaxy is
\begin{equation}
 t_{\rm cool}(r) = 4.69 \left(\frac{n_e(r)}{10^{-3}~\mbox{cm}^{-3}}\right)^{-1} \left(\frac{T(r)}{10^{6}~\mbox{K}}\right)^{1/2} \mbox{Gyr}\ ,
 \label{eqn:tcool}
\end{equation}
where $n_e(r)$ is the electron number density and $T(r)$ is the gas temperature. (In reality the plasma cooling curve rises below $T \sim 10^6$~K, so this is an upper limit to $t_{\rm cool}(r)$.) Gas inside $r_{\rm cool}$ rapidly cools and condenses, forming stars. Meanwhile, ram pressure reduces $r_{\rm ram}$ from its initial value of the galaxy's virial radius. We can define the time $t_{\rm se}$ at which strangulation (removal of hot gas) ends when $r_{\rm ram}(t_{\rm se}) = r_{\rm cool}(t_{\rm se})$. If we neglect ram pressure stripping of the cooled gas inside this radius, and if $t_{\rm se} \lesssim t_{\rm ff} \sim 2$~Gyr, then $f_{\rm cold}$ is simply the ratio of the gas mass enclosed within $r_{\rm cool}(t_{\rm se})$ and the total initial gas mass.

\begin{figure*}
  \begin{center}
    \subfigure[Cooling timescale]
    {\includegraphics[width=3.4in]{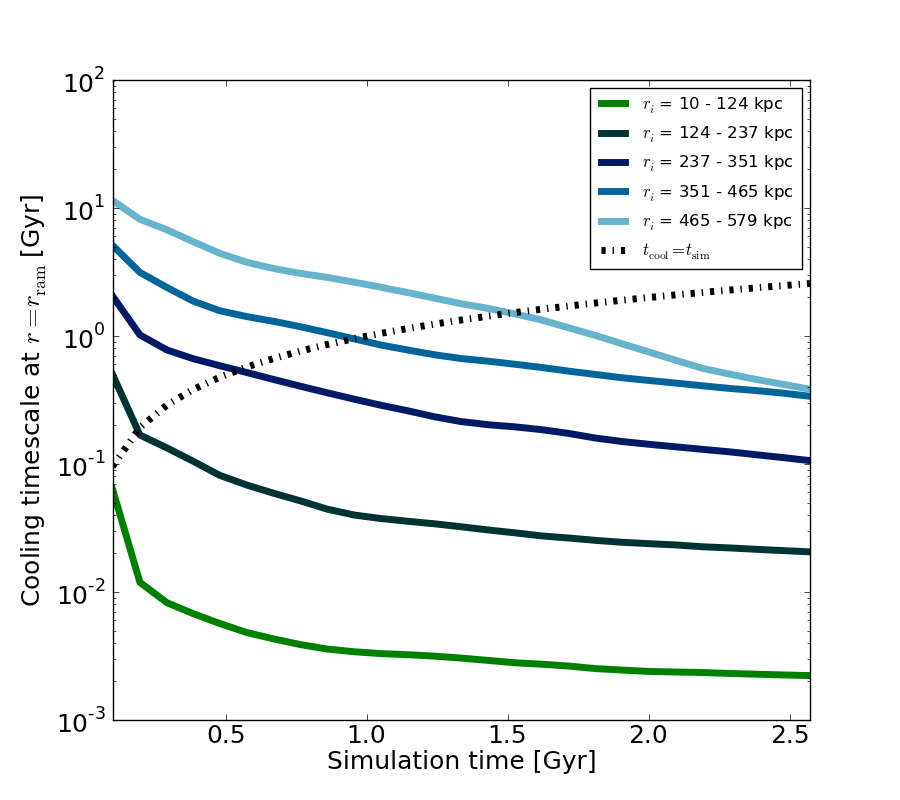}\label{fig:tcool_tsim}}
    \subfigure[Fraction of hot gas lost]
    {\includegraphics[width=3.4in]{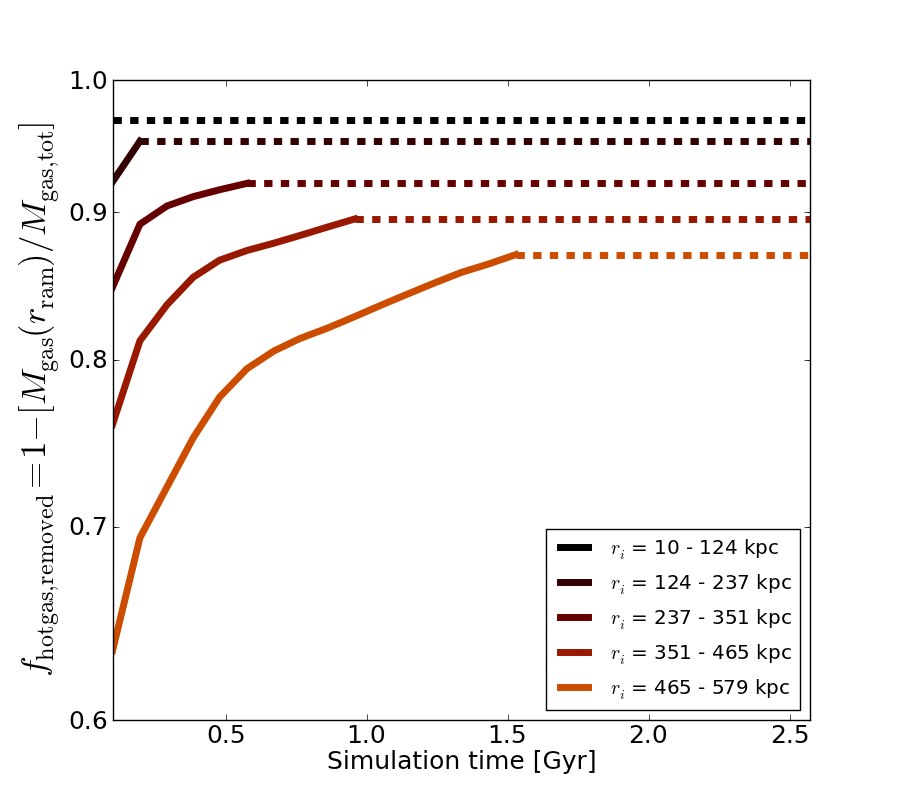}\label{fig:fhotgas}}
    \caption{Left: The evolution of the gas cooling timescale at the average $r_{\rm ram}$ for galaxies in each of five group-centric radial bins in the isolated group. The black dotted line indicates $t_{\rm cool}(r_{\rm ram}(t)) = t$. Right: The fraction of gas lost by galaxies in different radial bins due to strangulation. After $t = t_{\rm se}$ in each bin, the remaining gas has cooled and strangulation ends. The total fractions of gas removed by $t = t_{\rm se}$ are indicated using horizontal lines. \label{fig:strang_tcool}}
  \end{center}  
\end{figure*}

To determine whether strangulation operates quickly enough to affect star formation in group galaxies near a cluster, in Figure~\ref{fig:tcool_tsim} we plot the cooling time for our model galaxy ensemble in different radial bins within the isolated group. For each radial bin at a given time, the cooling time is computed by evaluating equation~\ref{eqn:tcool} at the average ram pressure radius $r_{\rm ram}$ for that bin. Where each curve intersects the line $t_{\rm cool} = t$ indicates the point beyond which all of the hot gas may be considered removed. (Note that since equation~\ref{eqn:tcool} overestimates the cooling time, in reality these intersections should come at earlier times.) Even in the outermost radial bin, all hot gas is removed well before the group reaches the cluster's virial radius in the merger simulation. Thus we argue that strangulation should affect the amount of cold gas available for star formation in groups just outside a cluster's virial radius. (Note that here we are neglecting the replenishment of hot gas through, for example, supernova feedback.)

In Figure~\ref{fig:fhotgas} we plot the fraction of hot gas that is removed as a function of time for model galaxies in the isolated group. Because strangulation operates most rapidly for galaxies near the center of the group, $r_{\rm ram}$ reaches the cooling radius most rapidly for these galaxies, and therefore they lose the most gas (97\% of their original amount). We see that even the outermost galaxies in the group lose 87\% of their gas before strangulation ends. Thus strangulation is very effective; even well outside a cluster's virial radius, galaxies in infalling groups may have less than 15\% of the cold gas they would otherwise have had in the field. This is somewhat less than the more detailed simulation results of \citet{McCarthy08} would suggest ($\sim 2-5\times$ reduction in hot gas), but given the uncertainties in our model it is reasonably consistent.

Combining this value for $f_{\rm cold}$ with our range of values for $f_{\rm group}$, we thus find that $f_{\rm q}$ should lie between 0.5 and 0.75. Clearly this is a crude estimate, but it shows that strangulation in infalling galaxy groups can plausibly explain the observed SFR suppression in regions just outside cluster virial radii. To more accurately calculate this suppression requires that we actually track gas cooling in galaxy halos; in particular, including stellar feedback and realistic stripping of cold gas should be important for improving the upper bound on $f_{\rm q}$. The lower bound depends on improved identification of galaxies in cosmological $N$-body simulations.
We leave a more detailed observational comparison, including suppression as a function of cluster-centric radius and the fraction of galaxies undergoing any star formation, for future work.

\subsubsection{Galaxy merging in the infalling group}

Our calculated values of $t_{\rm merg}$ and $t_{\rm coll}$, as seen in Figure~\ref{fig:tcoll_tmerg_gal}, show that the isolated group's merger and collision timescales are comparable to the group's dynamical timescale of $\sim 2.3 \times 10^3$ Myr\footnote{$t_{\rm dyn} = \sqrt{3\pi/32 G \overline{\rho}}$, $\overline{\rho} = 200 \rho_{\rm crit}$}. The isolated cluster's merger timescale is about twice as large. Because the group's merger timescale is comparable to or smaller than the amount of time required for the group itself to fall through the cluster's outskirts, it is reasonable to expect that many group galaxies will have undergone at least one merger by the time the group reaches the pericenter of its orbit about the cluster. 

Galaxy mergers have a more complex range of outcomes than the competition between strangulation and radiative cooling discussed in the previous section. Our simple galaxy particle model thus cannot address the minimum fraction of cluster galaxies that should be early-type or gas-poor because of major mergers in a previous group environment. However, given the expected prevalence of mergers between group members during group infall and the range of estimates of $f_{\rm group}$, at most half of cluster galaxies should have undergone `classic' pre-processing as members of groups during group infall. Since the large majority of cluster galaxies are of early type, this suggests that processes other than, or in addition to, major mergers prior to cluster infall must be responsible.

\subsection{The impact of the merger and post-merger evolution}

\subsubsection{Galaxy-galaxy interaction rates}
\label{sec:ggint}

As seen in the velocity space structure of the merging group and cluster in the cosmological simulation (Figures~\ref{fig:fig4} and~\ref{fig:fig5}), groups that merge with clusters can remain coherent in velocity space over timescales longer than those for which they remain gravitationally bound (Figure~\ref{fig:fig3}). This phenomenon of `dynamical coherence' or `coherence of substructure' has been studied in previous numerical simulations by \citet{White10} and \citet{Cohn12}. In our idealized simulation, we calculate the timescale over which this coherence holds, and we see that the velocities of the group's components remain coherent until after the group makes its first pericentric passage and moves to the apocenter of its orbit. The exact coherence period depends on a galaxy's initial position within the group; the group's core components alone are coherent up to the second pericentric passage at $t \simeq$ 4 Gyr, as seen in Figures~\ref{fig:vcgccc} and~\ref{fig:vcgcch}. 

The average number of galaxies available for collision, as well as the average relative galaxy velocity, increases dramatically during the group's first pericentric passage within the cluster. This in turn leads to a significant decrease in collision timescales during the first passage for both group and cluster galaxies. On the other hand, there is no corresponding effect in the merger timescales of cluster galaxies, and even for group galaxies the decrease in merger timescale is modest. This is because the mean difference between the velocities of group and cluster particles is almost twice the overall velocity dispersion (as seen in Figure~\ref{fig:figvc1}). Therefore, although the group galaxies see an average increase in local density due to the presence of cluster galaxies during this time, the velocities of these galaxies do not satisfy $v_{\rm rel} < 3\sigma_{\rm gal}$, and thus these galaxies cannot merge with group galaxies. 

\begin{figure*}
  \begin{center}
    \subfigure[Group]
    {\includegraphics[width=3.4in]{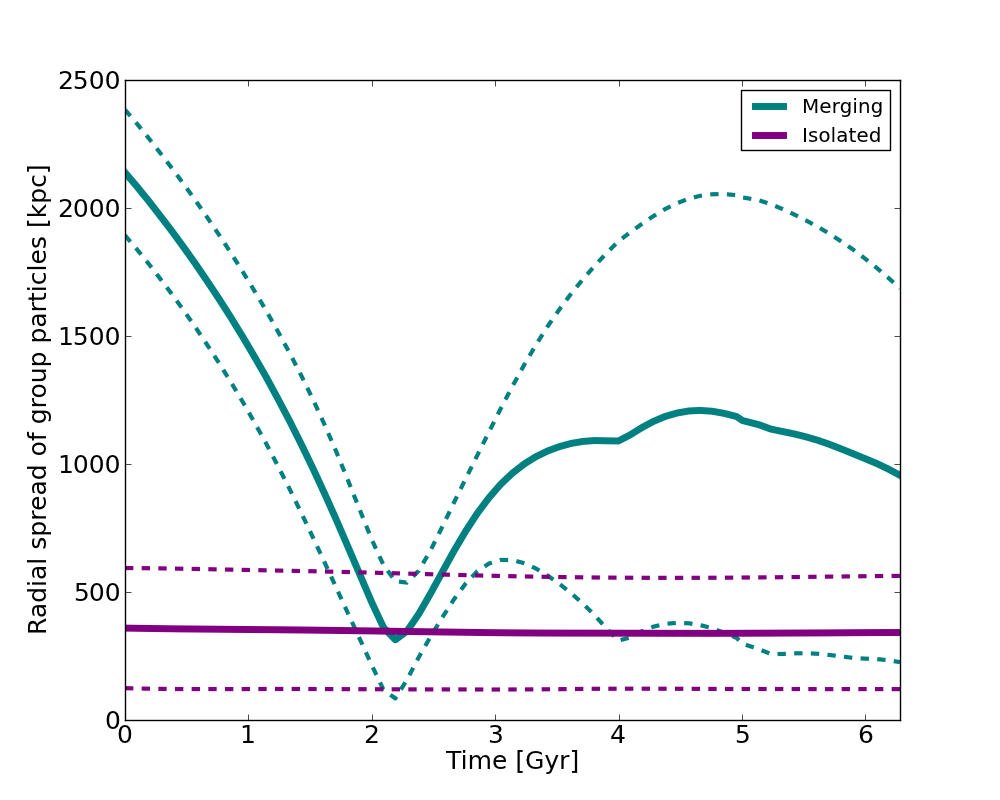}\label{fig:group_meanr}}
    \subfigure[Cluster]
    {\includegraphics[width=3.4in]{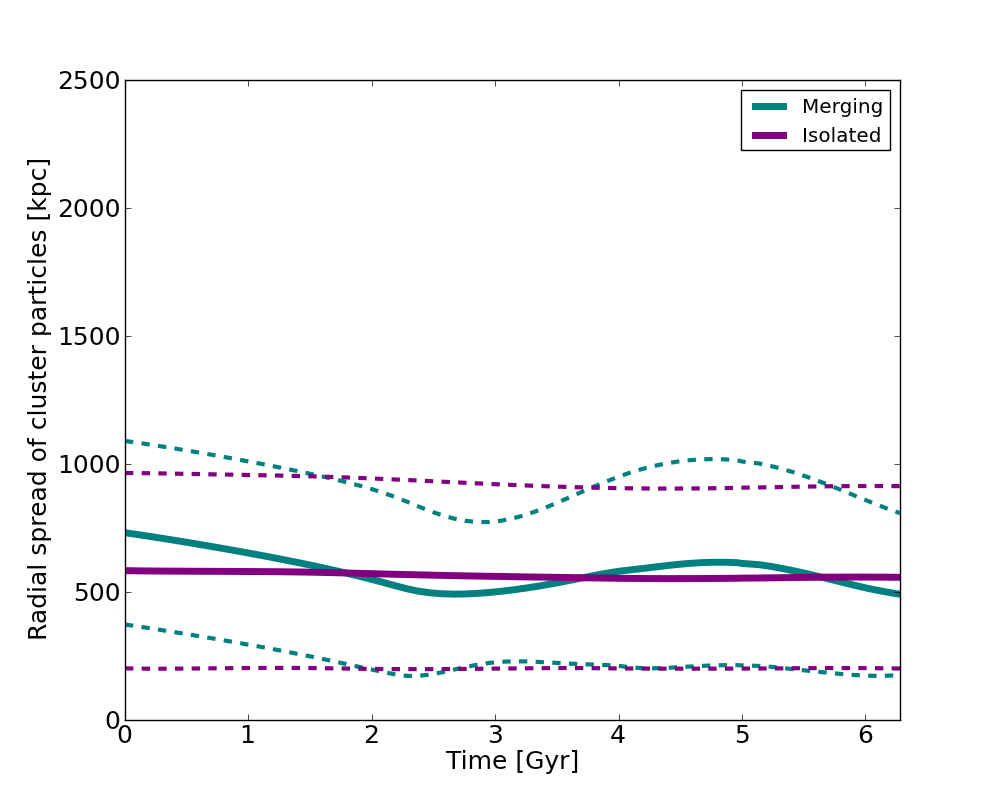}\label{fig:cluster_meanr}}
    \caption{The radial distribution of group and cluster particles with respect to the merging and isolated systems' centers of mass. Solid lines correspond to the mean radial distance from the center of mass, and dashed lines correspond to the $1\sigma$ spread in radial distances. Teal lines correspond to the radial distribution of merging group or cluster particles, and purple lines correspond to those in an isolated halo. \label{fig:gc_meanr}}
  \end{center}  
\end{figure*}

Despite the extended period of velocity coherence, at late times the group galaxies have much larger collision and merger timescales than the cluster galaxies. This is because at late times the group galaxies, on average, live in lower density environments compared to the cluster galaxies. This is illustrated in Figure~\ref{fig:gc_meanr}, which shows the mean and $1\sigma$ spread in radial distances of group and cluster particles (calculated with respect to the center of mass) in both the merging and isolated systems. Thus intra-group velocity coherence should only allow for enhanced merger rates inside the cluster for a short time near the group's first pericentric passage.

Analyses of cosmological simulations by \citet{Wetzel09} and \citet{Angulo09} have shown that subhalos of infalling groups sometimes merge with the central halo of their original host group rather than becoming cluster satellites. \citet{Wetzel09} constructed halo and subhalo merger trees from a cosmological simulation and analyzed subhalo merger rates. They found that subhalo merger rates decrease with redshift, and estimated that a $10^{11} - 10^{12} ~\mbox{M}_{\odot}$ subhalo undergoes $\sim 0.6$ mergers per Gyr at redshift 5 and $\sim 0.05$ mergers per Gyr at a redshift of 0.6. Their analysis was independent of host halo mass. \citet{Angulo09} also analyzed subhalo-subhalo merger rates in a cosmological simulation. They found that a subhalo (of mass $M_{\rm sub} \simeq 0.01 M_{\rm host}$) in a host halo of mass $10^{14} ~\mbox{M}_{\odot}$ has roughly a $\sim 10\%$ probability of undergoing a merger within a Hubble time, and that satellite-satellite mergers are as likely as satellite-central mergers. The parameters of our model group and cluster galaxies are $R_{200} = 100$~kpc and $M_{200} = 1.7 \times 10^{11} ~\mbox{M}_{\odot}$, and their merger timescales in the isolated group and cluster are 3~Gyr and 6~Gyr respectively, corresponding to merger rates of 0.33~Gyr$^{-1}$ and 0.17~Gyr$^{-1}$. However, dynamical friction, which can drive the merger of a merged group's satellites with its central galaxy, is not accounted for in our calculation. We do not have actual galaxies, but rather galaxy particles, which are dark matter particles tracing the orbits of galaxies. Our calculation does not account for tidal truncation and decreased cross sections, nor velocity bias of galaxies with respect to dark matter. These latter effects will lead to increased merger and collision timescales; therefore our results place lower limits on these timescales under ideal conditions. 

\subsubsection{Strangulation and tidal truncation}

The increased ram pressure on the infalling group during each of its pericentric passages results in an increase in gas stripping and a decrease in stripping radius, as seen in Figure~\ref{fig:req_mgroup}. At the first pericentric passage, the merger shock-driven increase in ram pressure is strong enough to remove practically all of the group galaxies' diffuse hot gas: the upper limit on the stripping radius for a group galaxy after the pericentric passage is less than 1~kpc. This effect extends to galaxies that are at large halo-centric distances. Thus, galaxies that were recently swept up by groups just before cluster infall can also be stripped of their hot gas by the increased shock-driven ram pressure. The merger and the infall shock also result in an increase in the average ram pressure on the cluster's galaxies, as seen in Figure~\ref{fig:pram_merg_iso}, particularly those in the central regions of the cluster, during successive pericentric passages of the group. This results in increased gas removal from cluster galaxies (as seen on comparing Figures~\ref{fig:req_mcluster} and \ref{fig:req_icluster}), particularly during the group's first pericentric passage. The largest value of the stripping radius of all merging cluster galaxies is only $\sim 1.5$~kpc at the end of the simulation, compared to a few kpc for those in the isolated cluster, since the latter galaxies are not subject to any significant periods of increased ram pressure.

The merging group's galaxies are subject to increased tidal truncation relative to the isolated group's galaxies (Figures~\ref{fig:rtid_mgroup} and \ref{fig:rtid_igroup}). The overall increase in the background halo density as the group falls into the cluster's deep potential well and moves past the cluster's center results in a decrease in tidal truncation radii. Notably, the tidal radius does not decrease significantly after the first pericentric passage, since the only significant large-scale background density enhancement occurs when the group core passes through the cluster core. Our calculation, however, does not account for any possible recapturing of material by galaxies that travel out of their orbital pericenters. The group-cluster merger does not affect the tidal truncation radius of the cluster galaxies to the same extent as it does the group's galaxies (Figure~\ref{fig:rtid_migroup_micluster}). This is because a smaller fraction of the cluster's galaxies (compared to most of the infalling group's galaxies, which are on radial orbits, even after being unbound from the group) feel the effect of the increased local density when the group passes through the cluster's center on its first pericentric passage.

We note here that although our galaxy models are relatively crude (uniform population of galaxies, galactic response to and re-equilibration following stripping and tidal truncation not accounted for), our estimates of strangulation and tidal truncation rates compare favorably with observations. Late-type spiral and early-type elliptical galaxies in the field can have hot gaseous halos extending up to tens of kpc or even $\sim 100$~kpc (\citealt{Forman85}, \citealt{Li07}, \citealt{Anderson11}, \citealt{Li13}, \citealt{Anderson13}). Recent studies also show the presence of X-ray gas halos around galaxies in groups and clusters, and these halos are smaller than those in the field. \citet{Vikhlinin01} studied the X-ray coronae of the two dominant galaxies in the Coma cluster and found that these galaxies had X-ray emitting coronae of $\sim 3$~kpc. \citet{Sun07}, from a study of X-ray coronae in 179 galaxies in 25 clusters, showed that most early-type galaxies had hot X-ray halos extending out to $\sim 1.5 - 4$~kpc, and diffuse X-ray emission was detected in $\sim 35\%$ of late-type galaxies. \citet{Jeltema08} studied the hot gas content of 13 galaxy groups and detected X-ray halos in more than 80\% of luminous group galaxies. They also found that a higher fraction of group galaxies have detectable hot gas halos than cluster galaxies, and that group and cluster galaxies have fainter X-ray halos compared to field galaxies. These results are consistent with a scenario where groups are less efficient at strangulation than clusters; however galaxies in both groups and clusters will be gas-poor compared to field galaxies. 

Comparing our estimates of truncation radius to observations is not as straightforward due to observational difficulties in accurately estimating the radii of dark matter halos and subhalos. However, recent estimates using gravitational lensing have made some progress. \citet{Okabe13} detected 32 subhalos in the Coma cluster using Subaru/Suprime-Cam and estimated the truncation radius of these subhalos. They found that subhalo mass and truncation radius tends to decrease with decreasing halo-centric radius, as expected in a model where tidal stripping is most effective in dense cluster cores. \citet{Gillis13} studied satellite galaxies in galaxy groups in the CFHTLens survey and found that galaxies in high-density environments are less massive than those in low-density environments by a factor of 0.65, and that this factor can be as low as 0.41 for satellite galaxies. For satellite galaxy masses of $\sim 5.9 \times 10^{11} ~\mbox{M}_{\odot}$, they estimate tidal truncation radii of $\sim 40 \pm 21$~kpc. The truncation radii of our satellite galaxy models in the group and cluster, which have masses of $1.7 \times 10^{11} ~\mbox{M}_{\odot}$, are $\sim 20 - 50$~kpc for the isolated group and $\sim 10 - 50$~kpc in the isolated cluster. 

\subsection{Limitations and uses of our models}

Our simulations do not consider the evolution of actual galaxies in groups and clusters, but rather tag randomly selected particles with model galaxies and examine the environment experienced by these model galaxies along the corresponding particle trajectories. However, subhalos (and galaxies) in clusters have a velocity bias with respect to the dark matter, as mentioned in \S\ref{sec:vcoh}. \citet{Diemand04} showed that the velocity bias $b$ of galaxies in clusters can range from an average of $\sim 1.12 \pm 0.04$ to greater than $1.3$ in the centers of clusters. Galaxies that are on average faster than the dark matter particles in a cluster will have smaller collision and merger timescales than we have measured. However, the effect, if any, of velocity bias on velocity coherence within merging subclusters is less obvious. We leave an investigation of it for future work.

Our model galaxies also do not experience dynamical friction as would be expected for real galaxies. We therefore expect the true rate of mergers of satellite and central galaxies to be higher than we predict. Ram pressure stripping and tidal truncation should also be more effective for a larger number of galaxies since galaxies should experience higher densities than typical dark matter particles.

Our estimates of strangulation due to ram pressure do not consider any additional input of gas from a galaxy after removal. In reality, galaxies may have outflows that can replenish the gaseous halo. Our calculation does not account for the cold gaseous disk component of disk galaxies that fall ino clusters. The removal of cold gas from disk galaxies due to ram pressure will depend on the inclination angle with respect to the galaxy's orbit (\citealt{Roediger06}). Additionally, removal of cold gas will have a more immediate impact on star formation rates. In fact, some observations show that ram pressure stripping due to a cluster can briefly \emph{enhance} star formation: \citet{Owers12} studied 4 galaxies in a merging cluster and found star-forming knots in gas tails stripped from the galaxies. Interestingly, these galaxies line up with a shock front, suggesting that the enhanced ram pressure due to the merger shock could both strip these galaxies of gas and enhance their star formation rates. 

An additional limitation of our models is that they assume a uniform population of galaxies. Real galaxies in clusters encompass a range of masses, morphologies, and gas fractions. As discussed earlier, a decrease (increase) in galaxy cross sections will lead to increased (decreased) collision and merger times. However, the effect of a positive velocity bias for galaxies will lead to a decrease in interaction times. The relative importance of these seemingly opposite effects will affect real galaxy-galaxy collision and merger rates. In future work we will consider realistic subhalos in clusters and will account for these effects.

\section{Summary and conclusions}
\label{sec:conclusions}
We studied a group-cluster merger in a cosmological simulation and performed an idealized controlled resimulation of the same merger to understand the importance of and the role played by the group environment in the evolution of cluster galaxies. We have quantified pre-processing due to the group environment on galaxies before cluster infall, post-processing in the coherent bound environment of the group within the cluster, and the impact of the merger on the cluster itself. 

Our cosmological simulation indicated that infalling groups appear to be tidally distorted by the massive cluster and are stretched out as they fall along cosmological filaments. However, an idealized resimulation involving an infalling spherically symmetric group showed that the tidal distortion of the group's density profile is not significant within the group's virial radius. We also found that infalling groups can sweep up some field galaxies in the vicinity of the cluster, and these can consequently undergo a brief pre-processing period. We find that most of the merging group's outer halo particles and subhalos are gravitationally unbound from the group and bound to the cluster before the group's first pericentric passage. These include the group's most recently accreted satellites. However, these stripped components are still coherent in velocity space even after being gravitationally unbound, and they orbit within the cluster on radial orbits.

To quantitatively study the importance of pre-processing and post-processing, we simulated a group and a cluster with collisionless dark matter particles and adiabatic gas initially in hydrostatic equilibrium. We allowed these halos to evolve in equilibrium and also to merge. We showed that pre-processing can play an important role in the evolution of galaxies that end up in clusters. In particular, ram pressure on group galaxies can be strong enough to strip their hot gas halos and inhibit star formation even before their host group has passed inside the virial radius of a cluster. Galaxy-galaxy mergers within groups are about twice as frequent as in clusters, and the merger timescale inside an infalling group is comparable to the amount of time required for the group to fall $\sim 2-3 \times$ the cluster's virial radius. Thus even a recently accreted group galaxy can undergo a merger event before the group enters the cluster. Tides in the group are not as effective in truncating the radii of group galaxies as they are within the cluster.

The merger has several effects on both the group and cluster. The velocities of the merging group's components are coherent past the group's first pericentric passage. After one orbital period, the group galaxies' velocity dispersion reaches a steady value comparable to that of the cluster galaxies, suggesting that the group has become virialized within the cluster. When the infalling group on its radial orbit reaches its pericenter near the cluster's potential minimum, the increased local galaxy density leads to an increase in galaxy-galaxy collision and merger rates. However, after the pericentric passage, the group's galaxies are on average in lower density environments and consequently have longer merger and collision timescales.  The merger also affects the cluster itself. There is an increase in the cluster's galaxy-galaxy collision rates as the dense group passes through the cluster. The merger rate of cluster galaxies is not affected during the group's pericentric passage because of the high relative velocities of the group and cluster galaxies.

We also see that the merger shock due to the infalling group leads to an increase in ram pressure on the group's galaxies and consequently a significant decrease in the stripping radii of their hot gaseous halos. This strangulation can inhibit future star formation within the cluster. Although there are periodic episodes of increased ram pressure on the group's components corresponding to the group's pericentric passages, these cannot cause further strangulation as the group galaxies have already lost most of their hot gas. There is also an increase in the ram pressure on the cluster galaxies due to the merger and the merger shock: most of the ram pressure stripping of the gaseous halos of cluster galaxies happens when the group initially falls into the cluster and passes through the center of the cluster. The increased local density as the group's galaxies pass through the cluster's center during their radial orbits results in the tidal truncation of their halos. However, the merger does not modify the truncation radii of the cluster galaxies. 

The primary purpose of this paper is to quantify the relative importance of group and cluster environments before and during a merger on the evolution of their galaxies. We have shown that interaction rates can be enhanced during a merger, but only up to the first pericentric passage. We have also calculated a timescale for velocity coherence of galaxies in an infalling group; in combination with an estimate of group-cluster merger rates, this can be used to estimate the possibility of detection of substructure in velocity space within clusters. We also show that a group-cluster merger can affect cluster galaxies themselves: galaxies in clusters that undergo one or more major mergers in their evolutionary history can be subject to more transformation processes than those in clusters that evolve quiescently. Our simple estimates of stripping radii and tidal truncation radii agree well with observations. The increase in ram pressure due to a merger shock and the consequently enhanced stipping of gas will have observational consequences, as in the recent paper by \citet{Owers12}. Observations of the gas in galaxies in clusters that are undergoing major mergers, especially those that are aligned with shock features, can help in further understanding the effect of a merger on gas bound to galaxies. 

\section*{Acknowledgments}

The simulations presented here were carried out using the NSF XSEDE Kraken system at the National Institute for Computational Sciences under allocation TG-AST040034N. FLASH was developed largely by the DOE-supported ASC/Alliances Center for Astrophysical Thermonuclear Flashes at the University of Chicago. We thank Paul Sutter for use of his cosmological simulation data and the anonymous referee for several suggestions that greatly improved the paper.

\bibliography{ms}

\begin{thebibliography}{}

\bibitem[\protect\citeauthoryear{{Aguerri} \& {S{\'a}nchez-Janssen}}{{Aguerri}
  \& {S{\'a}nchez-Janssen}}{2010}]{Aguerri10}
{Aguerri} J.~A.~L.,  {S{\'a}nchez-Janssen} R.,  2010, A\&A, 521, A28

\bibitem[\protect\citeauthoryear{{Anderson} \& {Bregman}}{{Anderson} \&
  {Bregman}}{2011}]{Anderson11}
{Anderson} M.~E.,  {Bregman} J.~N.,  2011, ApJ, 737, 22

\bibitem[\protect\citeauthoryear{{Anderson}, {Bregman} \& {Dai}}{{Anderson}
  et~al.}{2013}]{Anderson13}
{Anderson} M.~E.,  {Bregman} J.~N.,    {Dai} X.,  2013, ApJ, 762, 106

\bibitem[\protect\citeauthoryear{{Andrade-Santos}, {Nulsen}, {Kraft}, {Forman},
  {Jones}, {Churazov} \& {Vikhlinin}}{{Andrade-Santos}
  et~al.}{2013}]{Andrade-Santos13}
{Andrade-Santos} F.,  {Nulsen} P.~E.~J.,  {Kraft} R.~P.,  {Forman} W.~R.,
  {Jones} C.,  {Churazov} E.,    {Vikhlinin} A.,  2013, ApJ, 766, 107

\bibitem[\protect\citeauthoryear{{Angulo}, {Lacey}, {Baugh} \&
  {Frenk}}{{Angulo} et~al.}{2009}]{Angulo09}
{Angulo} R.~E.,  {Lacey} C.~G.,  {Baugh} C.~M.,    {Frenk} C.~S.,  2009, MNRAS,
  399, 983

\bibitem[\protect\citeauthoryear{{Balogh}, {Navarro} \& {Morris}}{{Balogh}
  et~al.}{2000}]{Balogh00}
{Balogh} M.~L.,  {Navarro} J.~F.,    {Morris} S.~L.,  2000, ApJ, 540, 113

\bibitem[\protect\citeauthoryear{{Barnes} \& {Hernquist}}{{Barnes} \&
  {Hernquist}}{1992}]{Barnes92}
{Barnes} J.~E.,  {Hernquist} L.,  1992, ARA\&A, 30, 705

\bibitem[\protect\citeauthoryear{{Bekki}}{{Bekki}}{1999}]{Bekki99}
{Bekki} K.,  1999, ApJL, 510, L15

\bibitem[\protect\citeauthoryear{{Berrier}, {Stewart}, {Bullock}, {Purcell},
  {Barton} \& {Wechsler}}{{Berrier} et~al.}{2009}]{Berrier09}
{Berrier} J.~C.,  {Stewart} K.~R.,  {Bullock} J.~S.,  {Purcell} C.~W.,
  {Barton} E.~J.,    {Wechsler} R.~H.,  2009, ApJ, 690, 1292

\bibitem[\protect\citeauthoryear{{Binney} \& {Tremaine}}{{Binney} \&
  {Tremaine}}{2008}]{Binney08}
{Binney} J.,  {Tremaine} S.,  2008, {Galactic Dynamics: Second Edition}.
Princeton University Press

\bibitem[\protect\citeauthoryear{{Br{\"u}ggen} \& {De Lucia}}{{Br{\"u}ggen} \&
  {De Lucia}}{2008}]{Bruggen08}
{Br{\"u}ggen} M.,  {De Lucia} G.,  2008, MNRAS, 383, 1336

\bibitem[\protect\citeauthoryear{{Buote} \& {Tsai}}{{Buote} \&
  {Tsai}}{1995}]{Buote95}
{Buote} D.~A.,  {Tsai} J.~C.,  1995, ApJ, 452, 522

\bibitem[\protect\citeauthoryear{{Cavagnolo}, {Donahue}, {Voit} \&
  {Sun}}{{Cavagnolo} et~al.}{2009}]{Cavagnolo09}
{Cavagnolo} K.~W.,  {Donahue} M.,  {Voit} G.~M.,    {Sun} M.,  2009, ApJS, 182,
  12

\bibitem[\protect\citeauthoryear{{Chandrasekhar}}{{Chandrasekhar}}{1933}]{Chandrasekhar33}
{Chandrasekhar} S.,  1933, MNRAS, 93, 449

\bibitem[\protect\citeauthoryear{{Coe}, {Ben{\'{\i}}tez}, {Broadhurst} \&
  {Moustakas}}{{Coe} et~al.}{2010}]{Coe10}
{Coe} D.,  {Ben{\'{\i}}tez} N.,  {Broadhurst} T.,    {Moustakas} L.~A.,  2010,
  ApJ, 723, 1678

\bibitem[\protect\citeauthoryear{{Cohn}}{{Cohn}}{2012}]{Cohn12}
{Cohn} J.~D.,  2012, MNRAS, 419, 1017

\bibitem[\protect\citeauthoryear{{Colella} \& {Woodward}}{{Colella} \&
  {Woodward}}{1984}]{Colella84}
{Colella} P.,  {Woodward} P.~R.,  1984, Journal of Computational Physics, 54,
  174

\bibitem[\protect\citeauthoryear{{Col{\'{\i}}n}, {Klypin} \&
  {Kravtsov}}{{Col{\'{\i}}n} et~al.}{2000}]{Colin00}
{Col{\'{\i}}n} P.,  {Klypin} A.~A.,    {Kravtsov} A.~V.,  2000, ApJ, 539, 561

\bibitem[\protect\citeauthoryear{{De Lucia}, {Weinmann}, {Poggianti},
  {Arag{\'o}n-Salamanca} \& {Zaritsky}}{{De Lucia} et~al.}{2012}]{DeLucia12}
{De Lucia} G.,  {Weinmann} S.,  {Poggianti} B.~M.,  {Arag{\'o}n-Salamanca} A.,
    {Zaritsky} D.,  2012, MNRAS, 423, 1277

\bibitem[\protect\citeauthoryear{{Diemand}, {Moore} \& {Stadel}}{{Diemand}
  et~al.}{2004}]{Diemand04}
{Diemand} J.,  {Moore} B.,    {Stadel} J.,  2004, MNRAS, 352, 535

\bibitem[\protect\citeauthoryear{{Dressler}}{{Dressler}}{1980}]{Dressler80}
{Dressler} A.,  1980, ApJ, 236, 351

\bibitem[\protect\citeauthoryear{{Dressler} \& {Shectman}}{{Dressler} \&
  {Shectman}}{1988}]{Dressler88}
{Dressler} A.,  {Shectman} S.~A.,  1988, AJ, 95, 985

\bibitem[\protect\citeauthoryear{Dubey, Reid \& Fisher}{Dubey
  et~al.}{2008}]{Dubey08}
Dubey A.,  Reid L.~B.,    Fisher R.,  2008, Physica Scripta, 2008, 014046

\bibitem[\protect\citeauthoryear{{Dubois}, {Pichon}, {Devriendt}, {Silk},
  {Haehnelt}, {Kimm} \& {Slyz}}{{Dubois} et~al.}{2013}]{Dubois13}
{Dubois} Y.,  {Pichon} C.,  {Devriendt} J.,  {Silk} J.,  {Haehnelt} M.,  {Kimm}
  T.,    {Slyz} A.,  2013, MNRAS, 428, 2885

\bibitem[\protect\citeauthoryear{{Eddington}}{{Eddington}}{1916}]{Eddington16}
{Eddington} A.~S.,  1916, MNRAS, 76, 572

\bibitem[\protect\citeauthoryear{{Einasto}, {Tago}, {Saar}, {Nurmi}, {Enkvist},
  {Einasto}, {Hein{\"a}m{\"a}ki}, {Liivam{\"a}gi}, {Tempel}, {Einasto},
  {Mart{\'{\i}}nez}, {Vennik} \& {Pihajoki}}{{Einasto}
  et~al.}{2010}]{Einasto10}
{Einasto} M.,  {Tago} E.,  {Saar} E.,  {Nurmi} P.,  {Enkvist} I.,  {Einasto}
  P.,  {Hein{\"a}m{\"a}ki} P.,  {Liivam{\"a}gi} L.~J.,  {Tempel} E.,  {Einasto}
  J.,  {Mart{\'{\i}}nez} V.~J.,  {Vennik} J.,    {Pihajoki} P.,  2010, A\&A,
  522, A92

\bibitem[\protect\citeauthoryear{{Fitchett} \& {Webster}}{{Fitchett} \&
  {Webster}}{1987}]{Fitchett87}
{Fitchett} M.,  {Webster} R.,  1987, ApJ, 317, 653

\bibitem[\protect\citeauthoryear{{Forman}, {Jones} \& {Tucker}}{{Forman}
  et~al.}{1985}]{Forman85}
{Forman} W.,  {Jones} C.,    {Tucker} W.,  1985, ApJ, 293, 102

\bibitem[\protect\citeauthoryear{{Fryxell}, {Olson}, {Ricker}, {Timmes},
  {Zingale}, {Lamb}, {MacNeice}, {Rosner}, {Truran} \& {Tufo}}{{Fryxell}
  et~al.}{2000}]{Fryxell00}
{Fryxell} B.,  {Olson} K.,  {Ricker} P.,  {Timmes} F.~X.,  {Zingale} M.,
  {Lamb} D.~Q.,  {MacNeice} P.,  {Rosner} R.,  {Truran} J.~W.,    {Tufo} H.,
  2000, ApJS, 131, 273

\bibitem[\protect\citeauthoryear{{Gill}, {Knebe} \& {Gibson}}{{Gill}
  et~al.}{2004}]{Gill04}
{Gill} S.~P.~D.,  {Knebe} A.,    {Gibson} B.~K.,  2004, MNRAS, 351, 399

\bibitem[\protect\citeauthoryear{{Gillis}, {Hudson}, {Erben}, {Heymans},
  {Hildebrandt}, {Hoekstra}, {Kitching}, {Mellier}, {Miller} et~al.,}{{Gillis}
  et~al.}{2013}]{Gillis13}
{Gillis} B.~R.,  {Hudson} M.~J.,  {Erben} T.,  {Heymans} C.,  {Hildebrandt} H.,
   {Hoekstra} H.,  {Kitching} T.~D.,  {Mellier} Y.,  {Miller} L.,    et~al.,
  2013, MNRAS

\bibitem[\protect\citeauthoryear{{Gnedin}}{{Gnedin}}{2003a}]{Gnedin03b}
{Gnedin} O.~Y.,  2003a, ApJ, 589, 752

\bibitem[\protect\citeauthoryear{{Gnedin}}{{Gnedin}}{2003b}]{Gnedin03a}
{Gnedin} O.~Y.,  2003b, ApJ, 582, 141

\bibitem[\protect\citeauthoryear{{G{\'o}mez}, {Nichol}, {Miller}, {Balogh},
  {Goto}, {Zabludoff}, {Romer}, {Bernardi}, {Sheth}, {Hopkins}, {Castander},
  {Connolly}, {Schneider}, {Brinkmann}, {Lamb}, {SubbaRao} \&
  {York}}{{G{\'o}mez} et~al.}{2003}]{Gomez03}
{G{\'o}mez} P.~L.,  {Nichol} R.~C.,  {Miller} C.~J.,  {Balogh} M.~L.,  {Goto}
  T.,  {Zabludoff} A.~I.,  {Romer} A.~K.,  {Bernardi} M.,  {Sheth} R.,
  {Hopkins} A.~M.,  {Castander} F.~J.,  {Connolly} A.~J.,  {Schneider} D.~P.,
  {Brinkmann} J.,  {Lamb} D.~Q.,  {SubbaRao} M.,    {York} D.~G.,  2003, ApJ,
  584, 210

\bibitem[\protect\citeauthoryear{{Gunn} \& {Gott} III}{{Gunn} \&
  {Gott}}{1972}]{Gunn72}
{Gunn} J.~E.,  {Gott} III J.~R.,  1972, ApJ, 176, 1

\bibitem[\protect\citeauthoryear{{Hoyle}, {Masters}, {Nichol}, {Jimenez} \&
  {Bamford}}{{Hoyle} et~al.}{2012}]{Hoyle12}
{Hoyle} B.,  {Masters} K.~L.,  {Nichol} R.~C.,  {Jimenez} R.,    {Bamford}
  S.~P.,  2012, MNRAS, 423, 3478

\bibitem[\protect\citeauthoryear{{Jeltema}, {Binder} \& {Mulchaey}}{{Jeltema}
  et~al.}{2008}]{Jeltema08}
{Jeltema} T.~E.,  {Binder} B.,    {Mulchaey} J.~S.,  2008, ApJ, 679, 1162

\bibitem[\protect\citeauthoryear{{Kazantzidis}, {Magorrian} \&
  {Moore}}{{Kazantzidis} et~al.}{2004}]{Kazantzidis04}
{Kazantzidis} S.,  {Magorrian} J.,    {Moore} B.,  2004, ApJ, 601, 37

\bibitem[\protect\citeauthoryear{{Knollmann} \& {Knebe}}{{Knollmann} \&
  {Knebe}}{2009}]{Knollmann09}
{Knollmann} S.~R.,  {Knebe} A.,  2009, ApJS, 182, 608

\bibitem[\protect\citeauthoryear{{Kraft}, {Jones}, {Nulsen} \&
  {Hardcastle}}{{Kraft} et~al.}{2006}]{Kraft06}
{Kraft} R.~P.,  {Jones} C.,  {Nulsen} P.~E.~J.,    {Hardcastle} M.~J.,  2006,
  ApJ, 640, 762

\bibitem[\protect\citeauthoryear{{Kwan}, {Bhattacharya}, {Heitmann} \&
  {Habib}}{{Kwan} et~al.}{2013}]{Kwan13}
{Kwan} J.,  {Bhattacharya} S.,  {Heitmann} K.,    {Habib} S.,  2013, ApJ, 768,
  123

\bibitem[\protect\citeauthoryear{{Larson}, {Tinsley} \& {Caldwell}}{{Larson}
  et~al.}{1980}]{Larson80}
{Larson} R.~B.,  {Tinsley} B.~M.,    {Caldwell} C.~N.,  1980, ApJ, 237, 692

\bibitem[\protect\citeauthoryear{{Lewis}, {Balogh}, {De Propris}, {Couch},
  {Bower}, {Offer} et~al.,}{{Lewis} et~al.}{2002}]{Lewis02}
{Lewis} I.,  {Balogh} M.,  {De Propris} R.,  {Couch} W.,  {Bower} R.,  {Offer}
  A.,    et~al., 2002, MNRAS, 334, 673

\bibitem[\protect\citeauthoryear{{Li} \& {Wang}}{{Li} \& {Wang}}{2013}]{Li13}
{Li} J.-T.,  {Wang} Q.~D.,  2013, MNRAS, 428, 2085

\bibitem[\protect\citeauthoryear{{Li}, {Wang} \& {Hameed}}{{Li}
  et~al.}{2007}]{Li07}
{Li} Z.,  {Wang} Q.~D.,    {Hameed} S.,  2007, MNRAS, 376, 960

\bibitem[\protect\citeauthoryear{{Lu}, {Gilbank}, {McGee}, {Balogh} \&
  {Gallagher}}{{Lu} et~al.}{2012}]{Lu12}
{Lu} T.,  {Gilbank} D.~G.,  {McGee} S.~L.,  {Balogh} M.~L.,    {Gallagher} S.,
  2012, MNRAS, 420, 126

\bibitem[\protect\citeauthoryear{{Luki{\'c}}, {Heitmann}, {Habib}, {Bashinsky}
  \& {Ricker}}{{Luki{\'c}} et~al.}{2007}]{Lukic07}
{Luki{\'c}} Z.,  {Heitmann} K.,  {Habib} S.,  {Bashinsky} S.,    {Ricker}
  P.~M.,  2007, ApJ, 671, 1160

\bibitem[\protect\citeauthoryear{{MacNeice}, {Olson}, {Mobarry}, {de
  Fainchtein} \& {Packer}}{{MacNeice} et~al.}{2000}]{MacNeice00}
{MacNeice} P.,  {Olson} K.~M.,  {Mobarry} C.,  {de Fainchtein} R.,    {Packer}
  C.,  2000, Computer Physics Communications, 126, 330

\bibitem[\protect\citeauthoryear{{Markevitch}, {Ponman}, {Nulsen}
  et~al.,}{{Markevitch} et~al.}{2000}]{Markevitch00}
{Markevitch} M.,  {Ponman} T.~J.,  {Nulsen} P.~E.~J.,    et~al., 2000, ApJ,
  541, 542

\bibitem[\protect\citeauthoryear{{Mastropietro}, {Moore}, {Mayer},
  {Debattista}, {Piffaretti} \& {Stadel}}{{Mastropietro}
  et~al.}{2005}]{Mastropietro05}
{Mastropietro} C.,  {Moore} B.,  {Mayer} L.,  {Debattista} V.~P.,  {Piffaretti}
  R.,    {Stadel} J.,  2005, MNRAS, 364, 607

\bibitem[\protect\citeauthoryear{{McCarthy}, {Frenk}, {Font}, {Lacey}, {Bower},
  {Mitchell}, {Balogh} \& {Theuns}}{{McCarthy} et~al.}{2008}]{McCarthy08}
{McCarthy} I.~G.,  {Frenk} C.~S.,  {Font} A.~S.,  {Lacey} C.~G.,  {Bower}
  R.~G.,  {Mitchell} N.~L.,  {Balogh} M.~L.,    {Theuns} T.,  2008, MNRAS, 383,
  593

\bibitem[\protect\citeauthoryear{{McGee}, {Balogh}, {Bower}, {Font} \&
  {McCarthy}}{{McGee} et~al.}{2009}]{McGee09}
{McGee} S.~L.,  {Balogh} M.~L.,  {Bower} R.~G.,  {Font} A.~S.,    {McCarthy}
  I.~G.,  2009, MNRAS, 400, 937

\bibitem[\protect\citeauthoryear{{Moore}, {Katz}, {Lake}, {Dressler} \&
  {Oemler}}{{Moore} et~al.}{1996}]{Moore96}
{Moore} B.,  {Katz} N.,  {Lake} G.,  {Dressler} A.,    {Oemler} A.,  1996,
  Nature, 379, 613

\bibitem[\protect\citeauthoryear{{Moore}, {Lake} \& {Katz}}{{Moore}
  et~al.}{1998}]{Moore98}
{Moore} B.,  {Lake} G.,    {Katz} N.,  1998, ApJ, 495, 139

\bibitem[\protect\citeauthoryear{{Moore}, {Lake}, {Quinn} \& {Stadel}}{{Moore}
  et~al.}{1999}]{Moore99}
{Moore} B.,  {Lake} G.,  {Quinn} T.,    {Stadel} J.,  1999, MNRAS, 304, 465

\bibitem[\protect\citeauthoryear{{Navarro}, {Frenk} \& {White}}{{Navarro}
  et~al.}{1997}]{Navarro97}
{Navarro} J.~F.,  {Frenk} C.~S.,    {White} S.~D.~M.,  1997, ApJ, 490, 493

\bibitem[\protect\citeauthoryear{{O'Hara}, {Mohr}, {Bialek} \&
  {Evrard}}{{O'Hara} et~al.}{2006}]{OHara06}
{O'Hara} T.~B.,  {Mohr} J.~J.,  {Bialek} J.~J.,    {Evrard} A.~E.,  2006, ApJ,
  639, 64

\bibitem[\protect\citeauthoryear{{Okabe}, {Futamase}, {Kajisawa} \&
  {Kuroshima}}{{Okabe} et~al.}{2013}]{Okabe13}
{Okabe} N.,  {Futamase} T.,  {Kajisawa} M.,    {Kuroshima} R.,  2013, arXiv
  1304.2399

\bibitem[\protect\citeauthoryear{{Okabe}, {Okura} \& {Futamase}}{{Okabe}
  et~al.}{2010}]{Okabe10}
{Okabe} N.,  {Okura} Y.,    {Futamase} T.,  2010, ApJ, 713, 291

\bibitem[\protect\citeauthoryear{{Owers}, {Couch}, {Nulsen} \&
  {Randall}}{{Owers} et~al.}{2012}]{Owers12}
{Owers} M.~S.,  {Couch} W.~J.,  {Nulsen} P.~E.~J.,    {Randall} S.~W.,  2012,
  ApJL, 750, L23

\bibitem[\protect\citeauthoryear{{Poole}, {Babul}, {McCarthy}, {Sanderson} \&
  {Fardal}}{{Poole} et~al.}{2008}]{Poole08}
{Poole} G.~B.,  {Babul} A.,  {McCarthy} I.~G.,  {Sanderson} A.~J.~R.,
  {Fardal} M.~A.,  2008, MNRAS, 391, 1163

\bibitem[\protect\citeauthoryear{{Poole}, {Fardal}, {Babul}, {McCarthy},
  {Quinn} \& {Wadsley}}{{Poole} et~al.}{2006}]{Poole06}
{Poole} G.~B.,  {Fardal} M.~A.,  {Babul} A.,  {McCarthy} I.~G.,  {Quinn} T.,
  {Wadsley} J.,  2006, MNRAS, 373, 881

\bibitem[\protect\citeauthoryear{{Postman} \& {Geller}}{{Postman} \&
  {Geller}}{1984}]{Postman84}
{Postman} M.,  {Geller} M.~J.,  1984, ApJ, 281, 95

\bibitem[\protect\citeauthoryear{{Prada}, {Klypin}, {Cuesta}, {Betancort-Rijo}
  \& {Primack}}{{Prada} et~al.}{2012}]{Prada12}
{Prada} F.,  {Klypin} A.~A.,  {Cuesta} A.~J.,  {Betancort-Rijo} J.~E.,
  {Primack} J.,  2012, MNRAS, 423, 3018

\bibitem[\protect\citeauthoryear{{Rasmussen}, {Mulchaey}, {Bai}, {Ponman},
  {Raychaudhury} \& {Dariush}}{{Rasmussen} et~al.}{2012}]{Rasmussen12}
{Rasmussen} J.,  {Mulchaey} J.~S.,  {Bai} L.,  {Ponman} T.~J.,  {Raychaudhury}
  S.,    {Dariush} A.,  2012, ApJ, 757, 122

\bibitem[\protect\citeauthoryear{{Richard}, {Smith}, {Kneib}, {Ellis},
  {Sanderson}, {Pei}, {Targett}, {Sand}, {Swinbank}, {Dannerbauer}, {Mazzotta},
  {Limousin}, {Egami}, {Jullo}, {Hamilton-Morris} \& {Moran}}{{Richard}
  et~al.}{2010}]{Richard10}
{Richard} J.,  {Smith} G.~P.,  {Kneib} J.-P.,  {Ellis} R.~S.,  {Sanderson}
  A.~J.~R.,  {Pei} L.,  {Targett} T.~A.,  {Sand} D.~J.,  {Swinbank} A.~M.,
  {Dannerbauer} H.,  {Mazzotta} P.,  {Limousin} M.,  {Egami} E.,  {Jullo} E.,
  {Hamilton-Morris} V.,    {Moran} S.~M.,  2010, MNRAS, 404, 325

\bibitem[\protect\citeauthoryear{{Richstone}}{{Richstone}}{1976}]{Richstone76}
{Richstone} D.~O.,  1976, ApJ, 204, 642

\bibitem[\protect\citeauthoryear{{Ricker}}{{Ricker}}{2008}]{Ricker08}
{Ricker} P.~M.,  2008, ApJS, 176, 293

\bibitem[\protect\citeauthoryear{{Ricker} \& {Sarazin}}{{Ricker} \&
  {Sarazin}}{2001}]{Ricker01}
{Ricker} P.~M.,  {Sarazin} C.~L.,  2001, ApJ, 561, 621

\bibitem[\protect\citeauthoryear{{Roediger} \& {Br{\"u}ggen}}{{Roediger} \&
  {Br{\"u}ggen}}{2006}]{Roediger06}
{Roediger} E.,  {Br{\"u}ggen} M.,  2006, MNRAS, 369, 567

\bibitem[\protect\citeauthoryear{{Roettiger}, {Loken} \& {Burns}}{{Roettiger}
  et~al.}{1997}]{Roettiger97}
{Roettiger} K.,  {Loken} C.,    {Burns} J.~O.,  1997, ApJS, 109, 307

\bibitem[\protect\citeauthoryear{{Sijacki}, {Springel}, {Di Matteo} \&
  {Hernquist}}{{Sijacki} et~al.}{2007}]{Sijacki07}
{Sijacki} D.,  {Springel} V.,  {Di Matteo} T.,    {Hernquist} L.,  2007, MNRAS,
  380, 877

\bibitem[\protect\citeauthoryear{{Springel}, {White}, {Jenkins}, {Frenk},
  {Yoshida}, {Gao}, {Navarro}, {Thacker}, {Croton}, {Helly}, {Peacock}, {Cole},
  {Thomas}, {Couchman}, {Evrard}, {Colberg} \& {Pearce}}{{Springel}
  et~al.}{2005}]{Springel05}
{Springel} V.,  {White} S.~D.~M.,  {Jenkins} A.,  {Frenk} C.~S.,  {Yoshida} N.,
   {Gao} L.,  {Navarro} J.,  {Thacker} R.,  {Croton} D.,  {Helly} J.,
  {Peacock} J.~A.,  {Cole} S.,  {Thomas} P.,  {Couchman} H.,  {Evrard} A.,
  {Colberg} J.,    {Pearce} F.,  2005, Nature, 435, 629

\bibitem[\protect\citeauthoryear{{Sun}, {Jones}, {Forman}, {Vikhlinin},
  {Donahue} \& {Voit}}{{Sun} et~al.}{2007}]{Sun07}
{Sun} M.,  {Jones} C.,  {Forman} W.,  {Vikhlinin} A.,  {Donahue} M.,    {Voit}
  M.,  2007, ApJ, 657, 197

\bibitem[\protect\citeauthoryear{{Sutter} \& {Ricker}}{{Sutter} \&
  {Ricker}}{2010}]{Sutter10}
{Sutter} P.~M.,  {Ricker} P.~M.,  2010, ApJ, 723, 1308

\bibitem[\protect\citeauthoryear{{Vikhlinin}, {Burenin}, {Ebeling}, {Forman},
  {Hornstrup}, {Jones}, {Kravtsov}, {Murray}, {Nagai}, {Quintana} \&
  {Voevodkin}}{{Vikhlinin} et~al.}{2009}]{Vikhlinin09}
{Vikhlinin} A.,  {Burenin} R.~A.,  {Ebeling} H.,  {Forman} W.~R.,  {Hornstrup}
  A.,  {Jones} C.,  {Kravtsov} A.~V.,  {Murray} S.~S.,  {Nagai} D.,  {Quintana}
  H.,    {Voevodkin} A.,  2009, ApJ, 692, 1033

\bibitem[\protect\citeauthoryear{{Vikhlinin}, {Markevitch}, {Forman} \&
  {Jones}}{{Vikhlinin} et~al.}{2001}]{Vikhlinin01}
{Vikhlinin} A.,  {Markevitch} M.,  {Forman} W.,    {Jones} C.,  2001, ApJL,
  555, L87

\bibitem[\protect\citeauthoryear{{Wetzel}, {Cohn} \& {White}}{{Wetzel}
  et~al.}{2009}]{Wetzel09}
{Wetzel} A.~R.,  {Cohn} J.~D.,    {White} M.,  2009, MNRAS, 395, 1376

\bibitem[\protect\citeauthoryear{{White}, {Cohn} \& {Smit}}{{White}
  et~al.}{2010}]{White10}
{White} M.,  {Cohn} J.~D.,    {Smit} R.,  2010, MNRAS, 408, 1818

\bibitem[\protect\citeauthoryear{{Yang}, {Ricker} \& {Sutter}}{{Yang}
  et~al.}{2009}]{Yang09}
{Yang} H.-Y.~K.,  {Ricker} P.~M.,    {Sutter} P.~M.,  2009, ApJ, 699, 315

\bibitem[\protect\citeauthoryear{{Yang}, {Mo} \& {van den Bosch}}{{Yang}
  et~al.}{2008}]{Yang08}
{Yang} X.,  {Mo} H.~J.,    {van den Bosch} F.~C.,  2008, ApJ, 676, 248

\bibitem[\protect\citeauthoryear{{Zabludoff} \& {Mulchaey}}{{Zabludoff} \&
  {Mulchaey}}{1998}]{Zabludoff98}
{Zabludoff} A.~I.,  {Mulchaey} J.~S.,  1998, ApJ, 496, 39

\bibitem[\protect\citeauthoryear{{ZuHone}}{{ZuHone}}{2011}]{ZuHone11}
{ZuHone} J.~A.,  2011, ApJ, 728, 54

\end{thebibliography}

\end{document}